\documentclass[aps,twocolumn,10pt,prc,floatfix,showpacs,preprintnumbers,amsmath,amssymb,nofootinbib,superscriptaddress]{revtex4-1}

\makeatletter
\def\doauthor#1#2#3{%
  \ignorespaces #1 \unskip \@listcomma
  \begingroup
    #3%
  \@if@empty{#2}{\endgroup{}{}}{\endgroup{\comma@space}{}\frontmatter@footnote{#2}}%
  \space \@listand
}%
\makeatother

\usepackage{graphicx}%
\usepackage{dcolumn}%
\usepackage{bm}%
\usepackage[mathlines]{lineno}%
\usepackage[dvipsnames]{xcolor}
\usepackage[normalem]{ulem}

\usepackage{multirow}
\usepackage{booktabs}
\usepackage{diagbox}
\usepackage{comment}

\usepackage[utf8]{inputenc}
\usepackage[T1]{fontenc}
\usepackage{etoolbox}
\usepackage{mathtools,slashed}
\usepackage{cases}
\usepackage{lipsum}
\usepackage{placeins}
\usepackage{xspace}
\usepackage{siunitx}

\usepackage{caption}
\DeclareCaptionJustification{plainjust}{\leftskip=0pt\rightskip=0pt\parfillskip=0pt plus 1fil\relax}
\usepackage{chngcntr} %
\usepackage{apptools} %

\usepackage[pdfstartview=FitH,
            CJKbookmarks=true,
            bookmarksnumbered=true,
            bookmarksopen=true,
            colorlinks,
            linkcolor=blue,
            anchorcolor=blue,
            citecolor=blue,
            urlcolor=blue,
            ]{hyperref}

\newcommand\figsubref[2]{\hyperref[#1]{\ref*{#1}#2}}
\usepackage{orcidlink}

\newcommand{\Xs}{X_1,\ldots,X_n}
\newcommand{\Ys}{Y_1,\ldots,Y_m}

\newcommand{\defs}{Definition~\ref{def:syn}\xspace}

\newcommand{\bP}{\ensuremath{\bar{P}}\xspace}
\newcommand{\D}{\ensuremath{D}}
\newcommand{\Pf}{\ensuremath{P_{X_1X_2Y}}\xspace}
\newcommand{\Pff}{\ensuremath{P_{X_1X_2Y_1Y_2}}\xspace}

\newcommand\rtr{\ensuremath{{R}\!\rightarrow\!{R}}\xspace}
\newcommand\rtx{\ensuremath{{R}\!\rightarrow\!{U}_1}\xspace}
\newcommand\rty{\ensuremath{{R}\!\rightarrow\!{U}_2}\xspace}
\newcommand\rts{\ensuremath{{R}\!\rightarrow\!{S}}\xspace}
\newcommand\xtr{\ensuremath{{U}_1\!\rightarrow\!{R}}\xspace}
\newcommand\xtx{\ensuremath{{U}_1\!\rightarrow\!{U}_1}\xspace}
\newcommand\xty{\ensuremath{{U}_1\!\rightarrow\!{U}_2}\xspace}
\newcommand\xts{\ensuremath{{U}_1\!\rightarrow\!{S}}\xspace}
\newcommand\ytr{\ensuremath{{U}_2\!\rightarrow\!{R}}\xspace}
\newcommand\ytx{\ensuremath{{U}_2\!\rightarrow\!{U}_1}\xspace}
\newcommand\yty{\ensuremath{{U}_2\!\rightarrow\!{U}_2}\xspace}
\newcommand\yts{\ensuremath{{U}_2\!\rightarrow\!{S}}\xspace}
\newcommand\str{\ensuremath{{S}\!\rightarrow\!{R}}\xspace}
\newcommand\stx{\ensuremath{{S}\!\rightarrow\!{U}_1}\xspace}
\newcommand\sty{\ensuremath{{S}\!\rightarrow\!{U}_2}\xspace}
\newcommand\sts{\ensuremath{{S}\!\rightarrow\!{S}}\xspace}

\newcommand\rti{\ensuremath{{R}\!\rightarrow\!{U}_i}\xspace}
\newcommand\rtj{\ensuremath{{R}\!\rightarrow\!{U}_j}\xspace}
\newcommand\itr{\ensuremath{{U}_i\!\rightarrow\!{R}}\xspace}
\newcommand\iti{\ensuremath{{U}_i\!\rightarrow\!{U}_i}\xspace}
\newcommand\xti{\ensuremath{{U}_1\!\rightarrow\!{U}_i}\xspace}
\newcommand\itx{\ensuremath{{U}_i\!\rightarrow\!{U}_1}\xspace}
\newcommand\ity{\ensuremath{{U}_i\!\rightarrow\!{U}_2}\xspace}
\newcommand\itj{\ensuremath{{U}_i\!\rightarrow\!{U}_j}\xspace}
\newcommand\its{\ensuremath{{U}_i\!\rightarrow\!{S}}\xspace}

\newcommand\sti{\ensuremath{{S}\!\rightarrow\!{U}_i}\xspace}
\newcommand\stj{\ensuremath{{S}\!\rightarrow\!{U}_j}\xspace}

\newcommand\rtmi{\ensuremath{{R}\!\rightarrow\!{U}_{-i}}\xspace}

\newcommand\xtmi{\ensuremath{{U}_1\!\rightarrow\!{U}_{-i}}\xspace}

\newcommand{\phiid}{\ensuremath{\Phi\mathrm{ID}}\xspace}
\newcommand{\phiidig}{\ensuremath{\phiid^{\text{IG}}}\xspace}

\newcommand{\tp}{t^{\prime}}

\newtheorem{definition}{Definition}

\newcommand\yti{\ensuremath{{U}_2\!\rightarrow\!{U}_i}\xspace}
\newcommand{\rowstrut}{\vphantom{\Large (X_{1})}}
\newcommand{\DAllS}{\D(\Pff\|\bar{P}_0)}
\newcommand{\DSstar}{\D(\Pff\|P_*)}
\newcommand{\DPstarPi}{\D(P_*\|\bar{P}_i)}
\newcommand{\DPiPzero}{\D(\bar{P}_i\|\bar{P}_0)}
\newcommand{\DAllT}{\D(\Pff\|\bar{P}_0)}
\newcommand{\DSstarT}{\D(\Pff\|P^*)}
\newcommand{\DPstarPI}{\D(P^*\|\bar{P}^i)}
\newcommand{\DPIPzero}{\D(\bar{P}^i\|\bar{P}_0)}
\newcommand{\DAllFour}{\D(\Pff\|\bar{P}_0)}
\newcommand{\DFourStar}{\D(\Pff\|P^*_*)}
\newcommand{\DPstarPij}{\D(P^*_*\|\bar{P}_i^j)}
\newcommand{\DPijPzero}{\D(\bar{P}_i^j\|\bar{P}_0)}

\begin{document}

\title{An information-geometric formulation of distributed computation}

\newcommand{\eqcontr}{\thanks{These authors contributed equally to this work. \\ Corresponding author: 
\href{mailto:a.liardi@imperial.ac.uk}{a.liardi@imperial.ac.uk}}}
\author{Alberto Liardi\,\orcidlink{0000-0003-2247-8710}}\eqcontr
\affiliation{Department of Computing, Imperial College London, UK}

\author{George Blackburne\,\orcidlink{0000-0002-6289-915X}}\eqcontr
\affiliation{Department of Experimental Psychology, University College London, UK}
\affiliation{Department of Computing, Imperial College London, UK}

\author{Pedro A.M. Mediano\,\orcidlink{0000-0003-1789-5894}}
\affiliation{Department of Computing, Imperial College London, UK}
\affiliation{Division of Psychology and Language Sciences, University College London, UK}

\begin{abstract}
Understanding how complex systems perform computations, i.e.\ how they store, transmit, and transform information across their components, is a major goal across the natural sciences. However, characterising genuinely collective computations across multiple inputs and outputs that cannot be reduced to contributions from individual components remains an open problem. In this work, we introduce \phiidig, a geometric theory of information dynamics that captures irreducible collective phenomena and enables a complete decomposition of multi-input multi-output operations into interpretable computational modes, providing a complete informational profile of a system.
We first validate our framework to identify and quantify distinct computational modes in discrete and Gaussian systems with known ground truth. We then apply it to uncover the computational architecture of complex systems across synthetic and real-world scenarios, including chaotic maps, where distinct universality classes are associated with characteristic modes of computation; brain anatomy of over 100 mammalian species, revealing an evolutionary shift towards more distributed computation in the human neocortex; and financial markets, where collective computations recover the modular organisation of market sectors.
Overall, our results show how complex systems can be quantitatively investigated through multidimensional computational portraits derived from the geometry of their underlying probability distributions.
\end{abstract}

\maketitle %

\section{Introduction}

How does nature compute? From Earth's climate to the human brain and the global economy, complex systems -- whether evolved or engineered -- seem to function as engines of distributed computation~\cite{Chu2018-vl, barack2021two, Crutchfield1995-ep}. They adaptively store, transmit, and transform information across interacting spatiotemporal scales, producing global behaviours that are qualitatively richer than what any subpart of the system could achieve alone~\cite{jensen2022complexity}.

While understanding computation in systems we have designed is comparatively easier, this task is far more difficult in those whose structure and underlying rules are unknown. This challenge is exacerbated by the decentralised, stochastic, and continuous dynamics usually observed in natural complex systems~\cite{Piccinini2015-il, Mitchell2012-xo}, and by the fact that even elementary fixed dynamical systems can be viewed as performing an infinite set of computations~\cite{newman2003structure, Wolpert2026-dp, israeli2004computational, shalizi2001computational}. 
Central to this difficulty, and yet fundamental to their functioning, is the property of irreducibility: the global behaviour of distributed systems cannot be reproduced by any proper subset of its components~\cite{tumer2012collectives, malone2015handbook}. No agent acting in isolation, and no collection of agents ignoring the rest, can replicate what the whole achieves~\cite{anderson1972more, krakauer2020information}. The structural signature of this irreducibility lies in the statistical organisation of the system's components: its parts must be bound together by dependencies that do not reduce to pairwise or lower-order interactions. These \textit{higher-order}, or \textit{synergistic}, dependencies precisely constitute the internal structure that makes collective computation possible and richer than local dynamics. However, these are perhaps the hardest to quantify and are often bereft of rigorous formalisms for their conceptualisation and quantification~\cite{liardi2026statistical}.
Therefore, a crucial scientific question is how we can best characterise such irreducible structures and reveal the distributed information architecture of complex systems~\cite{Crutchfield1994-dg, lizier2012local}.

A promising approach to this challenge is offered by information geometry, the study of families of probability distributions as smooth geometric objects endowed with a differential structure~\cite{amari2016information}. 
By treating distributions as points on statistical manifolds, information geometry provides model-agnostic tools that enable the separation of different statistical features in a mathematically rigorous manner. 
Within this field, for example, multivariate dependencies can be decomposed into interaction orders by projecting distributions onto submanifolds defined by constraints on marginals or sufficient statistics, yielding a clear interpretation of higher-order interactions as deviations from lower-order models~\cite{amari2001information}.

Previous studies have applied this approach to quantify synergy in both undirected~\cite{amari2001information} and directed settings~\cite{olbrich2015information, perrone2016hierarchical, niu2019measure}, but they either only provide an overall quantification of synergy across the whole system~\cite{perrone2016hierarchical} or are restricted to scenarios with only a few variables~\cite{olbrich2015information, niu2019measure}. Moreover, although previous works have investigated hierarchical levels of dependencies~\cite{perrone2016hierarchical, schneidman2003network, schneidman2003synergy, schneidman2006weak, rosas2016understanding}, disentangling interdependencies of the same order has received little attention. As a result, existing frameworks conflate qualitatively different higher-order dependencies, not discriminating among which specific components these structures are shared. 

A complementary perspective to distributed computation comes from the study of information dynamics~\cite{lizier2012local}, and in particular of information decomposition approaches~\cite{williams2010nonnegative, mediano2025toward}. These split the information that inputs carry about outputs into redundant, unique, and synergistic ``atoms'' yielding a comprehensive taxonomy of the space of computation in multivariate systems.
Although they have been successful in characterising the computational nature of systems ranging from biological and artificial neural networks~\cite{beer2015information, wibral2017partial, ince2017measuring, tax2017partial, proca2024synergistic, luppi2022synergistic, varley2023partial, luppi2024information}, to financial markets~\cite{rajpal2023quantifying}, physical models~\cite{finn2018quantifying, rosas2018information, liardi2025simple}, and ecological systems~\cite{rajpal2025information, sas2025improved}, they remain underdetermined, not providing a unique formalism. Many competing measures exist, and the choice among them rests on operational or axiomatic grounds rather than on the intrinsic structure of the system under study~\cite{liardi2026mathematical}.

In this work, we merge these two lines of research and overcome the limitations outlined above, introducing a complete geometric characterisation of the statistical relationship between sets of sources and targets. In doing so, we provide a hierarchical decomposition of the information structure of the system. 
First, we formalise a general multi-input multi-output information-geometric quantification of higher-order statistical structures applicable to multivariate systems of arbitrary size, differentiating not only across interaction orders but also within the same order (Secs.~\ref{sec:IG_3var}-\ref{sec:IG_4var}). Next, we detail the natural integration of this approach with information-theoretic decomposition schemes, providing finer-grained insights into the statistical architecture of complex systems and bridging the gap between geometrical structure and distributed computation (Sec.~\ref{sec:IG_IT}). We then validate the framework across a variety of synthetic and real-world scenarios, demonstrating its flexibility and descriptive power across physical, neural, and financial systems (Sec.~\ref{sec:results_empirical}). 
Finally, we conclude with a discussion of how this technique may serve as a mathematically principled approach for characterising computation in natural systems (Sec.~\ref{sec:discussion}).

\section{Geometric foundations of distributed computation} \label{sec:theo_results}

To understand the ways in which complex systems compute information, we must be able to characterise the structure of dependencies between a system's multiple inputs (sources) and multiple outputs (targets).
In what follows, we illustrate and resolve a fundamental difficulty in this endeavour: how to operationalise from first principles the interrelations between dependencies at every order of interaction. 

We first consider a simple system of three variables to establish the information-geometric foundations of our approach and develop intuition for our geometric formalisation of synergy. Then, we extend these concepts to characterise synergistic interdependencies in larger systems. Finally, we show how these constructs can be interpreted within a unified information-theoretic framework, completing the connection with modes of distributed computation.

\subsection{An information-geometric definition of synergy} \label{sec:IG_3var}

To get some intuition on the geometry of the problem and the related notion of synergy we propose, we start by considering the simplest directed system composed of two source variables $X_1,X_2$ and a target $Y$, and indicate with $P_{X_1 X_2 Y}:=P(X_1,X_2,Y)$ the underlying probability distribution of the system. In information geometry, this distribution is viewed as a point in the statistical manifold $M$ which contains all possible distributions for the system $(X_1,X_2,Y)$. More formally, this manifold can be represented as an exponential family, whose coordinates explicitly encode the statistical interactions between subsets of variables~\cite{amari2016information}. In $\theta$-coordinates, $M$ can be parametrised as
\begin{equation} \label{eq:M3d}
    \begin{split}
& \quad M = \{ {P_{{X_1}{X_2}Y}} \text{ s.t. } \\ %
& \begin{aligned}[t]
    & {P_{{X_1}{X_2}Y}} = \exp ( {\theta _{X_1}}{F_{X_1}}({x_1}) + {\theta _{X_2}}{F_{X_2}}({x_2}) + {\theta _{Y}}{F_{Y}}(y) + \\      
& \quad + {\theta _{X_1X_2}}{F_{X_1X_2}}({x_1},\;{x_2}) + {\theta _{X_1Y}}{F_{X_1Y}}({x_1},\;y)+ \\
& \quad + {\theta _{X_2Y}}{F_{X_2Y}}({x_2},\;y) + \\ 
& \quad + {\theta _{X_1X_2Y}}{F_{X_1X_2Y}}({x_1},\;{x_2},\;y) - \psi (\theta ) ) \} \,,
    \end{aligned}
    \end{split}
\end{equation}
where $F$ represent the sufficient statistics associated with each subset of variables, $\theta$ are the corresponding natural parameters that quantify the strength of these interdependencies, and $\psi$ is the normalising factor. 
This representation makes it explicit that the full manifold $M$ contains distributions with all possible interactions between sources and target, including pairwise and genuinely three-way dependencies.

Since we aim to study directed interdependencies between $(X_1, X_2)$ and $Y$, it is natural to explore how these dependencies change when specific pairwise source-target interactions are selectively removed. For this purpose, let us first consider the following submanifolds of $M$ obtained by constraining the corresponding interaction parameters to zero
\begin{equation}
\begin{aligned}
E_0 & :=\{P\in M\text{ s.t. } \theta_{X_1Y}=\theta_{X_2Y}=\theta_{X_1X_2Y}=0\} \\
E_i & :=\{P\in M\text{ s.t. } \theta_{X_{-i}Y}=\theta_{X_1X_2Y}=0\} \,,
\end{aligned}
\end{equation}
where we indicated with $-i$ all source variables except for $i$ (which, in this case, is only one).
Here, $E_0$ is the space of distributions in which the sources are independent of the target, while $E_i$ represents that in which source $X_{-i}$ carries no information about $Y$ beyond what is already conveyed by $X_i$, i.e.\ $X_{-i}\perp Y|X_i$. 
We remark that, unlike decompositions of undirected quantities where projections are defined by systematically eliminating interaction orders~\cite{amari2001information}, this construction focuses solely on input-output dependencies, aiming to isolate how each source contributes information about $Y$~\cite{niu2019measure}. 
Then, as is common in information geometry, we quantify how much information is lost when either dependency is removed by measuring the Kullback-Leibler (KL) divergence\footnote{With a small abuse of language, we will also refer to the Kullback-Leibler divergence as a ``distance'', despite it not being formally so, as it is common practice in the field of information geometry.} between the starting distribution $\Pf$ and its projection onto the corresponding submanifold.
Denoting as $\bP_0$ and $\bP_i$ such projections onto $E_0$ and $E_i$, respectively, the nesting $E_0\subseteq E_i\subseteq M$ yields the following additive decomposition via the Pythagorean relation~\cite{amari2016information}:  
\begin{equation} \label{eq:dkl_proj1}
\begin{split}
    \underbrace{\D(\Pf||\bar{P}_0)}_{\scriptstyle M\,\hookrightarrow\, E_0} = & \underbrace{\D(\Pf||\bar{P}_i)}_{\scriptstyle M\,\hookrightarrow\, E_i} + \underbrace{\D(\bar{P}_i||\bar{P}_0)}_{\scriptstyle E_i\,\hookrightarrow\, E_0} \,.
\end{split}
\end{equation}
In information-theoretic terms, Eq.~\eqref{eq:dkl_proj1} precisely corresponds to the well-known chain rule of mutual information~\cite{cover1999elements} 
\begin{equation} \label{eq:mi_chainrule_3var}
    I(X_1,X_2;Y) = I(X_{-i};Y|X_i) + I(X_i;Y) \,. %
\end{equation}
Although these decompositions provide some insights into how source-target dependencies are structured, they do not directly estimate the synergistic contributions, i.e.\ the dependencies of $Y$ on the joint configuration of $(X_1,X_2)$ that cannot be attributed to either source alone. 
In fact, while Eqs.~\eqref{eq:dkl_proj1}-\eqref{eq:mi_chainrule_3var} make it clear that synergistic dependencies must be contained in the first term of the RHS, as the second only involves $X_i$, lower-order dependencies induced by $X_{-i}$ might also be present there. 
In the following, we investigate how these can be disentangled.

Geometrically, synergy therefore corresponds to statistical dependencies that persist after removing all single-source explanatory pathways. The projections $\bar{P}_1$ and $\bar{P}_2$ implement precisely such removal, with $\bP_i$ eliminating all direct dependencies between $X_{-i}$ and $Y$ and thus retaining only those mediated by $X_i$. Hence, since distributions obtained as normalised combinations of $\bP_1$ and $\bP_2$, i.e.\ of the form $\bP_1^t\bP_2^{1-t}$, admit an explanation in terms of single-source mechanisms alone, genuine higher-order dependencies can only be associated with deviations from this class of models. Following this reasoning, we define synergy\footnote{A seemingly natural alternative would be to define synergy by imposing only $\theta_{X_1X_2Y}=0$. However, this approach would not be compatible with a decomposition of mutual information, since the corresponding projection generally modifies the induced pairwise marginals, and therefore does not lead to a projection orthogonal to the manifold $E_0$. Consequently, it does not yield an additive splitting of the form $\D(\Pf||P^*)+\D(P^*||\bP_0)$.} as the distance between the true distribution and the closest such combination of the conditional-independence projections~\cite{niu2019measure}:
\begin{equation} \label{eq:syndef_PID}
    S(X_1,X_2;Y) := \min_{t\in\mathbb{R}} \D\left(\Pf\big|\big|\,\mathcal{Z}\bar{P}_1^t \cdot\bar{P}_2^{1-t}\right) \,,
\end{equation}
where $t$ weights the projections, and $\mathcal{Z}$ is a normalising constant. 
This intuitive construction has a precise geometric foundation. In information geometry, distributions on $M$ can be connected by two dual notions of paths: $e$-geodesics, which are linear in the natural $\theta$ coordinates and correspond to exponential-family interpolations, and $m$-geodesics, which are linear in the probabilities and correspond to mixture interpolations~\cite{amari2016information}. The family in Eq.~\eqref{eq:syndef_PID} is precisely the $e$-geodesic through $\bP_1$ and $\bP_2$, and the distribution minimising Eq.~\eqref{eq:syndef_PID} is the unique point on it at which the $m$-geodesic from $\Pf$ meets it orthogonally~\cite{niu2019measure}. Analogously to Eq.~\eqref{eq:dkl_proj1}, this orthogonality implies that $\D(\Pf||\bP_i)$ admits a Pythagorean decomposition into two additive contributions, providing an explicit disentanglement of synergistic and residual lower-order dependencies. Denoting $P^*$ as the distribution minimising Eq.~\eqref{eq:syndef_PID}, this yields a refined hierarchical decomposition of Eq.~\eqref{eq:dkl_proj1}:
\begin{equation}\label{eq:3var_decomp_S}
\begin{split}
    & \overbrace{\D(\Pf||\bar{P}_0)}^{\Large  I(X_1,X_2;Y)} = \\
    & \quad \underbrace{\D(\Pf||P^*)}_{\text{\normalfont synergy\vphantom{d}}} + \underbrace{\D(P^*||\bP_i) + \D(\bar{P}_i||\bar{P}_0)}_{\text{\normalfont pairwise dependencies}} \,,\\[-2.5em]
    & \quad \underbrace{\phantom{\underbrace{\D(\Pf||P^*)}_{\Large \!S} + \D(P^*||\bP_i)}}_{\Large I(X_{-i};Y|X_i)}
    \phantom{{}+{}} \underbrace{\phantom{\vphantom{\underbrace{\D(\Pf||P^*)}_{\Large S} + \D(P^*||\bP_i)}\D(\bar{P}_i||\bar{P}_0)}}_{\Large I(X_i;Y)} \phantom{\,,}
\end{split}
\end{equation}
where the first term isolates the genuinely synergistic contribution, while the remaining terms capture the dependencies attributable to lower-order interactions.

This approach gives us the premises to formalise synergy in more general terms:
\begin{definition} \label{def:syn}
Synergistic contributions are quantified as the KL divergence between the system's distribution and the closest distribution along the $e$-geodesic connecting the corresponding conditional-independence projections.
\end{definition}

With this definition at hand, we can now analyse how this translates to larger systems comprising an arbitrary number of inputs and outputs. 

\subsection{Orders of synergy in multivariate systems} \label{sec:IG_4var}

Consider a generic system $(\Xs,\Ys)$ with $n$ sources and $m$ targets and underlying probability distribution $P$. The concept of \textit{higher-order dependencies} in such a system becomes more nuanced, as there is a combinatorially large number of sets of variables that present synergistic interactions. 
For this reason, we differentiate between \textit{orders of synergy}, depending on the number of sources and targets that the interaction entails. For instance, a subsystem with $n'$ sources and $m'$ targets will contain synergies up to order $n'+ m'$, with smaller subsystems containing increasingly smaller orders of synergy. This reasoning continues until we reach the minimal three-variable set-up $(X_i,X_j, Y_k)$ or $(X_i, Y_j, Y_k)$ we just explored, where the only remaining synergy is of order 3. %
In all of these cases, each synergy can be obtained by applying Definition~\ref{def:syn} to appropriate submanifolds where specific sources and targets are marginalised, and particular interactions are set to zero. 

In the following, we formalise these ideas more concretely. For ease of notation and clarity of exposition, we focus on the simplest fully multivariate case with two sources $X_1, X_2$ and two targets $Y_1, Y_2$. However, the extension to larger settings is straightforward, albeit it requires the treatment of a combinatorially increasing number of terms. 

Let $\Pff$ be the distribution of the system under examination, and $M$ the manifold containing all probability distributions over $(X_1,X_2,Y_1,Y_2)$. Now consider the submanifolds ${E_i^j \subseteq M}$ in which we set to zero all interactions between variables $X_{-i}$ and $Y_{-j}$, so that the remaining dependencies are exclusively mediated by $X_i$ and $Y_j$:
\begin{equation}
\begin{split}
    E_i^j & := \{ \Pff\in M \text{ s.t. } \\ & \theta_{X_{-i}Y_{-j}}
    = \theta_{X_1X_2Y_{-j}} = \theta_{X_{-i}Y_1Y_2} = \theta_{X_1X_2Y_1Y_2}=0\} \,,
\end{split}
\end{equation}
with ${i,j = 1,2}$. In other words, $E_i^j$ represents the space where variables $X_{-i}$ and $Y_{-j}$ are conditionally independent of the rest of the system, i.e.\ ${X_{-i}\perp Y_{-j}|X_i, Y_j}$.
We then quantify the statistical contributions, or rather, the removal, of each component by measuring the distance between $\Pff$ and its projections onto these submanifolds, which we indicate as $\bP^j_i$.
We can hence perform a procedure analogous to the one in Sec.~\ref{sec:IG_3var}, now with the probability distribution $\Pff$ having four projections rather than only two, each corresponding to a submanifold $E_i^j$. As before, each projection $\bP_i^j$ removes exactly one specific explanatory pathway, corresponding to the interaction between source $X_{-i}$ and target $Y_{-j}$, while leaving all remaining dependencies intact. Consequently, any statistical dependence that can be represented as a combination of these four projections is explainable in terms of pairwise or conditionally mediated interactions, without requiring purely joint source-target effects. 
Thus, using \defs, the minimum distance between \Pff and the interpolation among the $\bP_i^j$ defines synergy of order 4, as it captures the statistical dependency lost once any of the four source-target interactions is removed:
\begin{align} \label{eq:s4_def}
    & \begin{aligned}[t]  & S_4(X_1,X_2;Y_1,Y_2) := \\ & \qquad
    \min_{r,s,t\,\in\,\mathbb{R}^3}\D \Big(\Pff \big|\big| P_*^*(r,s,t) \Big) \end{aligned} \\ \label{eq:s4_def_2}
    & P_*^*(r,s,t) = \mathcal{Z}{\left(\bP^1_1\right)}^{1-r-s-t}\,{\left(\bP^2_1\right)}^r\,{\left(\bP^1_2\right)}^s\,{\left(\bP^2_2\right)}^{t} \,.
\end{align}

As anticipated above, there are now also synergies of order 3 present in the subsystems $(X_1,X_2,Y_i)$, and $(X_i,Y_1,Y_2)$, $i=1,2$. For each of these, we can calculate the marginal probability distributions $P_{X_1 X_2 Y_i}$, $P_{X_i Y_1 Y_2}$ and then perform the procedure outlined in Sec.~\ref{sec:IG_3var}. As an example, for the system $(X_1,Y_1,Y_2)$ we have:
\begin{align} \label{eq:S3_syn_example}
    \begin{split} 
    & S_3(X_1; Y_1, Y_2) := \\ & \quad \min_{t\in\mathbb{R}} \D\Big(P_{X_1Y_1Y_2}\big|\big|\,\mathcal{Z}\bar{P}_{1,\,X_1Y_1Y_2}^t \cdot \bar{P}_{2,\,X_1Y_1Y_2}^{1-t}\Big) \,,
    \end{split} 
\end{align}
where $\bar{P}_{i,\,X_1Y_1Y_2}$ represents the projection on the submanifold where $Y_{-i}\perp X_1|Y_i$. 
This provides a decomposition of the dependencies of $(X_1,Y_1,Y_2)$ formally analogous to that of Eq.~\eqref{eq:3var_decomp_S}, with similar procedures holding for the other three-variable subsystems.
Thus, this technique allows one to consistently classify how synergies in the system are distributed, providing a scalable technique that can be employed in large complex systems. 

However, this approach does not enable a fully hierarchical decomposition. That is, unlike the simpler scenario encountered in Eq.~\eqref{eq:3var_decomp_S}, the total dependency of the system is not just the sum of all synergies and lower-order structures
\begin{equation*}
    \D(\Pff||\bar{P}_0) \ne S_4 + \sum S_3 + \substack{\text{\small pairwise} \\ \text{\small dependencies}}\,.
\end{equation*}
This is not a shortcoming of the framework, but a direct consequence of the granularity at which we are proposing to investigate the system. 
The reason for this is that synergistic structures of the same order are not orthogonal to each other as they may overlap across components.
Geometrically, this corresponds to the fact that the submanifolds of the marginal probability distributions $P_{X_1 X_2 Y_i}$, $P_{X_i Y_1 Y_2}$ are not nested in each other (see Supplementary Material S1). 

Thus, the last piece of the puzzle is to understand such relations between synergies, which we achieve by combining our approach with a multivariate information-theoretic decomposition framework. As we will see in the next section, this naturally leads to a complete characterisation of the statistical interdependencies of the system, disentangling different modes of computation.

\begin{figure*}[ht]
    \centering
    \includegraphics[width=1\linewidth]{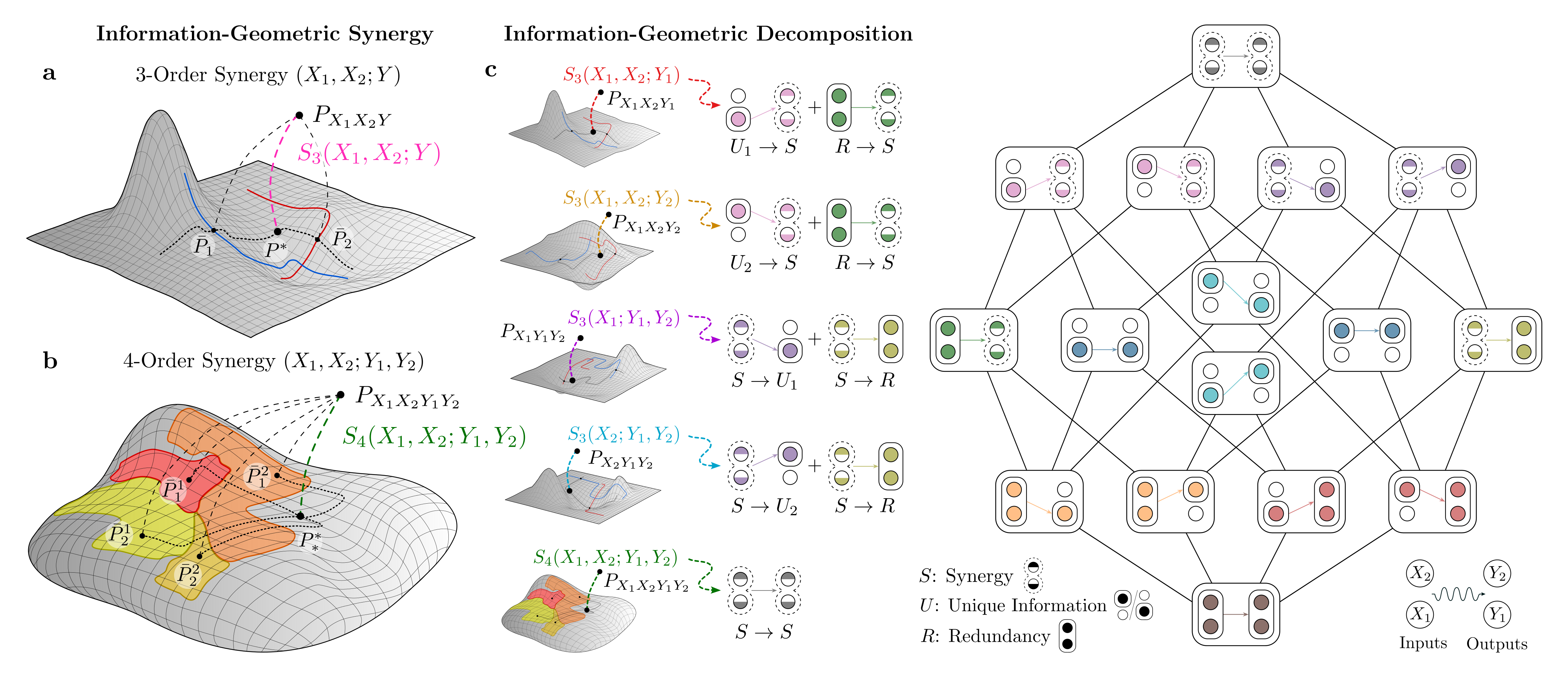}
    \caption{\textbf{Mapping input-output relations via information geometry and information-theoretic decomposition.} \textbf{a} Representation of the order-three information-geometric synergy. \textbf{b} Representation of the order-four information-geometric synergy. \textbf{c} Connections between information-geometric synergies and \phiid atoms (left), and \phiid lattice (right).}
    \label{fig:scheme}
\end{figure*}

\subsection{Decomposing synergistic dependencies} \label{sec:IG_IT} %
We now aim to investigate the relationships among these various synergies of the same order, providing a comprehensive perspective on how they contribute to the system's statistical structure.
To achieve this, we employ Integrated Information Decomposition (\phiid), an information-theoretic approach which allows one to decompose statistical dependencies into multiple interpretable informational modes~\cite{mediano2025toward}. 

Given $n$ sources $\Xs$ and $m$ targets $\Ys$, \phiid decomposes the mutual information $I(\Xs; \Ys)$ into three informational components: redundancy, which quantifies the shared information that the sources provide about the target; unique information, the information that only one source contributes to the prediction of the target; and synergy, the information that only the joint knowledge of the sources can provide.
More precisely, \phiid tracks how redundancy, unique information, and synergy between the sources affect redundancy, unique information, and synergy within the targets.
In the case of $n=m=2$, \phiid involves quantities such as $\rtx$, where redundant information between the sources $X_1,X_2$ becomes unique to the target $Y_1$ (hence often called \textit{erasure}), or $\str$, where synergistic information in the sources becomes redundant in the targets. This approach provides a set of 16 atoms that make up the mutual information between source-target subsets of $(X_1,X_2,Y_1,Y_2)$, which can be naturally organised in a lattice (Fig.~\figsubref{fig:scheme}{c}). For instance, the mutual information $I(X_i;Y_j)$ will contain terms that are redundant in sources and targets, and those unique to source $X_i$ and target $Y_j$:
\begin{equation} \label{eq:single_MI_ij}
    I(X_i; Y_j) = \rtr + \rtj + \itr + \itj \,.
\end{equation}
Similarly, the mutual information $I(X_i; Y_1,Y_2)$ contains redundant contributions, unique to source $X_i$ and to the targets, and synergistic in the targets $Y_1,Y_2$, i.e.\
\begin{equation} \label{eq:phiid_Ix1y1y2}
    I(X_i; Y_1, Y_2) = 
  \begin{aligned}[t]
    &\rtr + \rtx + \rty + \itr +\\
    & \kern-1em + \itx + \ity + \rts + \its \,.
  \end{aligned} 
\end{equation}  

A limitation of \phiid is that its solution relies on an arbitrary choice of what constitutes redundant information, whose definition has given rise to an abundance of proposals and no overall agreement~\cite{williams2010nonnegative, harder2013bivariate, griffith2014intersection, bertschinger2014quantifying, griffith2014quantifying, griffith2015quantifying, barrett2015exploration, goodwell2017temporal, ince2017measuring, james2018unique, kay2018exact, james2018uniquekey, finn2018pointwise, niu2019measure, kay2024partial, sigtermans2020partial, sigtermans2020towards, makkeh2021introducing, schick2021partial, ehrlich2024partial, kolchinsky2022novel, venkatesh2022partial, mages2023decomposing, lyu2024explicit, liardi2026mathematical}. 
Here, we approach the problem from the opposite end: rather than postulating what redundant information should be, we build the decomposition from synergy, using the geometric definition introduced above. This way, the additional ingredient that \phiid requires is no longer an axiomatic or operational notion of shared information, but a quantity grounded in the geometry of the statistical manifold. Moreover, \defs uniquely determines all \phiid atoms in systems with any number of sources and targets, without requiring any modification or ad hoc extension of the framework as the system grows~\cite{mediano2025toward}.
As we see below, this connection between information geometry and \phiid explains why a fully geometric hierarchical decomposition cannot be achieved, while providing an alternative route to characterising interdependencies.

\begin{figure*}[ht]
    \centering
    \includegraphics[width=\textwidth]{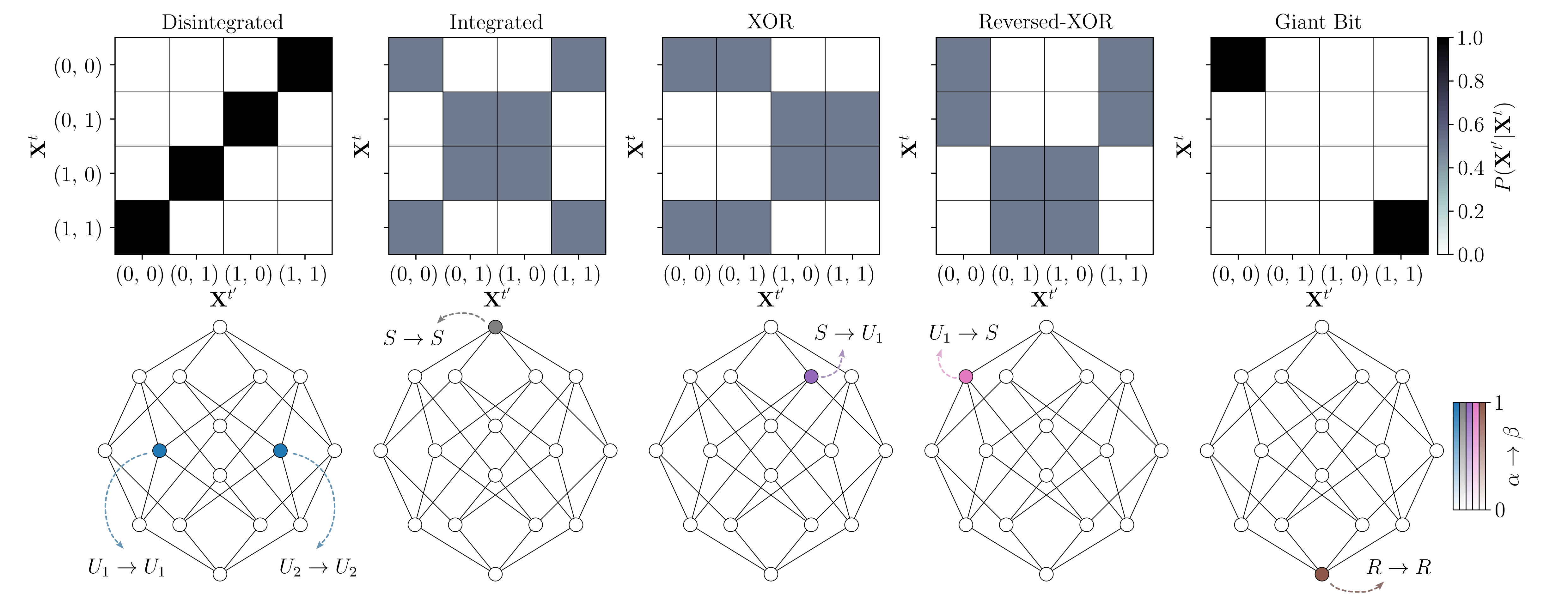}
    \caption{\textbf{\phiidig accurately captures computational modes in discrete systems.}
    Transition probability matrices for five different logic gates, and corresponding values of the \phiidig atoms computed on them. 
    \label{fig:discrete_tpms}
    }
\end{figure*}

Consider, for instance, the subsystem $(X_1,Y_1,Y_2)$. Taking the difference between Eq.~\eqref{eq:phiid_Ix1y1y2} ($X_i=X_1$) and Eq.~\eqref{eq:single_MI_ij} (with source $X_1$ and target $Y_i$) and using the mutual information chain rule, we obtain that 
\begin{equation}
    I(X_1;Y_{-i}|Y_i) = \rtmi + \xtmi + \rts + \xts \label{eq:cond_mi_3var} \,.
\end{equation}
Comparing Eq.~\eqref{eq:cond_mi_3var} to Eq.~\eqref{eq:3var_decomp_S}, we recognise that there are two synergistic atoms contributing to the synergy of order three. This leads us to the following identification: %
\begin{equation}
    S_3(X_1;Y_1,Y_2) = \rts + \xts \,.
\end{equation}
Applying the same reasoning to the other 3-variable subsystems, we obtain the remaining order-three synergies:
\begin{align}
    S_3(X_2;Y_1,Y_2) & = \rts + \yts \\
    S_3(X_1,X_2;Y_1) & = \str + \stx \\
    S_3(X_1,X_2;Y_2) & = \str + \sty \,.
\end{align}
Thus, this approach not only allows us to represent synergistic structures in terms of specific information atoms, but also explains that synergies of the same order are not orthogonal to each other due to the presence of redundancy. In particular, $\rts$ precisely represents a synergistic dependency in the targets which is present both in the subset $(X_1, Y_1, Y_2)$ and in $(X_2, Y_1, Y_2)$, thus being \textit{redundant} in the sources. Analogously, $\str$ represents synergistic structure in the sources present when either target is considered, and is thus present in both $S_3(X_1, X_2; Y_i)$.

Finally, the characterisation of order-four synergy is more straightforward. Since synergy at order four refers to the higher-order dependencies that must be present only at the level of the whole system, and not in any subsystem, this must correspond to the most synergistic \phiid atom:
\begin{equation}
    S_4(X_1,X_2;Y_1,Y_2) = \sts \,.
\end{equation}
We refer to Supplementary Material S1 for a more formal derivation.

Importantly, this representation of \phiid atoms in terms of information-geometric synergy allows us to completely determine the system of equations of \phiid, and thus to explicitly compute all informational atoms. We call this approach \phiidig.
Hence, not only does \phiid help us formally understand the relationships between synergies, but it also provides a fine-grained decomposition of the interdependencies between the parts of a system, including modes of dependencies complementary to higher-order structures.  

Although here we specifically focused on the scenario of two sources and two targets, this approach is more general and can be applied to systems of any size (see Sec.~\ref{sec:discussion} and Supplementary Material S1).

\section{Results} \label{sec:results_empirical}

\begin{figure*}[ht]
    \centering
    \includegraphics[scale=0.85]{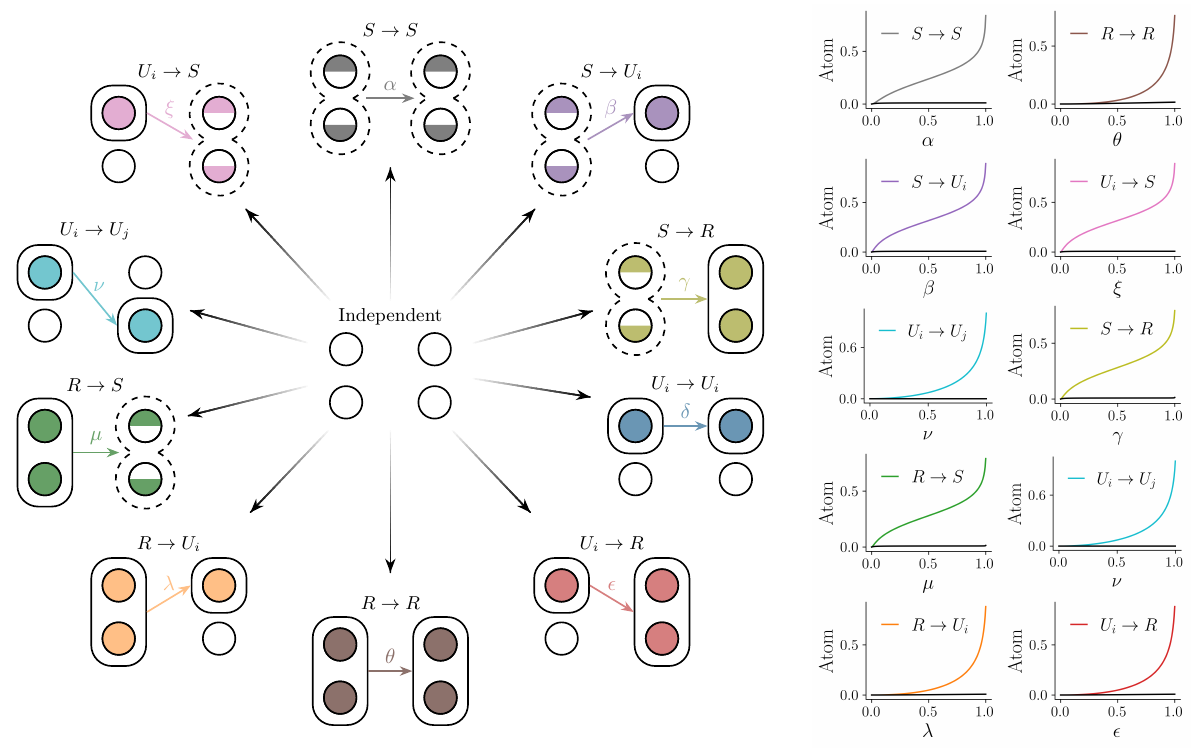}
    \caption{\textbf{\phiidig accurately captures computational modes in Gaussian systems.} \textbf{a} Schematic representation of the ten toy models considered, each characterised by a specific computational trait. The strength of each mode is determined by a coupling parameter that interpolates from a fully independent baseline toward the target interdependency structure. Dashed outlines denote synergistic structures ($S$), solid boxes denote redundant ones ($R$), and single-circle boxes denote unique contributions ($U_i$, $U_j$). \textbf{b} \phiidig atoms as a function of the coupling parameter for each toy model. The coloured line tracks the atom the system was designed to maximise; black lines show the remaining atoms.}
    \label{fig:gauss_toy}
\end{figure*}

\subsection{Computation via state transitions}

Since we aim to characterise computation in complex dynamical systems, we begin by testing our approach in a controlled setup using known Boolean functions. Specifically, we consider a bivariate system where the input states at time $t$ ($X_i^t$) evolve into output states at time $t^\prime$ ($X_i^{t^\prime}$) according to a transition probability matrix ${T_{ij}=P(\mathbf{X}^{t^\prime}=j \mid \mathbf{X}^t=i)}$.
To systematically probe different computational modes, we manipulate relationships between past and future states by manually constructing $T_{ij}$ with prescribed logical structures, each designed to exhibit specific informational traits. We then derive the underlying probability distribution and apply \phiidig to characterise the computational structure between inputs and outputs (see Sec.~\ref{sec:methods}). 

Following Ref.~\cite{varley2023decomposing}, we first examine the behaviour of a ``disintegrated'' system, where the future is determined by a flip of each source, and an ``integrated'' one, where the parity of both past and future states is kept constant (Fig.~\ref{fig:discrete_tpms}). As one would expect, the disintegrated system presents modes which are purely characterised by unique information, i.e.\ information which is passed along each source without cross-transfers. In other words, the dynamics of these modes are completely determined by their own past. On the contrary, the integrated system is characterised by interdependencies that cannot be reduced to the knowledge of any component in isolation. Hence, the highest-order synergy $\sts$ appears as the sole computational mode.

We then use a standard XOR gate, in which the parity of the sources determines the future state of $X_1$, while $X_2$'s future is random. This correctly provides the unique functional mode $\stx$. Similarly, we construct its time-reversed counterpart ``Reversed-XOR'', which provides the correct informational trait $\xts$. 
Finally, we conclude with the ``Giant bit'' system, which deterministically alternates between the all-1's and all-0's configurations~\cite{james2017multivariate}. As expected, the only computational mode for such a scenario is fully redundant information $\rtr$. 
We can devise analogous systems for all remaining atoms, obtaining in all cases a perfect correspondence between the expected and observed results (see Supplementary Material S3 Fig.~S2). 

\begin{figure*}[ht]
    \centering
    \includegraphics[width=\textwidth]{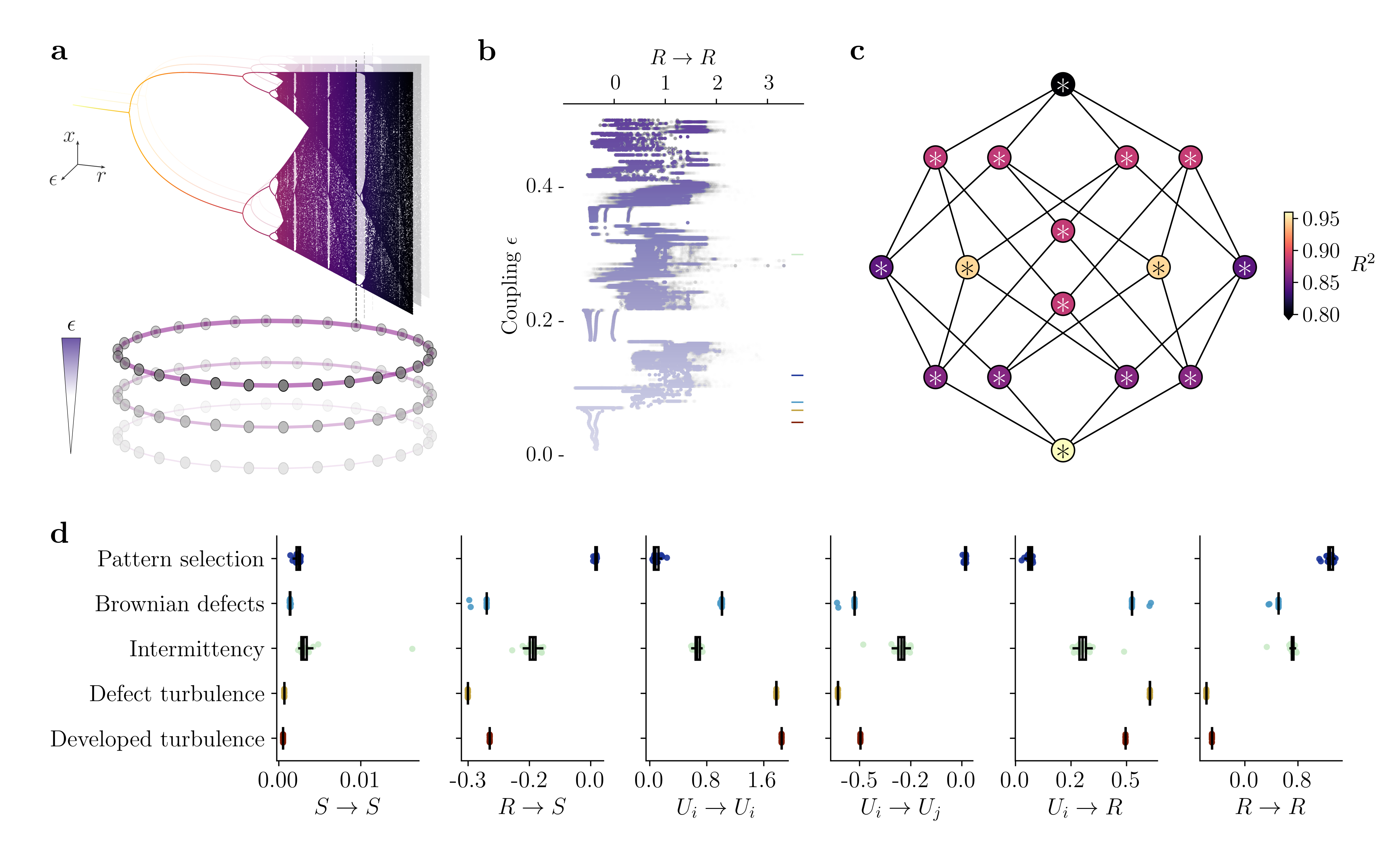}
    \caption{\textbf{Universality classes exhibit distinct computational profiles.} 
    \textbf{a} Coupled map lattice on a ring, consisting of locally chaotic logistic maps interacting through nearest-neighbour diffusion dynamics with coupling strength $\epsilon$ (see Sec.~\ref{sec:methods}).
    \textbf{b} Phase transitions in \rtr as a function of the coupling strength $\epsilon$.
    \textbf{c} \phiidig lattice representation of the sensitivity of each mode of information dynamics to discriminating dynamical regimes, quantified by the marginal $R^2$ of the fixed effects in the LME model.
    \textbf{d} Profiles of information dynamics in each dynamical regime, including: Pattern selection (PS, $\epsilon=0.12$), Brownian motion with defects (BMWD, $\epsilon=0.08$), Spatiotemporal intermittency II (STI, $\epsilon=0.3$), Defective turbulence (DT, $\epsilon=0.068$), Fully developed turbulence (FDT, $\epsilon=0.05$). Points indicate average global values for each realisation. 
    * indicates $p_{FDR}<10^{-15}$ FDR-corrected with the Benjamini–Hochberg procedure.
    }
    \label{fig:chaos}
\end{figure*}

\subsection{Statistical motifs of computation in continuous systems}

We now validate that \phiidig also accurately identifies and tracks distinct computational modes in continuous Gaussian systems. To show this, we construct a suite of Gaussian toy models, each designed to maximise one of the ten \phiid atoms that are distinct up to relabelling of the variables, and then apply \phiidig (Fig.~\figsubref{fig:gauss_toy}{a}). These atoms comprise fully synergistic (\sts), redundant (\rtr), unique (\iti, \itj), or mixed (\str, \sti, \rti, \itr, \its, \rts) contributions. More specifically, we start from a fully independent covariance matrix and interpolate toward the target covariance that instantiates a specific statistical interdependency~\cite{liardi2025scalable}, with each interpolation mediated by a specific scalar coupling parameter. 
Results show that, in each case, the \phiidig atom corresponding to the system's target interdependency grows monotonically with the coupling, while all other atoms remain near zero (Fig.~\figsubref{fig:gauss_toy}{b}). This confirms that the framework decomposes the system's information dynamics into genuine, functionally distinct and non-overlapping contributions. 
These findings demonstrate both the specificity and sensitivity of \phiidig as a measure of distributed computation in continuous settings: accurately identifying \emph{what kind} of information dynamics are occurring, and faithfully tracking \emph{how strongly} they are expressed.  

\begin{figure*}[ht]
    \centering
    \includegraphics[width=\textwidth]{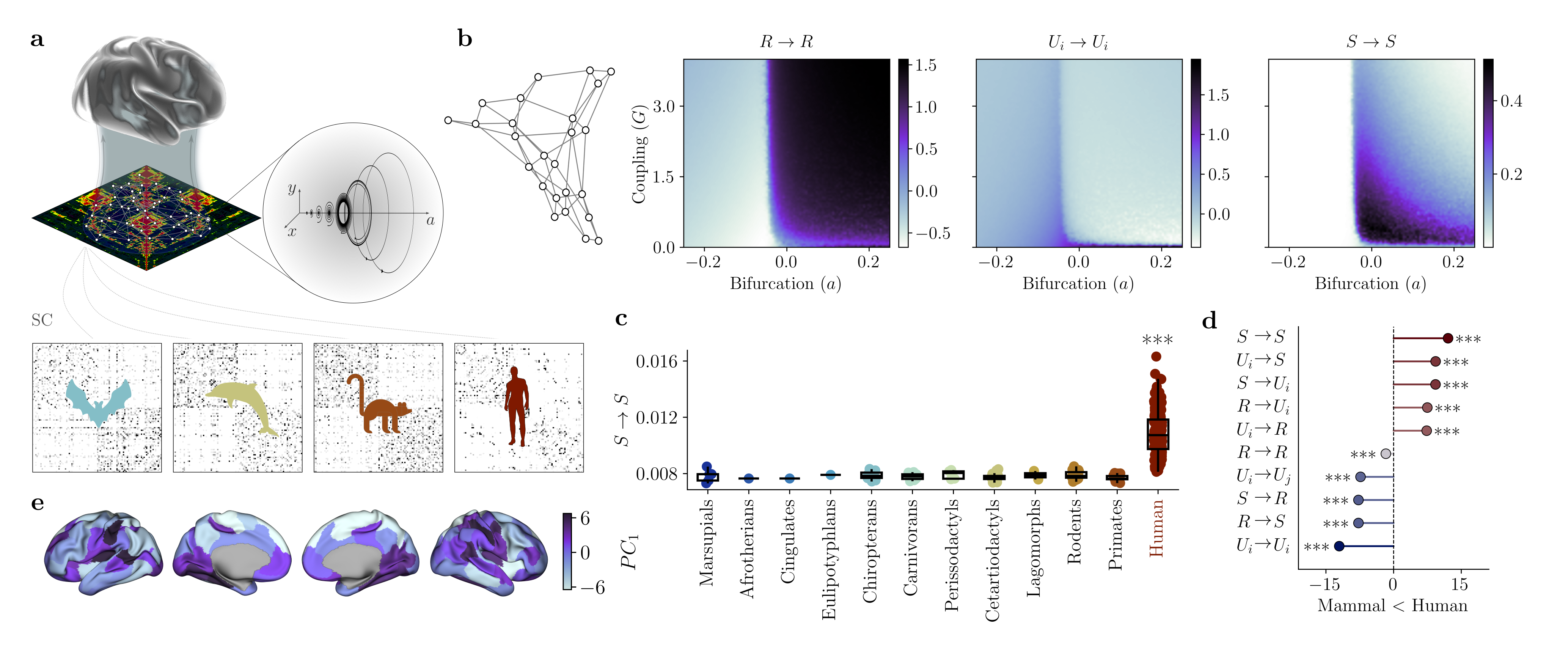}
    \caption{\textbf{Evolution of human cortical wiring promoted greater distributed computation.}
    \textbf{a} Phenomenological whole-brain computational model, based on networks of coupled Stuart-Landau oscillators around Hopf bifurcation (see Sec.~\ref{sec:methods}), with specific structural connectivities (SC) shown for bats, dolphins, lemurs, and humans.
    \textbf{b} Effects of local and global interdependencies on distributed information storage. State-space of the global coupling ($G$) and bifurcation ($a$) parameters, in a model with Watts-Strogatz small-world network topology, for \rtr (left), \iti (middle), and \sts (right).
    \textbf{c} Distribution of order-four synergy (\sts) across mammals and humans, with each point indicating the average global values for each mammal's connectome.
    \textbf{d} Deviation of human neural computation from the mammalian distribution, expressed in across-species standard deviation.
    \textbf{e} Low-dimensional ($PC_1$) organisation of computation in simulated human brain activity, with higher values indicating stronger unique storage (\iti) and weaker persistent redundancy (\rtr) and synergy (\sts). Topographic gradient from medial occipital, lateral sensory and temporal regions (high $PC_1$) to dorsomedial frontal and paracentral cortex (low $PC_1$).
    }
    \label{fig:hopf}
\end{figure*}

\subsection{Computational profile of universality classes}

To understand how distributed computation occurs in complex spatiotemporal systems, we apply \phiidig to coupled map lattices (CMLs)~\cite{kaneko1985spatiotemporal, kaneko1989pattern, kaneko1993theory} (Fig.~\figsubref{fig:chaos}{a}), analysing the computational structure between past states (sources) and future states (targets), i.e.\ the computations performed by the system's dynamics (see Sec.~\ref{sec:methods}). 
CMLs are dynamical processes characterised by internal chaotic dynamics and external diffusive coupling between nearest-neighbours, and are known to exhibit drastic phase transitions as non-linear functions of the coupling strength $\epsilon$~\cite{jost2001spectral}. 
To determine how these transitions shape the system's computational structure, we sweep $\epsilon$ over $[0.01, 0.5]$, a range spanning multiple universality classes, and then perform \phiidig in each scenario. 
We observe pronounced, non-linear changes in the information dynamics, with marked changes in the expected ranges of $0.07$--$0.17$ and $0.37$--$0.50$, and stability maintained in $0.01$--$0.06$ and $0.17$--$0.21$~\cite{kaneko1989pattern, kaneko1993theory, jost2001spectral} (Fig.~\figsubref{fig:chaos}{b}). Crucially, these changes are not merely variations in the overall magnitude of information transfer: they are accompanied by systematic changes in how information is distributed across the different \phiidig atoms, revealing transitions between distinct modes of distributed computation (Supplementary Material S3 Fig.~S3).

Beyond being dynamically distinct, several of these regimes also correspond to different universality classes~\cite{kaneko1989pattern}. Following previous work~\cite{santoro2023higher, kaneko1989pattern}, we focus on five of them: pattern selection (PS: $\epsilon=0.12$), where quasi-periodic spatial patterns emerge from a chaotic background; Brownian motion with defects (BMWD: $\epsilon=0.08$), where localised defects randomly walk across the lattice; spatiotemporal intermittency II (STI: $\epsilon=0.3$), where laminar patches and chaotic bursts coexist and compete; defective turbulence (DT: $\epsilon=0.068$), where turbulent structures collide and annihilate; and fully developed turbulence (FDT: $\epsilon=0.05$), where incoherent, high-dimensional chaos dominates.

To quantify the sensitivity of each mode of information dynamics to differences in regime, we employ a linear mixed-effects (LME) model for each \phiid atom, with the atom value as the dependent variable, regime as a fixed effect, and random intercepts for each simulation and each node within it, with time-series half as a fixed covariate. 
All atoms showed statistically significant shifts across regimes (all $p_{FDR}<10^{-15}$), though with a range of effect sizes (Fig.~\figsubref{fig:chaos}{c}), with \rtr explaining the largest share of variance ($R^2=0.96$, $\chi^2=2.59\cdot10^{4}$), \iti yielding the largest Wald statistic ($\chi^2=9.06\cdot10^{4}$, $R^2=0.95$), and \sts being the least regime-dependent ($R^2=0.03$) (Supplementary Material S4 Table~S1).
Thus, the most effective point of distinction between regimes of spatiotemporal chaos is how redundant information is stored across sources and targets. 

Beyond discriminating regimes, \phiidig delivers interpretable computational profiles of each of them (Fig.~\figsubref{fig:chaos}{d}). Order-four synergy \sts is highest in STI ($Z=4.99$, $p_{FDR}=6.0\cdot10^{-7}$), consistent with the coexistence of ordered and chaotic phases, where resolving which phase will locally dominate next requires integrating information from multiple sites. Unique-to-redundant transfer \itr peaks in DT ($Z=4.86$, $p_{FDR}=1.2\cdot10^{-6}$), where colliding defects copy information into a shared, redundant outcome. Unique storage \iti is highest in FDT ($Z=5.40$, $p_{FDR}=6.8\cdot10^{-8}$), where spatial correlations decay rapidly and the information determining each site's future is generated, and remains, local. \rtr, \itj, and \rts are most pronounced in PS ($Z=5.40$, $p_{FDR}\leq7.6\cdot10^{-8}$): persistent redundancy reflects its long-range spatial correlations and \itj its alternation between two sub-lattice states, while \rts plausibly reflects residual chaotic fluctuations converting part of this redundancy into local synergy. In contrast, BMWD does not maximise any atom, consistent with its sparse and weakly interacting defects. Each of these maxima is significant against every other regime (two-sided Mann--Whitney $U$ test, FDR-corrected; statistics refer to the weakest of the four contrasts, with all remaining contrasts reported in Supplementary Material S4 Tables~S2 and S3), and is not driven by the total information carried by each system, as it is robust under normalisation~\cite{liardi2024null} (Supplementary Material S3 Fig.~S4).

Overall, these findings complement previous work characterising coupled map lattice regimes along an order-to-chaos axis~\cite{santoro2023higher}, showing that \phiidig can additionally resolve each regime into a multidimensional profile of computational modes.

\subsection{Neural computation across mammalian phylogeny}

To shed light on the behaviour of more non-linear oscillatory systems, we turn to coupled Stuart–Landau oscillators (i.e.\ the normal form of Hopf bifurcation; Fig.~\figsubref{fig:hopf}{a}), owing to their extensive use both in statistical physics~\cite{landau1944problem, stuart1960non, matthews1990phase}, and, more recently, as large-scale brain models in computational neuroscience~\cite{deco2017dynamics, deco2021rare, deco2019awakening}.

We first explore information dynamics in the state-space of the coupling $G \in [0, 4]$ and the bifurcation parameter $a \in [-0.25, 0.25]$. Here, $G$ globally scales the weights of the connectivity matrix $C$, and $a$ controls the critical behaviour of the local dynamics (see Sec.~\ref{sec:methods}). We consider $C$ as the fixed topology of a randomly generated Watts-Strogatz small-world network~\cite{watts1998collective} with $n=25$ nodes, $k=5$ nearest neighbours, subject to rewiring probability $p=0.5$ (Fig.~\figsubref{fig:hopf}{b}). 
Computing \phiidig between the past and future states of the system, we observe, as expected, a distinctive phase boundary in distributed computation around the bifurcation point ($a\approx0$). On the other hand, strongly dampened and noise-driven subcritical dynamics ($a\lesssim 0$) exhibit information dynamics that remain largely unaffected by the coupling $G$, in contrast to critical and supercritical regimes ($a\gtrsim0$) which show substantial changes. More specifically, we find that \rtr grows with $a$ and $G$, whereas \sts increases for higher $a$ and lower $G$, clearly marking the phase transition point. In other words, synchronised oscillators produce persistent redundancy, while structured fluctuations produce persistent synergy. On the other hand, as expected, \iti is maximised when dynamics are either sufficiently decoupled and/or fluctuating ($G\lesssim0.1$ and $a\lesssim0$).

With these tools in hand, we are then able to broach a long-standing scientific question: does the anatomical wiring of the human brain, irrespective of variations in other biophysical factors, contribute to our distinctive information processing capabilities compared to other mammals~\cite{van2016comparative, changeux2021connectomic, pang2022evolutionary}? To test this, we exploit previously collected structural connectomes from humans and 123 different mammalian species~\cite{assaf2020conservation, suarez2022connectomics}, setting them as the connectivity matrix $C$ for the Hopf model above. We then globally fix the remaining parameters $a=-0.05$ and $G=10^{-3}$, which have been shown to faithfully reproduce BOLD brain activity~\cite{deco2017dynamics, deco2021revisiting}, simulate brain dynamics, and finally apply \phiidig between past and future states. Distributed computation in human brain dynamics differs markedly from that of other mammalian species (Fig.~\figsubref{fig:hopf}{d}) (Mann--Whitney $U$ $p_{FDR} < 10^{-19}$ for all atoms; Supplementary Material S4 Table~S4). The separation is near-total, with a rank-biserial effect size of $|r_{rb}| \geq 0.99$ for nine of the ten distinct atoms, indicating that essentially no non-human connectome overlaps the human range, the sole exception being \rtr ($|r_{rb}|=0.62$). These findings are not driven by the choice of bifurcation parameter, as the differences persist across a range of values of $a$ (Supplementary Material S3 Fig.~S8 and S4 Table~S6). 
Specifically, we find that human brains present far greater highest-order synergy (\sts), exceeding the mammalian mean by $12.1$ mammalian standard deviations ($Z=14.69$, $p_{FDR}=8.6\cdot10^{-49}$) (Fig.~\figsubref{fig:hopf}{c}), as well as stronger interplay between unique and synergistic information (\its, \sti, both $9.3$ standard deviations), and between unique and redundant information (\itr, \rti) (see Supplementary Material S3 Fig.~S6 and S4 Table~S4). Simultaneously, they also display significantly less redundant storage (\rtr), unique storage and transfer (\iti, \itj), and transformation between redundant and synergistic information (\rts, \str).
Overall, these results point to a fundamental evolutionary reorganisation of the information processing profile prescribed by mammalian brain wiring, with the human cortex being endowed with capacity for greater distributed and higher-order forms of computation, at the expense of more local, redundant, and uniquely mediated processes.

Moreover, we examine the low-dimensional topographic organisation of information dynamics in the simulated human brain via PCA, treating brain regions as samples and the z-scored regional maps of the atoms as features (Fig.~\figsubref{fig:hopf}{e}). The first principal component ($PC_1$) explains $95.4\%$ of the variance and is shared by all atoms, whose regional maps all correlate strongly with it ($|r|>0.78$, and $|r|>0.94$ for every atom except \rtr): regions with higher $PC_1$ values display stronger unique storage (\iti, $r\approx0.95$) and weaker persistent redundancy (\rtr, $r=-0.78$) and synergy (\sts, $r=-0.98$) (Supplementary Material S4 Table~S5). This gradient spans from postcentral, medial occipital, and posterior insular regions, which score highest, to paracentral, superior parietal, and dorsomedial frontal regions, which score lowest.

\subsection{Distributed computation in financial markets}

Finally, we turn our attention to the question of whether, and in what form, distributed computation manifests at the macroeconomic scale through the dynamics of financial markets. To this end, we analyse the time series of relative price fluctuations from global market indicators within the S\&P 500, calculating \phiidig between past and future states for every pair of market indicators in the dataset~\cite{fyahoo} and constructing a graph for each information atom, with companies as nodes and atomic contributions as edge weights.

\begin{figure*}[ht]
    \centering
    \includegraphics[width=\textwidth]{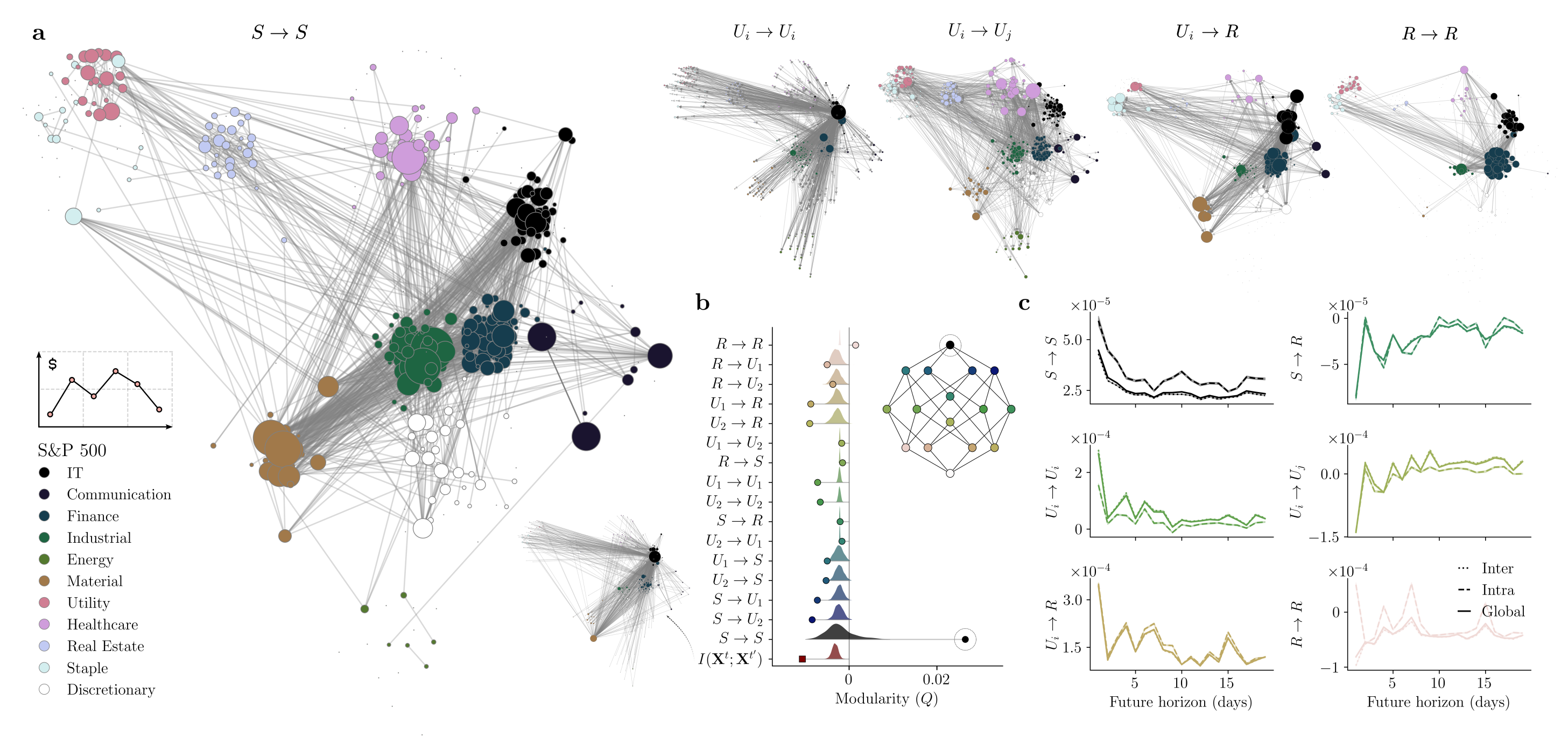}
    \caption{\textbf{Market sectors synergistically store information.}
    \textbf{a} Organisation of distributed computation in financial markets. Networks of information dynamics are plotted for \sts, \iti, \itj, \itr, \rtr (from left to right). Edges represent the inferred level of information processing between the connected financial entities. Nodes are coloured by sector and sized by degree centrality, with the top $10^3$ edges (min-max scaled) displayed. 
    \textbf{b} Modularity of the network of information dynamics for each computational mode, with total information sharing between sectors represented in red ($I(\mathbf{X}^t;\mathbf{X}^{t^\prime})$).
    Points represent empirical values, and non-parametric histograms represent null distributions for strength- and degree-preserving random networks.
    \textbf{c} Timescales of information dynamics within and between market sectors for different temporal lags. \textit{Inter} and \textit{Intra} refer to the information dynamics across and within sectors, respectively, while \textit{Global} denotes the aggregate information dynamics across the entire market.
    }
    \label{fig:finance}
\end{figure*}

The resulting networks of information dynamics across markets cast a striking lens on the operation and organisation of capital (Fig.~\figsubref{fig:finance}{a}). 
First, we observe that financial institutions uniquely copy (\itr) and erase (\rti) information more than most other sectors in the market ($p_{FDR}<0.05$, Mann--Whitney $U$ test; Supplementary Material S4 Table~S8). Using company-level values, only the information technology sector matches them on both measures, while utilities and consumer staples also have comparable \itr. %
Interestingly, we find that the energy sector appears to internally convert synergy into redundancy over time (\str), as well as be the recipient of unique information transfer (\itj) specifically from firms in the health care, utility, and consumer staples sectors. This suggests that it evolves by consolidation while being sensitive to foundational infrastructure and essential goods and services. 

We next determined the modularity ($Q$) of each network of information dynamics with respect to the predefined community structure given by sector divisions, comparing the results against strength- and degree-preserving random networks (Fig.~\figsubref{fig:finance}{b}). This enables us to inspect which forms of distributed computation most closely resemble the organisation of market sectors. 
Fascinatingly, while the total information of the system ($I(\mathbf{X}^t;\mathbf{X}^{t^\prime})$) shows far weaker modularity than null networks ($Q=-0.011$, $Z=-10.58$, $p<2\cdot10^{-3}$), we find that \sts displays significantly stronger modular structure ($Q=0.026$, $Z=9.23$, $p<2\cdot10^{-3}$; Supplementary Material S4 Table~S7).
In other words, synergistic dynamics arise within, as opposed to across, market sectors.
Finally, we explore how distributed computation may occur over multiple timescales by performing the \phiidig procedure with varying past-future temporal lag (Fig.~\figsubref{fig:finance}{c}). Notably, we observe that \sts decreases at longer timescales, whereas \str increases towards zero, suggesting that synergistic interactions dominant at short horizons are gradually absorbed into redundant structure shared across the market.

\section{Discussion} \label{sec:discussion}

\subsection{A geometric theory of information dynamics}

In this work, we asked how the distributed computation performed by a complex system can be captured from first principles via the geometry of its probability distribution. To answer this question, we combined information geometry, which provides a principled notion of distance and projection between probability distributions, with information decomposition, which characterises how information is shared, transferred, and integrated among a system's parts. 
The resulting framework, \phiidig, defines higher-order dependencies as geometric deviations from models explainable by lower-order mechanisms, resolves these dependencies both across and within interaction orders, and organises them into a complete taxonomy of computational modes.

Concretely, we construct submanifolds of conditional independence, each retaining only the dependencies mediated by a subset of components. Synergy is then the divergence between the full distribution and the closest distribution obtainable by combining these projections (\defs), quantifying the dependencies that cannot be explained without the joint contribution of all components.

Importantly, this formalism resolves higher-order structure both across and within interaction orders. Unlike approaches that quantify only a global irreducibility or arrange dependencies by order~\cite{perrone2016hierarchical, amari2001information, schneidman2003network, schneidman2006weak, rosas2016understanding}, \phiidig distinguishes synergies of the same order that reside on different subsets of components. In doing so, it clarifies that same-order synergies need not be orthogonal because their marginal submanifolds overlap. Therefore, it identifies not only how much irreducible structure a system contains, but where it resides and which components it binds together.

Moreover, \phiidig approaches information decomposition from synergy rather than redundancy. Synergy terms are supplied by geometric considerations, while the \phiid lattice serves only to organise their relationships, so that redundancy, unique information, and their transformations naturally emerge from the geometric constraints. 

\subsection{Benchmarking computational modes in discrete and continuous systems}

We first tested whether \phiidig recovers computational structure whose ground truth is known. In Boolean transition systems, each logical mechanism was assigned to the atom dictated by its construction~\cite{varley2023decomposing}: independent flips produced unique storage, parity dynamics produced \sts, XOR and its time-reversed counterpart produced \stx and \xts, respectively, and the giant-bit system produced \rtr. The distinction between the two XOR directions is particularly relevant: because they are time-reverses, they cannot be distinguished by time-symmetric measures of interdependence, whereas \phiidig recovers their directionality.

We obtained the same specificity in continuous systems using ten Gaussian models, each constructed to isolate a distinct atom~\cite{liardi2025scalable}. As the coupling to the target dependency increased, the corresponding atom grew monotonically while the remaining atoms stayed near zero. Together, these benchmarks show that \phiidig can identify both which mode of information transfer is present and how strongly it is expressed across discrete and continuous systems.

\subsection{Distinguishing modes of distributed computation in the natural world}

We next applied \phiidig to systems whose computational organisation is not known a priori. In coupled map lattices, the different universality classes of spatiotemporal chaos exhibited distinct computational signatures, rather than differing simply in the amount of information transferred. Redundant storage \rtr explained the largest share of variance across regimes, with each regime additionally favouring a specific mode, with pattern selection maximising \rtr, \itj and \rts, intermittency \sts, defective turbulence \itr, and fully developed turbulence \iti. Thus, regimes that might appear close along a single order-chaos axis instead perform qualitatively different computations.

In whole-brain models, varying the bifurcation and coupling parameters revealed a reorganisation of computation at criticality: synchronised regimes favoured redundancy, whereas structured fluctuations favoured synergy, in accordance with previous results in the literature~\cite{mediano2022integrated}. Moreover, an analysis of structural connectomes across 123 mammalian species revealed that human connectomes exhibit greater highest-order synergy \sts and stronger transformations between unique, redundant, and synergistic information, alongside reduced redundant and unique storage. This suggests that anatomical wiring itself contributes to the distinctive distributed computation of the human brain~\cite{van2016comparative, changeux2021connectomic, pang2022evolutionary}, complementing empirical evidence for enhanced synergistic organisation in human neural activity~\cite{luppi2022synergistic, luppi2024information}.

Finally, financial markets showed that distributed computation can reveal structure invisible to aggregate information measures. Total time-lagged information was surprisingly less modular than null networks, whereas \sts was significantly more modular with respect to the industrial sector partition. This indicates that market sectors are not primarily reflected in how much information is exchanged, but rather in where irreducible synergistic computation occurs. Over longer time lags, \sts decreased while \str increased, suggesting that short-horizon synergistic dynamics are progressively consolidated into redundant, market-wide structure.

\subsection{Computational portraits of complex systems}

Across these examples, it emerges that the informative object is not a scalar measure, but rather a computational portrait: a profile of the distinct ways in which a system stores, transfers, transforms, and integrates information. We argue that this view is valuable for an interpretable characterisation of complex systems for three main reasons.

First, it provides a multidimensional description of complex dynamics. Much of the literature in complexity science places explanatory weight on a single scalar measure, whether order-to-chaos, integration, or complexity indices of various kinds~\cite{langton1990computation, lopez1995statistical, tononi1998complexity, crutchfield2003regularities, tononi2004information, chialvo2010emergent, barrett2011practical, mediano2018measuring}.
Yet, these systems are known to display behaviour that can vary along different axes. Applications to the CMLs clearly demonstrate this point by showing that regimes with similar overall levels of disorder can perform qualitatively different computations, and the market results show that a system can be simultaneously unstructured in its aggregate information flow and highly structured in its synergy.

Secondly, computational portraits provide a clear operational interpretation. Each atom specifies how information present in one configuration of inputs is processed and transmitted to the outputs, yielding an explanatory vocabulary of information dynamics~\cite{Crutchfield1994-dg, williams2010nonnegative, lizier2012local, mediano2025toward}.

Finally, this approach is readily applicable to a variety of complex systems. Since the framework only requires a joint distribution over inputs and outputs, the same portrait can be computed for logic gates, coupled maps, brain models, and financial markets. This provides a common language for comparing systems that would otherwise be described by domain-specific measures. 

Viewed this way, \phiidig exploits the geometry of a system's probability distribution to provide a principled route from statistical dependence to structured descriptions of distributed computation, enabling the study of complex systems based on how they manipulate information.

\subsection{Limitations and future work}
A technical limitation of our theoretical edifice is that we require the statistical manifold to contain full-support, full-rank, exponential family distributions. %
However, in empirical settings, this restriction can often be alleviated using regularisation techniques such as epsilon-flooring. 
Another potential limitation is that, despite recent advances~\cite{jansma2025fast}, the \phiid framework quickly becomes computationally infeasible for large systems, with the number of atoms growing as the square of the Dedekind sequence~\cite{mediano2025toward}. Nevertheless, calculating the geometric synergies at each order is far more computationally feasible, as it only requires minimising the KL divergence over appropriate interpolations of the distributions. Thus, in these scenarios, the synergy terms $S_i$ can offer a coarser view of the system's higher-order structures. 
Future work could focus on devising a technique to extract meaningful information from the multitude of atoms produced by \phiid for larger systems.
Additionally, the present formulation of \phiidig for continuous systems is restricted to Gaussian settings. Future work could therefore extend its definition and application to non-Gaussian continuous systems, exploiting non-parametric techniques such as rejection sampling, kernel density estimation, and nearest-neighbour estimation.

More broadly, the framework opens several avenues beyond those above. Because \phiidig is defined directly on the geometry of the distribution, it is naturally suited to answer questions of emergence and coarse-graining, where one asks how the computational profile of a system changes under changes of description~\cite{hoel2013quantifying, rosas2019reconciling, barnett2023dynamical, varley2024emergence, hoel2025causal}. It could also be turned from a descriptive tool into a design tool: since the synergy terms are differentiable functions of the distribution's parameters, they may in principle be optimised directly, allowing systems, or learning algorithms, to be steered towards prescribed computational profiles~\cite{kaplanis2023learning, mcsharry2024learning}. 

\subsection{Conclusion} \label{sec:conclusion}
To conclude, we develop a geometric theory of information dynamics that characterises and quantifies the space of distributed computations in complex systems. In doing so, we provide theoretical tools to move beyond scalar summaries of such systems, yielding multidimensional computational portraits that can be used to understand, explain, and ultimately engineer the natural world.

\subsection*{Acknowledgments}
We thank Matteo Neri and Gustavo Menesse for the interesting discussions and useful feedback in the preparation of this manuscript. G.B. is supported by the Leverhulme Trust.

\section{Methods} \label{sec:methods}

\subsection{Transition probability matrices}

To validate the operation of \phiidig on discrete systems, we consider time-evolving Boolean systems characterised by transition probability matrices with specific computational traits.
We consider a bivariate system in which the variables at time $t$, denoted by $\mathbf{X}^t$, evolve into future states $\mathbf{X}^{t^\prime}$ according to a transition probability matrix $\mathbf{T}$ with entries
\begin{equation}
{T_{ij}=P(\mathbf{X}^{t^\prime}=j \mid \mathbf{X}^t=i)} \,. 
\end{equation}

For each TPM, we compute the stationary distribution $\pi$, defined as the unique probability vector satisfying $\pi = \pi \mathbf{T}$, which we calculate via least-squares estimation. The stationary distribution is then used to derive the joint distribution over past and future states simply as $P(\mathbf{X}^t, \mathbf{X}^{t'}) = \pi \cdot \mathbf{T}$, from which \phiidig can be computed.
To analyse different computational regimes, we manually design $T_{ij}$ with specific logical structures by imposing controlled dependencies between past and future states.

\subsection{Coupled-map lattice}

Simple deterministic equations, despite their apparent simplicity, can generate a remarkable variety of dynamical behaviours, ranging from stable fixed points and ordered states to large fluctuations, bifurcations, and chaos~\cite{may1976simple}. Within this family, coupled-map lattices (CMLs) are paradigmatic models of spatially extended non-linear systems, widely used to study the emergence of collective behaviour, spatiotemporal chaos, synchronisation, turbulence, and critical phenomena in statistical physics and complexity science~\cite{kaneko1985spatiotemporal, kaneko1989pattern, kaneko1993theory}. 
Here, we examine the behaviour of this class of systems to determine the computational profile of universality classes in complex dynamics. 
Specifically, we consider a discrete spacetime ring lattice of logistic maps ($N=100$), coupled to the nearest neighbours via diffusion dynamics. More formally, the system's evolution is governed by:
\begin{equation} 
\begin{split}
x_i(t+1) = & (1-\epsilon) f[x_i(t)] \,+ \\ & + \frac{\epsilon}{2} \left( f[x_{i-1}(t)] + f[x_{i+1}(t)] \right) \,,
\label{eq:cml}
\end{split}
\end{equation}
where $\epsilon\in[0,1]$ is the coupling strength and $f[x]=1-1.75x^2$ is the non-linear local map. Specifically, the quantity we analyse is the recorded activity
\begin{equation}
\tilde{x}_i(t) = x_i(t) + \eta_i(t) \,, 
\label{eq:cml_obs}
\end{equation}
with $\eta_i(t)\sim 10^{-3}\cdot\mathcal{N}(0,1)$. In other words, we consider the deterministic trajectory with a small amount of additive Gaussian observation noise. We remark that $\eta$ is purely observational: it is added to the stored trajectory and never re-enters the map, so the dynamics of Eq.~\eqref{eq:cml} are untouched by construction. We include this term for numerical reasons, to avoid the degenerate empirical distributions produced by perfect synchronisation~\cite{jost2001spectral} or clustering phenomena~\cite{belykh2000invariant} induced by the lattice's translational symmetry, which would make the information-geometric projections underlying \phiidig ill-defined. In most regimes, this has negligible effect, with its amplitude amounting to $0.20$--$0.22\%$ of the standard deviation of the noise-free lattice activity across all regimes.
However, in the case of pattern selection, it has more concrete consequences. Here, the noise-free dynamics settle onto a period-4 orbit, so that each node visits only four distinct activity values, each a quarter of the time and repeated to machine precision. The quantile bin edges of the nonparametric estimator then collapse onto those four values at every node, leaving most bins with zero width. Hence, the observational noise separates them.
We verify explicitly that the addition of the observational noise leaves the recorded activity distributions, their moments and the dynamical activity of every regime unchanged (Supplementary Material S5).

It is well established that varying the control parameter $\epsilon$ causes CMLs to exhibit dramatic phase transitions and enter into specific qualitative regimes of interest to diverse scientific fields~\cite{chazottes2005dynamics}. 
Here, we examine the canonical classes of pattern selection (PS: $\epsilon=0.12$), Brownian motion with defects (BMWD: $\epsilon=0.08$), spatiotemporal intermittency II (STI: $\epsilon=0.3$), defective turbulence (DT: $\epsilon=0.068$), and fully developed turbulence (FDT: $\epsilon=0.05$)~\cite{kaneko1989pattern, santoro2023higher}.

For each regime, we perform simulations with lattices of 100 nodes, over $2\cdot 10^6$ discrete steps after a $10^5$ step transient time, with 20 repeated iterations. For each realisation, we compute \phiidig between all node pairs, considering the nodes' past states as sources and future states as targets.
Because the stationary distributions of these maps are strongly non-Gaussian, being typically sharply peaked, bounded, or bimodal near the turning points of the local map, a Gaussian parametrisation is inappropriate in this setting. We therefore estimate \phiidig nonparametrically, discretising the past and future states of each pair of nodes and employing the discrete \phiid solver.
Full details of this procedure, together with its validation, are reported in Supplementary Material S5.

\subsection{Whole-brain model}
To understand distributed computation in models of spontaneous large-scale cortical activity, we turn to the well-explored setup of coupled Stuart-Landau oscillators, corresponding to the normal form of an Andronov-Hopf bifurcation~\cite{hopf1942abzweigung, andronov2013theory}: 
\begin{equation} \label{eq:Hopf}
\begin{aligned}
\frac{d x_i(t)}{d t} &= \left(a_i - x_i(t)^2 - y_i(t)^2\right) x_i(t) - \omega_i y_i(t) + \\
&\quad + G \sum_{j=1}^n C_{ij}(x_j(t) - x_i(t)) + \beta\eta_{x_i}(t) \\
\frac{d y_i(t)}{d t} &= \left(a_i - x_i(t)^2 - y_i(t)^2\right) y_i(t) + \omega_i x_i(t) + \\
&\quad + G \sum_{j=1}^n C_{ij}(y_j(t) - y_i(t)) + \beta\eta_{y_i}(t) \,,
\end{aligned}
\end{equation}
where $i = 1,\ldots,n$ indexes each brain region, $x_i(t)$ and $y_i(t)$ are the real and imaginary parts of the complex oscillator amplitude at node $i$, with the real part being the simulated brain activity. The bifurcation parameter $a_i$ controls the local dynamics, defining whether activity operates as a damped noise-driven fixed point ($a_i \ll 0$), akin to an Ornstein–Uhlenbeck process, a pure limit-cycle oscillation ($a_i > 0$), akin to Wilson–Cowan dynamics, or a critical stochastically structured fluctuation that mixes the two ($a_i \approx 0$). Local dynamics operate with frequency $f_i=\omega_i/2\pi$, which we set uniformly to \SI{.05}{\hertz} to emulate infra-slow BOLD signals from fMRI, and are subject to independent isotropic white noise $\eta_{x_i}(t), \eta_{y_i}(t) \sim \mathcal{N}(0,1)$ scaled by $\beta$, which we set to $0.01$. Interactions between regions are mediated through the connectivity matrix $C_{ij}$, via linear difference-based coupling, with inputs equally scaled by the global coupling strength $G$. The Hopf model has been used extensively in neuroscience to reproduce and probe the mechanisms of large-scale brain dynamics in health and disease~\cite{deco2026whole}. 
Since the best working point for fitting whole-brain neuroimaging dynamics has been shown to lie at the brink of the bifurcation, i.e.\ with a slightly negative value, we fix the Hopf parameters as $a=-0.05$ and $G=10^{-3}$~\cite{deco2017dynamics, deco2021revisiting}.  
Simulations were run at $dt=1$, with $10^3$ equilibration steps, with trials lasting $1200$ steps, with the mammalian runs narrow-band filtered between 0.008 and 0.09 Hz, as is typical in fMRI analysis to isolate BOLD from baseline drifts and physiological artefacts.
Since the simulated signals depart from full Gaussianity past the bifurcation, where the local dynamics become a limit-cycle oscillation, we employ a Gaussian copula approach for computing \phiidig. We report the details of this construction and the validation on the data under examination in Supplementary Material S5.
To characterise the topographic organisation of information dynamics in the simulated human brain, we z-scored the regional map of each atom, averaged across the 200 subjects, and performed a principal component analysis with brain regions as samples and atoms as features. The sign of $PC_1$ was chosen so that higher values indicate stronger \iti, and each atom's loading was quantified as the Pearson correlation between its z-scored map and the $PC_1$ map.

\subsection{Mammalian connectomes}

Full details on the MaMI database of mammalian connectomes can be found in the original publications~\cite{assaf2020conservation, suarez2022connectomics}, but are reiterated here in brief. 

Ex vivo diffusion tractography and anatomical MRI scans were acquired from 232 brains across 123 species, all sourced from incidental deaths and fixed in 10\% formaldehyde. A unified HARDI protocol was employed ($b = 1000 \text{ s/mm}^2$) using 60 or 64 gradient directions, with acquisition parameters adjusted to maintain an SNR $>20$ across varying brain volumes~\cite{assaf2020conservation}. Diffusion-weighted MRI (dwMRI) data were processed using ExploreDTI, including anisotropic smoothing and distortion correction, followed by fibre-orientation density function estimation via constrained spherical deconvolution~\cite{tournier2004direct}. Whole-brain tractography was performed with a 0.2 seed threshold, ensuring approximately 90\% of streamlines terminated in grey matter. Following the exclusion of cerebellar and external projection fibres, brains were parcellated into 100 nodes using a $k$-means clustering algorithm based on the similarity in connectivity profile between pairs of endpoints~\cite{suarez2022connectomics}. 

This uniform parcellation scheme allowed for the generation of sparse weighted structural connectivity matrices of the same size, providing a consistent reference space for cross-species comparison. Of the $122$ non-human species, eight were excluded from the statistical comparisons with the human connectomes: seven whose connectomes could not be matched to the taxonomic annotations of the database, and the common tree shrew, which does not belong to any of the $11$ mammalian orders considered. The comparisons therefore included $114$ species across $11$ orders.

\subsection{Human connectomes}

Full details on the human connectome acquisition and construction can be found in the original publications~\cite{domhof2021parcellation}, but are reiterated here in brief.

Empirical connectomes were extracted from 200 healthy, unrelated subjects (96 males, age $28.5 \pm 3.5$ years) from the HCP S1200 release~\cite{van2012human, van2013wu}. dwMRI data were preprocessed for bias field correction, tissue segmentation, and motion/eddy-current distortion correction using a combination of FreeSurfer, MRtrix3, and FSL~\cite{tournier2019mrtrix3, fischl2012freesurfer, jenkinson2012fsl}. Whole-brain tractography was reconstructed using a multi-shell, multi-tissue constrained spherical deconvolution algorithm. A probabilistic tracking algorithm was employed with $10^7$ streamlines and standardized parameters (step size $= 0.625$ mm, angle $= 45^{\circ}$, FOD threshold $= 0.06$). Finally, empirical structural connectivity was generated by transforming the Schaefer-100 parcellation~\cite{schaefer2018local} into native space and mapping streamlines to the resulting regions of interest to maintain the same network size as the mammalian connectomes.

\subsection{Communication model}

To allow robust comparison between connectomes and use in the Hopf model, we performed additional standardisation. First, anatomical networks $W$ were log-transformed and min-max scaled to the interval $[0, 1]$, yielding the weights $W'$. Then we applied the routing protocol communication model of shortest-path efficiency to convert structural connectivity to a more effective form of connectivity~\cite{seguin2020network}. Following the logic of the Floyd-Warshall algorithm, weights $W'_{ij}$ were mapped to lengths $d_{ij}$ using an inverse transform and the shortest path efficiency between nodes $i$ and $j$ was defined as the inverse of the shortest path length:
\begin{equation}
\Lambda_{ij}^* = \min_{\Omega_{ij}} \sum_{l=0}^{k} d_{u_l u_{l+1}}
\end{equation}
where $\Lambda_{ij}^*$ represents the minimum sum of edge lengths along all possible paths $\Omega_{ij} = \{i, u_1, \dots, u_k, j\}$ connecting the nodes, so that $E_{ij}=1/\Lambda_{ij}^*$. The resulting efficiency matrix was globally scaled by a factor of $0.1$ to define the final structural coupling matrix $C$ for simulations.

\subsection{Financial markets}

To evaluate the potential distributed computation in empirical non-biological complex systems, financial time-series data were retrieved for companies within the S\&P 500 from Yahoo Finance~\cite{fyahoo}. The constituent list and corresponding Global Industry Classification Standard (GICS) sectors were scraped, and daily adjusted closing prices were obtained for the period ranging from 2010 to 2026. Any assets containing missing values during the requested period were excluded from the analysis, leaving 421 elements. Multivariate time series were then transformed into log-returns for analysis of relative price fluctuations across distinct market sectors.
Since the considered log-returns are strongly heavy-tailed, we compute \phiidig under a Gaussian-copula model. Details and validation of this procedure are reported in Supplementary Material S5.

\bibliographystyle{ieeetr}
\bibliography{bib}

\providecommand{\noopsort}[1]{}\providecommand{\singleletter}[1]{#1}%
\begin{thebibliography}{100}

\bibitem{Chu2018-vl}
D.~Chu, M.~Prokopenko, and J.~C.~J. Ray, ``Computation by natural systems,''
  {\em Interface Focus}, vol.~8, no.~6, p.~20180058, 2018.

\bibitem{barack2021two}
D.~L. Barack and J.~W. Krakauer, ``Two views on the cognitive brain,'' {\em
  Nature Reviews Neuroscience}, vol.~22, no.~6, pp.~359--371, 2021.

\bibitem{Crutchfield1995-ep}
J.~P. Crutchfield and M.~Mitchell, ``The evolution of emergent computation,''
  {\em Proc. Natl. Acad. Sci. U.S.A.}, vol.~92, no.~23, pp.~10742--10746, 1995.

\bibitem{jensen2022complexity}
H.~J. Jensen, {\em Complexity science: the study of emergence}.
\newblock Cambridge University Press, 2022.

\bibitem{Piccinini2015-il}
G.~Piccinini, {\em Physical computation: A mechanistic account}.
\newblock Oxford University Press, 2015.

\bibitem{Mitchell2012-xo}
M.~Mitchell, ``Biological computation,'' {\em Comput. J.}, vol.~55, no.~7,
  pp.~852--855, 2012.

\bibitem{newman2003structure}
M.~E. Newman, ``The structure and function of complex networks,'' {\em SIAM
  review}, vol.~45, no.~2, pp.~167--256, 2003.

\bibitem{Wolpert2026-dp}
D.~H. Wolpert and J.~Korbel, ``What does it mean for a system to compute?,''
  {\em J. Phys. Complex}, vol.~7, no.~1, p.~015010, 2026.

\bibitem{israeli2004computational}
N.~Israeli and N.~Goldenfeld, ``Computational irreducibility and the
  predictability of complex physical systems,'' {\em Physical review letters},
  vol.~92, no.~7, p.~074105, 2004.

\bibitem{shalizi2001computational}
C.~R. Shalizi and J.~P. Crutchfield, ``Computational mechanics: Pattern and
  prediction, structure and simplicity,'' {\em Journal of statistical physics},
  vol.~104, no.~3, pp.~817--879, 2001.

\bibitem{tumer2012collectives}
K.~Tumer and D.~Wolpert, {\em Collectives and the design of complex systems}.
\newblock Springer Science \& Business Media, 2012.

\bibitem{malone2015handbook}
T.~W. Malone and M.~Bernstein, {\em Handbook of collective intelligence}.
\newblock MIT press, 2015.

\bibitem{anderson1972more}
P.~W. Anderson, ``More is different: Broken symmetry and the nature of the
  hierarchical structure of science.,'' {\em Science}, vol.~177, no.~4047,
  pp.~393--396, 1972.

\bibitem{krakauer2020information}
D.~Krakauer, N.~Bertschinger, E.~Olbrich, J.~C. Flack, and N.~Ay, ``The
  information theory of individuality,'' {\em Theory in Biosciences}, vol.~139,
  no.~2, pp.~209--223, 2020.

\bibitem{liardi2026statistical}
A.~Liardi, F.~E. Rosas, D.~Marinazzo, T.~F. Varley, and P.~A. Mediano,
  ``Statistical dark matter: Synergy is everywhere but is hard to capture,''
  {\em arXiv preprint arXiv:2610.03764}, 2026.

\bibitem{Crutchfield1994-dg}
J.~P. Crutchfield, ``The calculi of emergence: computation, dynamics and
  induction,'' {\em Physica D}, vol.~75, no.~1-3, pp.~11--54, 1994.

\bibitem{lizier2012local}
J.~T. Lizier, {\em The Local Information Dynamics of Distributed Computation in
  Complex Systems}.
\newblock Springer Science \& Business Media, 2012.

\bibitem{amari2016information}
S.-i. Amari, {\em Information geometry and its applications}.
\newblock Springer, 2016.

\bibitem{amari2001information}
S.-I. Amari, ``Information geometry on hierarchy of probability
  distributions,'' {\em IEEE transactions on information theory}, vol.~47,
  no.~5, pp.~1701--1711, 2001.

\bibitem{olbrich2015information}
E.~Olbrich, N.~Bertschinger, and J.~Rauh, ``Information decomposition and
  synergy,'' {\em Entropy}, vol.~17, no.~5, pp.~3501--3517, 2015.

\bibitem{perrone2016hierarchical}
P.~Perrone and N.~Ay, ``Hierarchical quantification of synergy in channels,''
  {\em Frontiers in Robotics and AI}, vol.~2, p.~35, 2016.

\bibitem{niu2019measure}
X.~Niu and C.~J. Quinn, ``A measure of synergy, redundancy, and unique
  information using information geometry,'' in {\em 2019 IEEE International
  Symposium on Information Theory (ISIT)}, pp.~3127--3131, IEEE, 2019.

\bibitem{schneidman2003network}
E.~Schneidman, S.~Still, M.~J. Berry, and W.~Bialek, ``Network information and
  connected correlations,'' {\em Physical review letters}, vol.~91, no.~23,
  p.~238701, 2003.

\bibitem{schneidman2003synergy}
E.~Schneidman, W.~Bialek, and M.~J. Berry, ``Synergy, redundancy, and
  independence in population codes,'' {\em Journal of Neuroscience}, vol.~23,
  no.~37, pp.~11539--11553, 2003.

\bibitem{schneidman2006weak}
E.~Schneidman, M.~J. Berry, R.~Segev, and W.~Bialek, ``Weak pairwise
  correlations imply strongly correlated network states in a neural
  population,'' {\em Nature}, vol.~440, no.~7087, pp.~1007--1012, 2006.

\bibitem{rosas2016understanding}
F.~Rosas, V.~Ntranos, C.~J. Ellison, S.~Pollin, and M.~Verhelst,
  ``Understanding interdependency through complex information sharing,'' {\em
  Entropy}, vol.~18, no.~2, p.~38, 2016.

\bibitem{williams2010nonnegative}
P.~L. Williams and R.~D. Beer, ``Nonnegative decomposition of multivariate
  information,'' {\em arXiv preprint arXiv:1004.2515}, 2010.

\bibitem{mediano2025toward}
P.~A. Mediano, F.~E. Rosas, A.~I. Luppi, R.~L. Carhart-Harris, D.~Bor, A.~K.
  Seth, and A.~B. Barrett, ``Toward a unified taxonomy of information dynamics
  via integrated information decomposition,'' {\em Proceedings of the National
  Academy of Sciences}, vol.~122, no.~39, p.~e2423297122, 2025.

\bibitem{beer2015information}
R.~D. Beer and P.~L. Williams, ``Information processing and dynamics in
  minimally cognitive agents,'' {\em Cognitive Science}, vol.~39, no.~1,
  pp.~1--38, 2015.

\bibitem{wibral2017partial}
M.~Wibral, V.~Priesemann, J.~W. Kay, J.~T. Lizier, and W.~A. Phillips,
  ``Partial information decomposition as a unified approach to the
  specification of neural goal functions,'' {\em Brain and Cognition},
  vol.~112, pp.~25--38, 2017.

\bibitem{ince2017measuring}
R.~A. Ince, ``Measuring multivariate redundant information with pointwise
  common change in surprisal,'' {\em Entropy}, vol.~19, no.~7, p.~318, 2017.

\bibitem{tax2017partial}
T.~M. Tax, P.~A. Mediano, and M.~Shanahan, ``The partial information
  decomposition of generative neural network models,'' {\em Entropy}, vol.~19,
  no.~9, p.~474, 2017.

\bibitem{proca2024synergistic}
A.~M. Proca, F.~E. Rosas, A.~I. Luppi, D.~Bor, M.~Crosby, and P.~A. Mediano,
  ``Synergistic information supports modality integration and flexible learning
  in neural networks solving multiple tasks,'' {\em PLOS Computational
  Biology}, vol.~20, no.~6, p.~e1012178, 2024.

\bibitem{luppi2022synergistic}
A.~I. Luppi, P.~A. Mediano, F.~E. Rosas, N.~Holland, T.~D. Fryer, J.~T.
  O’Brien, J.~B. Rowe, D.~K. Menon, D.~Bor, and E.~A. Stamatakis, ``A
  synergistic core for human brain evolution and cognition,'' {\em Nature
  neuroscience}, vol.~25, no.~6, pp.~771--782, 2022.

\bibitem{varley2023partial}
T.~F. Varley, M.~Pope, M.~Grazia, Joshua, and O.~Sporns, ``Partial entropy
  decomposition reveals higher-order information structures in human brain
  activity,'' {\em Proceedings of the National Academy of Sciences}, vol.~120,
  no.~30, p.~e2300888120, 2023.

\bibitem{luppi2024information}
A.~I. Luppi, F.~E. Rosas, P.~A. Mediano, D.~K. Menon, and E.~A. Stamatakis,
  ``Information decomposition and the informational architecture of the
  brain,'' {\em Trends in Cognitive Sciences}, 2024.

\bibitem{rajpal2023quantifying}
H.~Rajpal, C.~von Stengel, P.~A. Mediano, F.~E. Rosas, E.~Viegas, P.~A.
  Marquet, and H.~J. Jensen, ``Quantifying hierarchical selection,'' {\em arXiv
  preprint arXiv:2310.20386}, 2023.

\bibitem{finn2018quantifying}
C.~Finn and J.~T. Lizier, ``Quantifying information modification in cellular
  automata using pointwise partial information decomposition,'' in {\em
  Artificial Life Conference Proceedings}, pp.~386--387, MIT Press One Rogers
  Street, Cambridge, MA 02142-1209, USA journals-info~…, 2018.

\bibitem{rosas2018information}
F.~Rosas, P.~A. Mediano, M.~Ugarte, and H.~J. Jensen, ``An
  information-theoretic approach to self-organisation: Emergence of complex
  interdependencies in coupled dynamical systems,'' {\em Entropy}, vol.~20,
  no.~10, p.~793, 2018.

\bibitem{liardi2025simple}
A.~Liardi, M.~I. Sas, G.~Blackburne, W.~J. Knottenbelt, P.~A. Mediano, and
  H.~J. Jensen, ``Simple physical systems as a reference for multivariate
  information dynamics,'' {\em Chaos: An Interdisciplinary Journal of Nonlinear
  Science}, vol.~35, no.~10, 2025.

\bibitem{rajpal2025information}
H.~Rajpal, C.~v. Stengel, P.~A. Mediano, F.~E. Rosas, E.~Viegas, P.~A. Marquet,
  and H.~J. Jensen, ``Information dynamics and the emergence of high-order
  individuality in ecosystems,'' {\em Communications Biology}, vol.~8, no.~1,
  p.~1231, 2025.

\bibitem{sas2025improved}
M.~I. Sas, F.~E. Rosas, H.~Rajpal, D.~Bor, H.~J. Jensen, and P.~A. Mediano,
  ``Improved estimators of causal emergence for large systems,'' {\em arXiv
  preprint arXiv:2601.00013}, 2025.

\bibitem{liardi2026mathematical}
A.~Liardi, K.~J. Down, G.~Blackburne, M.~Neri, and P.~A. Mediano, ``The
  mathematical landscape of partial information decomposition: A comprehensive
  review of properties and measures,'' {\em arXiv preprint arXiv:2603.06678},
  2026.

\bibitem{cover1999elements}
T.~M. Cover and J.~A. Thomas, {\em Elements of Information Theory}.
\newblock John Wiley \& Sons, 1999.

\bibitem{harder2013bivariate}
M.~Harder, C.~Salge, and D.~Polani, ``Bivariate measure of redundant
  information,'' {\em Physical Review E}, vol.~87, no.~1, p.~012130, 2013.

\bibitem{griffith2014intersection}
V.~Griffith, E.~K. Chong, R.~G. James, C.~J. Ellison, and J.~P. Crutchfield,
  ``Intersection information based on common randomness,'' {\em Entropy},
  vol.~16, no.~4, pp.~1985--2000, 2014.

\bibitem{bertschinger2014quantifying}
N.~Bertschinger, J.~Rauh, E.~Olbrich, J.~Jost, and N.~Ay, ``Quantifying unique
  information,'' {\em Entropy}, vol.~16, no.~4, pp.~2161--2183, 2014.

\bibitem{griffith2014quantifying}
V.~Griffith and C.~Koch, ``Quantifying synergistic mutual information,'' in
  {\em Guided self-organization: inception}, pp.~159--190, Springer, 2014.

\bibitem{griffith2015quantifying}
V.~Griffith and T.~Ho, ``Quantifying redundant information in predicting a
  target random variable,'' {\em Entropy}, vol.~17, no.~7, pp.~4644--4653,
  2015.

\bibitem{barrett2015exploration}
A.~B. Barrett, ``Exploration of synergistic and redundant information sharing
  in static and dynamical gaussian systems,'' {\em Physical Review E}, vol.~91,
  no.~5, p.~052802, 2015.

\bibitem{goodwell2017temporal}
A.~E. Goodwell and P.~Kumar, ``Temporal information partitioning:
  Characterizing synergy, uniqueness, and redundancy in interacting
  environmental variables,'' {\em Water Resources Research}, vol.~53, no.~7,
  pp.~5920--5942, 2017.

\bibitem{james2018unique}
R.~G. James, J.~Emenheiser, and J.~P. Crutchfield, ``Unique information via
  dependency constraints,'' {\em Journal of Physics A: Mathematical and
  Theoretical}, vol.~52, no.~1, p.~014002, 2018.

\bibitem{kay2018exact}
J.~W. Kay and R.~A. Ince, ``Exact partial information decompositions for
  gaussian systems based on dependency constraints,'' {\em Entropy}, vol.~20,
  no.~4, p.~240, 2018.

\bibitem{james2018uniquekey}
R.~G. James, J.~Emenheiser, and J.~P. Crutchfield, ``Unique information and
  secret key agreement,'' {\em Entropy}, vol.~21, no.~1, p.~12, 2018.

\bibitem{finn2018pointwise}
C.~Finn and J.~T. Lizier, ``Pointwise partial information decompositionusing
  the specificity and ambiguity lattices,'' {\em Entropy}, vol.~20, no.~4,
  p.~297, 2018.

\bibitem{kay2024partial}
J.~W. Kay, ``A partial information decomposition for multivariate gaussian
  systems based on information geometry,'' {\em Entropy}, vol.~26, no.~7,
  p.~542, 2024.

\bibitem{sigtermans2020partial}
D.~Sigtermans, ``A partial information decomposition based on causal tensors,''
  {\em arXiv preprint arXiv:2001.10481}, 2020.

\bibitem{sigtermans2020towards}
D.~Sigtermans, ``Towards a framework for observational causality from time
  series: when shannon meets turing,'' {\em Entropy}, vol.~22, no.~4, p.~426,
  2020.

\bibitem{makkeh2021introducing}
A.~Makkeh, A.~J. Gutknecht, and M.~Wibral, ``Introducing a differentiable
  measure of pointwise shared information,'' {\em Physical Review E}, vol.~103,
  no.~3, p.~032149, 2021.

\bibitem{schick2021partial}
K.~Schick-Poland, A.~Makkeh, A.~J. Gutknecht, P.~Wollstadt, A.~Sturm, and
  M.~Wibral, ``A partial information decomposition for discrete and continuous
  variables,'' {\em arXiv preprint arXiv:2106.12393}, 2021.

\bibitem{ehrlich2024partial}
D.~A. Ehrlich, K.~Schick-Poland, A.~Makkeh, F.~Lanfermann, P.~Wollstadt, and
  M.~Wibral, ``Partial information decomposition for continuous variables based
  on shared exclusions: Analytical formulation and estimation,'' {\em Physical
  Review E}, vol.~110, no.~1, p.~014115, 2024.

\bibitem{kolchinsky2022novel}
A.~Kolchinsky, ``A novel approach to the partial information decomposition,''
  {\em Entropy}, vol.~24, no.~3, p.~403, 2022.

\bibitem{venkatesh2022partial}
P.~Venkatesh and G.~Schamberg, ``Partial information decomposition via
  deficiency for multivariate gaussians,'' in {\em 2022 IEEE International
  Symposium on Information Theory (ISIT)}, pp.~2892--2897, IEEE, 2022.

\bibitem{mages2023decomposing}
T.~Mages and C.~Rohner, ``Decomposing and tracing mutual information by
  quantifying reachable decision regions,'' {\em Entropy}, vol.~25, no.~7,
  p.~1014, 2023.

\bibitem{lyu2024explicit}
A.~Lyu, A.~Clark, and N.~Raviv, ``Explicit formula for partial information
  decomposition,'' in {\em 2024 IEEE International Symposium on Information
  Theory (ISIT)}, pp.~2329--2334, IEEE, 2024.

\bibitem{varley2023decomposing}
T.~F. Varley, ``Decomposing past and future: Integrated information
  decomposition based on shared probability mass exclusions,'' {\em Plos one},
  vol.~18, no.~3, p.~e0282950, 2023.

\bibitem{james2017multivariate}
R.~G. James and J.~P. Crutchfield, ``Multivariate dependence beyond shannon
  information,'' {\em Entropy}, vol.~19, no.~10, p.~531, 2017.

\bibitem{liardi2025scalable}
A.~Liardi, G.~Blackburne, H.~Rajpal, F.~E. Rosas, and P.~A. Mediano, ``A
  scalable estimator of high-order information in complex dynamical systems,''
  {\em arXiv preprint arXiv:2506.18498}, 2025.

\bibitem{kaneko1985spatiotemporal}
K.~Kaneko, ``Spatiotemporal intermittency in coupled map lattices,'' {\em
  Progress of Theoretical Physics}, vol.~74, no.~5, pp.~1033--1044, 1985.

\bibitem{kaneko1989pattern}
K.~Kaneko, ``Pattern dynamics in spatiotemporal chaos: Pattern selection,
  diffusion of defect and pattern competition intermettency,'' {\em Physica D:
  Nonlinear Phenomena}, vol.~34, no.~1-2, pp.~1--41, 1989.

\bibitem{kaneko1993theory}
K.~Kaneko, ``Theory and applications of coupled map lattices,'' {\em Nonlinear
  science: theory and applications}, 1993.

\bibitem{jost2001spectral}
J.~Jost and M.~P. Joy, ``Spectral properties and synchronization in coupled map
  lattices,'' {\em Physical Review E}, vol.~65, no.~1, p.~016201, 2001.

\bibitem{santoro2023higher}
A.~Santoro, F.~Battiston, G.~Petri, and E.~Amico, ``Higher-order organization
  of multivariate time series,'' {\em Nature physics}, vol.~19, no.~2,
  pp.~221--229, 2023.

\bibitem{liardi2024null}
A.~Liardi, F.~E. Rosas, R.~L. Carhart-Harris, G.~Blackburne, D.~Bor, and P.~A.
  Mediano, ``Null models for comparing information decomposition across complex
  systems,'' {\em PLOS Computational Biology}, vol.~21, no.~11, p.~e1013629,
  2025.

\bibitem{landau1944problem}
L.~D. Landau, ``On the problem of turbulence,'' in {\em Dokl. Akad. Nauk USSR},
  vol.~44, p.~311, 1944.

\bibitem{stuart1960non}
J.~T. Stuart, ``On the non-linear mechanics of wave disturbances in stable and
  unstable parallel flows part 1. the basic behaviour in plane poiseuille
  flow,'' {\em Journal of Fluid Mechanics}, vol.~9, no.~3, pp.~353--370, 1960.

\bibitem{matthews1990phase}
P.~C. Matthews and S.~H. Strogatz, ``Phase diagram for the collective behavior
  of limit-cycle oscillators,'' {\em Physical review letters}, vol.~65, no.~14,
  p.~1701, 1990.

\bibitem{deco2017dynamics}
G.~Deco, M.~L. Kringelbach, V.~K. Jirsa, and P.~Ritter, ``The dynamics of
  resting fluctuations in the brain: metastability and its dynamical cortical
  core,'' {\em Scientific reports}, vol.~7, no.~1, p.~3095, 2017.

\bibitem{deco2021rare}
G.~Deco, Y.~S. Perl, P.~Vuust, E.~Tagliazucchi, H.~Kennedy, and M.~L.
  Kringelbach, ``Rare long-range cortical connections enhance human information
  processing,'' {\em Current Biology}, vol.~31, no.~20, pp.~4436--4448, 2021.

\bibitem{deco2019awakening}
G.~Deco, J.~Cruzat, J.~Cabral, E.~Tagliazucchi, H.~Laufs, N.~K. Logothetis, and
  M.~L. Kringelbach, ``Awakening: Predicting external stimulation to force
  transitions between different brain states,'' {\em Proceedings of the
  National Academy of Sciences}, vol.~116, no.~36, pp.~18088--18097, 2019.

\bibitem{watts1998collective}
D.~J. Watts and S.~H. Strogatz, ``Collective dynamics of
  ‘small-world’networks,'' {\em nature}, vol.~393, no.~6684, pp.~440--442,
  1998.

\bibitem{van2016comparative}
M.~P. Van~den Heuvel, E.~T. Bullmore, and O.~Sporns, ``Comparative
  connectomics,'' {\em Trends in cognitive sciences}, vol.~20, no.~5,
  pp.~345--361, 2016.

\bibitem{changeux2021connectomic}
J.-P. Changeux, A.~Goulas, and C.~C. Hilgetag, ``A connectomic hypothesis for
  the hominization of the brain,'' {\em Cerebral cortex}, vol.~31, no.~5,
  pp.~2425--2449, 2021.

\bibitem{pang2022evolutionary}
J.~C. Pang, J.~K. Rilling, J.~A. Roberts, M.~P. Van Den~Heuvel, and L.~Cocchi,
  ``Evolutionary shaping of human brain dynamics,'' {\em Elife}, vol.~11,
  p.~e80627, 2022.

\bibitem{assaf2020conservation}
Y.~Assaf, A.~Bouznach, O.~Zomet, A.~Marom, and Y.~Yovel, ``Conservation of
  brain connectivity and wiring across the mammalian class,'' {\em Nature
  Neuroscience}, vol.~23, no.~7, pp.~805--808, 2020.

\bibitem{suarez2022connectomics}
L.~E. Suarez, Y.~Yovel, M.~P. van~den Heuvel, O.~Sporns, Y.~Assaf, G.~Lajoie,
  and B.~Misic, ``A connectomics-based taxonomy of mammals,'' {\em Elife},
  vol.~11, p.~e78635, 2022.

\bibitem{deco2021revisiting}
G.~Deco, D.~Vidaurre, and M.~L. Kringelbach, ``Revisiting the global workspace
  orchestrating the hierarchical organization of the human brain,'' {\em Nature
  human behaviour}, vol.~5, no.~4, pp.~497--511, 2021.

\bibitem{fyahoo}
``Yahoo finance.'' \url{https://finance.yahoo.com}.
\newblock Accessed: 2026-01-15.

\bibitem{mediano2022integrated}
P.~A. Mediano, F.~E. Rosas, J.~C. Farah, M.~Shanahan, D.~Bor, and A.~B.
  Barrett, ``Integrated information as a common signature of dynamical and
  information-processing complexity,'' {\em Chaos: An Interdisciplinary Journal
  of Nonlinear Science}, vol.~32, no.~1, 2022.

\bibitem{langton1990computation}
C.~G. Langton, ``Computation at the edge of chaos: Phase transitions and
  emergent computation,'' {\em Physica D: nonlinear phenomena}, vol.~42,
  no.~1-3, pp.~12--37, 1990.

\bibitem{lopez1995statistical}
R.~Lopez-Ruiz, H.~L. Mancini, and X.~Calbet, ``A statistical measure of
  complexity,'' {\em Physics letters A}, vol.~209, no.~5-6, pp.~321--326, 1995.

\bibitem{tononi1998complexity}
G.~Tononi, G.~M. Edelman, and O.~Sporns, ``Complexity and coherency:
  integrating information in the brain,'' {\em Trends in cognitive sciences},
  vol.~2, no.~12, pp.~474--484, 1998.

\bibitem{crutchfield2003regularities}
J.~P. Crutchfield and D.~P. Feldman, ``Regularities unseen, randomness
  observed: Levels of entropy convergence,'' {\em Chaos: An Interdisciplinary
  Journal of Nonlinear Science}, vol.~13, no.~1, pp.~25--54, 2003.

\bibitem{tononi2004information}
G.~Tononi, ``An information integration theory of consciousness,'' {\em BMC
  Neuroscience}, vol.~5, pp.~1--22, 2004.

\bibitem{chialvo2010emergent}
D.~R. Chialvo, ``Emergent complex neural dynamics,'' {\em Nature physics},
  vol.~6, no.~10, pp.~744--750, 2010.

\bibitem{barrett2011practical}
A.~B. Barrett and A.~K. Seth, ``Practical measures of integrated information
  for time-series data,'' {\em PLoS Computational Biology}, vol.~7, no.~1,
  p.~e1001052, 2011.

\bibitem{mediano2018measuring}
P.~A. Mediano, A.~K. Seth, and A.~B. Barrett, ``Measuring integrated
  information: Comparison of candidate measures in theory and simulation,''
  {\em Entropy}, vol.~21, no.~1, p.~17, 2018.

\bibitem{jansma2025fast}
A.~Jansma, P.~A. Mediano, and F.~E. Rosas, ``Fast m{\"o}bius transform: An
  algebraic approach to information decomposition,'' {\em Physical Review
  Research}, vol.~7, no.~3, p.~033049, 2025.

\bibitem{hoel2013quantifying}
E.~P. Hoel, L.~Albantakis, and G.~Tononi, ``Quantifying causal emergence shows
  that macro can beat micro,'' {\em Proceedings of the National Academy of
  Sciences}, vol.~110, no.~49, pp.~19790--19795, 2013.

\bibitem{rosas2019reconciling}
F.~E. Rosas, P.~A. Mediano, H.~J. Jensen, A.~K. Seth, A.~B. Barrett, R.~L.
  Carhart-Harris, and D.~Bor, ``Reconciling emergences: An
  information-theoretic approach to identify causal emergence in multivariate
  data,'' {\em PLoS Computational Biology}, vol.~16, no.~12, p.~e1008289, 2020.

\bibitem{barnett2023dynamical}
L.~Barnett and A.~K. Seth, ``Dynamical independence: discovering emergent
  macroscopic processes in complex dynamical systems,'' {\em Physical Review
  E}, vol.~108, no.~1, p.~014304, 2023.

\bibitem{varley2024emergence}
T.~F. Varley, O.~Sporns, N.~J. Stevenson, M.~G. Welch, M.~M. Myers,
  S.~Vanhatalo, and A.~Tokariev, ``Emergence of a synergistic scaffold in the
  brains of human infants,'' {\em bioRxiv}, pp.~2024--02, 2024.

\bibitem{hoel2025causal}
E.~Hoel, ``Causal emergence 2.0: Quantifying emergent complexity,'' {\em arXiv
  preprint arXiv:2503.13395}, 2025.

\bibitem{kaplanis2023learning}
C.~Kaplanis, P.~Mediano, and F.~Rosas, ``Learning causally emergent
  representations,'' in {\em NeurIPS 2023 workshop: Information-Theoretic
  Principles in Cognitive Systems}, 2023.

\bibitem{mcsharry2024learning}
D.~McSharry, C.~Kaplanis, F.~Rosas, and P.~A. Mediano, ``Learning diverse
  causally emergent representations from time series data,'' {\em Advances in
  Neural Information Processing Systems}, vol.~37, pp.~119547--119572, 2024.

\bibitem{may1976simple}
R.~M. May, ``Simple mathematical models with very complicated dynamics,'' {\em
  Nature}, vol.~261, no.~5560, pp.~459--467, 1976.

\bibitem{belykh2000invariant}
V.~Belykh, I.~Blykh, N.~Komrakov, and E.~Moseklide, ``Invariant manifolds and
  cluster synchronization in a family of locally coupled map lattices,'' {\em
  Discrete Dynamics in Nature and Society}, vol.~4, no.~3, pp.~245--256, 2000.

\bibitem{chazottes2005dynamics}
J.-R. Chazottes and B.~Fernandez, {\em Dynamics of coupled map lattices and of
  related spatially extended systems}, vol.~671.
\newblock Springer Science \& Business Media, 2005.

\bibitem{hopf1942abzweigung}
E.~Hopf, ``Abzweigung einer periodischen l{\"o}sung von einer station{\"a}ren
  l{\"o}sung eines differentialsystems,'' {\em Ber. Math.-Phys. Kl S{\"a}chs.
  Akad. Wiss. Leipzig}, vol.~94, pp.~1--22, 1942.

\bibitem{andronov2013theory}
A.~A. Andronov, A.~A. Vitt, and S.~E. Khaikin, {\em Theory of oscillators:
  Adiwes international series in physics}, vol.~4.
\newblock Elsevier, 2013.

\bibitem{deco2026whole}
G.~Deco and M.~L. Kringelbach, {\em Whole-brain modelling: Cartography of the
  Dynamics of Mind}.
\newblock Oxford University Press, 2026.

\bibitem{tournier2004direct}
J.-D. Tournier, F.~Calamante, D.~G. Gadian, and A.~Connelly, ``Direct
  estimation of the fiber orientation density function from diffusion-weighted
  mri data using spherical deconvolution,'' {\em Neuroimage}, vol.~23, no.~3,
  pp.~1176--1185, 2004.

\bibitem{domhof2021parcellation}
J.~W. Domhof, K.~Jung, S.~B. Eickhoff, and O.~V. Popovych,
  ``Parcellation-induced variation of empirical and simulated brain connectomes
  at group and subject levels,'' {\em Network Neuroscience}, vol.~5, no.~3,
  pp.~798--830, 2021.

\bibitem{van2012human}
D.~C. Van~Essen, K.~Ugurbil, E.~Auerbach, D.~Barch, T.~E. Behrens, R.~Bucholz,
  A.~Chang, L.~Chen, M.~Corbetta, S.~W. Curtiss, {\em et~al.}, ``The human
  connectome project: a data acquisition perspective,'' {\em Neuroimage},
  vol.~62, no.~4, pp.~2222--2231, 2012.

\bibitem{van2013wu}
D.~C. Van~Essen, S.~M. Smith, D.~M. Barch, T.~E. Behrens, E.~Yacoub,
  K.~Ugurbil, W.-M.~H. Consortium, {\em et~al.}, ``The wu-minn human connectome
  project: an overview,'' {\em Neuroimage}, vol.~80, pp.~62--79, 2013.

\bibitem{tournier2019mrtrix3}
J.-D. Tournier, R.~Smith, D.~Raffelt, R.~Tabbara, T.~Dhollander, M.~Pietsch,
  D.~Christiaens, B.~Jeurissen, C.-H. Yeh, and A.~Connelly, ``Mrtrix3: A fast,
  flexible and open software framework for medical image processing and
  visualisation,'' {\em Neuroimage}, vol.~202, p.~116137, 2019.

\bibitem{fischl2012freesurfer}
B.~Fischl, ``Freesurfer,'' {\em Neuroimage}, vol.~62, no.~2, pp.~774--781,
  2012.

\bibitem{jenkinson2012fsl}
M.~Jenkinson, C.~F. Beckmann, T.~E. Behrens, M.~W. Woolrich, and S.~M. Smith,
  ``Fsl,'' {\em Neuroimage}, vol.~62, no.~2, pp.~782--790, 2012.

\bibitem{schaefer2018local}
A.~Schaefer, R.~Kong, E.~M. Gordon, T.~O. Laumann, X.-N. Zuo, A.~J. Holmes,
  S.~B. Eickhoff, and B.~T. Yeo, ``Local-global parcellation of the human
  cerebral cortex from intrinsic functional connectivity mri,'' {\em Cerebral
  cortex}, vol.~28, no.~9, pp.~3095--3114, 2018.

\bibitem{seguin2020network}
C.~Seguin, Y.~Tian, and A.~Zalesky, ``Network communication models improve the
  behavioral and functional predictive utility of the human structural
  connectome,'' {\em Network Neuroscience}, vol.~4, no.~4, pp.~980--1006, 2020.

\bibitem{cover1991information}
T.~M. Cover and J.~A. Thomas, ``Information theory and the stock market,'' {\em
  Elements of Information Theory. Wiley Inc., New York}, pp.~543--556, 1991.

\bibitem{kraskov2004estimating}
A.~Kraskov, H.~St{\"o}gbauer, and P.~Grassberger, ``Estimating mutual
  information,'' {\em Physical Review E}, vol.~69, no.~6, p.~066138, 2004.

\end{thebibliography}

\clearpage
\newpage
\onecolumngrid
\appendix
\renewcommand{\appendixname}{Supplementary Material}
\setcounter{figure}{0}
\setcounter{table}{0}
\setcounter{equation}{0}
\setcounter{definition}{0}
\renewcommand{\thedefinition}{S\arabic{definition}}
\renewcommand{\theHdefinition}{S\arabic{definition}}
\renewcommand{\theequation}{S\arabic{section}-\arabic{equation}}
\renewcommand{\thesection}{S\arabic{section}}
\renewcommand{\thefigure}{S\arabic{figure}}
\renewcommand{\thetable}{S\arabic{table}}
\renewcommand{\theHequation}{S\arabic{section}-\arabic{equation}}
\renewcommand{\theHsection}{S\arabic{section}}
\renewcommand{\theHfigure}{S\arabic{figure}}
\renewcommand{\theHtable}{S\arabic{table}}
\renewcommand{\defs}{Definition~\ref{def:syn_app}\xspace}

\section{\phiidig details}
\label{app:sec:igphiid}

This section spells out the information-geometric construction presented in the main text, with every projection and every distribution written explicitly and additional information on the mathematical procedure. 
Mirroring the structure of Sec.~\ref{sec:IG_3var} of the manuscript, we first introduce the information-geometric (IG) notion of synergy for a single source--target pair. Then, we outline its extension to multiple sources and targets, offering insights into the emergence of different orders of synergy. Finally, we merge this framework with the information-theoretic approach \phiid, providing additional granularity in the study of computational modes in a complex system.

\subsection{Information-geometric synergy for a single target} \label{app:sec:3var}

We begin with the elementary case of two sources $X_1,X_2$ and a single target $Y$, which provides intuitions and the basis for every larger construction below. Let $M$ denote the manifold of all joint distributions $P_{X_1X_2Y}$, parametrised in $\theta$-coordinates as
\begin{equation} \label{app:eq:M3d}
    \begin{split}
M = & \{ {P_{{X_1}{X_2}Y}} \text{ s.t. } {P_{{X_1}{X_2}Y}} = \exp ( {\theta _{X_1}}{F_{X_1}}({x_1}) + {\theta _{X_2}}{F_{X_2}}({x_2}) + {\theta _{Y}}{F_{Y}}(y) + \\ 
& \quad + {\theta _{X_1X_2}}{F_{X_1X_2}}({x_1},\;{x_2}) + {\theta _{X_1Y}}{F_{X_1Y}}({x_1},\;y) + {\theta _{X_2Y}}{F_{X_2Y}}({x_2},\;y) \\  
& \quad + {\theta _{X_1X_2Y}}{F_{X_1X_2Y}}({x_1},\;{x_2},\;y) - \psi (\theta ) ) \} \,.
    \end{split}
\end{equation}
Here, $F$ denotes the sufficient statistic associated with each subset of variables, $\theta$ the corresponding natural parameter quantifying the strength of the corresponding interaction, and $\psi$ the log-partition function that normalises the distribution.
This representation makes explicit that the full manifold $M$ contains all possible interactions between the sources and the target, including pairwise and genuinely three-way dependencies.

To isolate which of these interactions are responsible for a given statistical dependency, we need a way of comparing distributions that lie in different regions of $M$, and, more specifically, of measuring how far the true distribution sits from a family of distributions in which a particular interaction has been removed. Information geometry supplies both a notion of distance for this purpose, the KL divergence, and two complementary notions of path along which such distances are naturally decomposed: $e$-geodesics, which are linear in the $\theta$-coordinates and correspond to exponential-family interpolations, and $m$-geodesics, which are linear in the probability itself and correspond to mixture interpolations. The two are dual to one another, and it is this duality that makes possible the orthogonal, Pythagorean decompositions used throughout this section.

Since our aim is to decompose the statistical structure between sources $(X_1,X_2)$ and the target $Y$, it is natural to ask how these dependences change when specific source–target interactions are selectively removed. Within this setting, this is simply achieved by constraining the corresponding interaction parameters to zero. This leads to the definition of the following submanifolds:
\begin{equation} \label{app:eq:E3var}
\begin{aligned}
E_0 & :=\{P\in M\text{ s.t. } \theta_{X_1Y}=\theta_{X_2Y}=\theta_{X_1X_2Y}=0\} \\
E_1 & :=\{P\in M\text{ s.t. } \theta_{X_2Y}=\theta_{X_1X_2Y}=0\} \\
E_2 & :=\{P\in M\text{ s.t. } \theta_{X_1Y}=\theta_{X_1X_2Y}=0\} \,.
\end{aligned}
\end{equation}
$E_0$ is the submanifold in which the sources are jointly independent of the target, while $E_i$ ($i=1,2$) is the submanifold in which $X_{-i}$ carries no information about $Y$ beyond what $X_i$ already conveys, i.e.\ $X_{-i}\perp Y\mid X_i$. 
In information geometry, the natural way to quantify how much information is lost when specific source-target interaction terms are set to zero is to measure the KL divergence between the true distribution \Pf and its projection onto the corresponding submanifold.
If the projection onto $E_0$ lies close to \Pf, the sources carry little information about the target overall; if instead the projection onto $E_i$ lies close to \Pf, the dependency between $(X_1,X_2)$ and $Y$ is well explained by $X_i$ alone, with $X_{-i}$ contributing little beyond what $X_i$ already conveys. The KL divergence to each submanifold thus quantifies precisely how much of the source-target dependency survives once the corresponding interaction is removed.

Since the projection of a distribution $P$ on a submanifold $N$ is given by 
\begin{equation}
    Q^* = \underset{Q\in N}{\arg\min} \D(P||Q) \,,
\end{equation}
we can directly calculate the projections of $P_{X_1X_2Y}$ onto each of these submanifolds, obtaining the following explicit closed forms
\begin{align} \label{app:eq:PID_probs_og}
    \bar{P}_0 & := \underset{Q\in E_0}{\arg\min}\,\D(P_{X_1X_2Y}\|Q) = P_{X_1X_2}P_Y \\
    \bar{P}_1 & := \underset{Q\in E_1}{\arg\min}\,\D(P_{X_1X_2Y}\|Q) = \frac{P_{X_1X_2}P_{X_1Y}}{P_{X_1}} \\
    \bar{P}_2 & := \underset{Q\in E_2}{\arg\min}\,\D(P_{X_1X_2Y}\|Q) = \frac{P_{X_1X_2}P_{X_2Y}}{P_{X_2}} \,.
\end{align}
Moreover, these manifolds are nested into each other, $E_0\subseteq E_i\subseteq M$, since every constraint defining $E_0$ also holds on $E_i$. Because $\bar{P}_0$ and $\bar{P}_i$ are $e$-projections of $\Pf$, the $e$-geodesic joining $\Pf$ to $\bar{P}_i$ is orthogonal, in the Fisher-information metric, to the $m$-geodesic joining $\bar{P}_i$ to $\bar{P}_0$; this orthogonality is precisely what licenses a Pythagorean relation among the corresponding KL divergences, giving the following decomposition of dependencies
\begin{equation} \label{app:eq:dkl_proj1}
\begin{split}
    \underbrace{\D(\Pf||\bar{P}_0)}_{\scriptstyle M\,\hookrightarrow\, E_0} = & \underbrace{\D(\Pf||\bar{P_i})}_{\scriptstyle M\,\hookrightarrow\, E_i} + \underbrace{\D(\bar{P}_i||\bar{P}_0)}_{\scriptstyle E_i\,\hookrightarrow\, E_0} \,.
\end{split}
\end{equation}
In classical information theory, this corresponds to the well-known chain rule of the mutual information~\cite{cover1991information}
\begin{equation} \label{eq:cond_mi_app}
    I(X_1,X_2;Y) = I(X_{-i};Y\mid X_i) + I(X_i;Y) \,.
\end{equation}
From Eq.~\eqref{app:eq:dkl_proj1} (and hence \eqref{eq:cond_mi_app} too), it is clear that higher-order interdependencies must be contained in the first term of the RHS, since the second term only provides pairwise dependencies between $X_i$ and $Y$. 
This is not yet the end of the story, however: $\D(\Pf\|\bar{P}_i)$ is not itself a clean measure of synergy, since it still mixes the irreducible contribution of $(X_1,X_2)$ jointly with any residual, lower-order dependency $X_{-i}$ contributes once $X_i$ has been accounted for. Separating the two requires one further geometric step, which we now describe.

The projections $\bar P_1$ and $\bar P_2$ each retain a single explanatory pathway: $\bar{P}_i$ keeps only the dependency mediated by $X_i$, having removed every direct link between $X_{-i}$ and $Y$. Consequently, any distribution reachable from $\bar{P}_1$ and $\bar{P}_2$ by an $e$-geodesic interpolation is, by construction, explainable as a combination of single-source mechanisms, with no need for a genuinely joint contribution from $(X_1,X_2)$. Synergy, being precisely the part of the dependency that a single source cannot supply, must therefore correspond to what remains once this entire family of single-source explanations has been exhausted -- that is, to the distance between $\Pf$ and the closest point on this interpolation~\cite{niu2019measure}:
\begin{definition} \label{def:syn_app}
Synergistic contributions are quantified as the KL divergence between the true distribution and the closest distribution along the $e$-geodesic connecting the corresponding conditional-independence projections.
\end{definition}
More formally, we have:
\begin{align} \label{app:eq:syndef_PID}
    t^* & := \underset{t\in\mathbb{R}}{\arg\min} \D\left(\Pf\big|\big|\,\mathcal{Z}\left(\bar{P}_1^t \cdot\bar{P}_2^{1-t}\right)\right) \\
    P^* & := \mathcal{Z}\left(\bar{P}_1^{t^*} \cdot\bar{P}_2^{1-{t^*}}\right) \\
    S(X_1,X_2;Y) & := \D(P_{X_1X_2Y}\|P^*) \,,
\end{align}
where $\mathcal{Z}$ is a normalising constant. 

Finally, we note that the intuition behind this notion of synergy can also be cast in terms of channels using the functional form of $\bP_i$, which allows us to explicitly express Eq.~\eqref{app:eq:syndef_PID} as
\begin{equation}
    S = \min_{t\in\mathbb{R}}\sum_{x_1, x_2, y}\!\! p(x_1, x_2, y) \log \left( \frac{p(y|x_1,x_2)}{\mathcal{Z}\,p(y|x_1)^{t} \, p(y|x_2)^{1-t}} \right) \,.
\end{equation}
This clarifies that synergy captures the divergence between the statistical structure contained by the full channel where both sources are present ($p(y|x_1,x_2)$) and that obtainable by the combination of single sources ($\mathcal{Z}\,p(y|x_1)^{t} \, p(y|x_2)^{1-{t}}$)~\cite{perrone2016hierarchical}. 
Since Ref.~\cite{niu2019measure} shows that $P^*$ is unique and that the corresponding $e$- and $m$-geodesics are orthogonal, the Pythagorean relation applied to Eq.~\eqref{app:eq:dkl_proj1} lets us decompose the mutual information terms orthogonally as
\begin{equation}\label{eq:3var_decomp_S_app}
\begin{split}
    \underbrace{\D(\Pf||\bar{P}_0)}_{\text{\normalfont all dependencies}} = 
    & \underbrace{\D(\Pf||P^*)}_{S(X_1,X_2;Y)} + \underbrace{\D(P^*||\bar{P_i}) + \D(\bar{P}_i||\bar{P}_0)}_{\text{\normalfont pairwise dependencies}} \,.\\[-1em]
    \underbrace{\rowstrut\phantom{\D(\Pf||\bar{P}_0)}}_{I(X_1,X_2;Y)} \phantom{=}
    & \underbrace{\rowstrut\phantom{\D(\Pf||P^*)} \phantom{{}+{}} \phantom{\D(P^*||\bar{P_i})}}_{I(X_{-i};Y|X_i)}
    \phantom{{}+{}} \underbrace{\rowstrut\phantom{\D(\bar{P}_i||\bar{P}_0)}}_{I(X_i;Y)} \phantom{\,,}
\end{split}
\end{equation}
where we made clear the nature of the contributions of each term. Importantly, the uniqueness of $P^*$ and the orthogonality between each term of the decomposition ensure that Eq.~\eqref{eq:3var_decomp_S_app} is valid for both $i=1,2$.

Synergy, in this construction, is therefore the KL divergence between the true distribution and the closest point on the $e$-geodesic joining the two conditional-independence projections (\defs). This is the main geometric ingredient this framework relies on: every larger construction below, whether applied to a marginal of a bigger system, a pair of variables held jointly, or a full double conditioning, is an analogous distance, evaluated between $\Pf$ and an interpolation among an appropriate set of conditional-independence projections.

\subsection{Orders of synergy for multiple sources and targets} \label{app:sec:4var}

Once more than two sources or two targets are involved, the notion of higher-order dependency becomes considerably richer: a system with $n$ sources $\Xs$ and $m$ targets $\Ys$ contains a combinatorially large number of subsets of variables, each capable of supporting its own irreducible, synergistic dependency. It is therefore natural to speak of \textit{orders of synergy}: a subsystem built from $n'$ sources and $m'$ targets can support synergy up to order $n'+m'$, and every subsystem obtained by dropping one or more of its variables supports synergies of correspondingly lower order, down to the minimal three-variable configurations $(X_i,X_j,Y_k)$ or $(X_i,Y_j,Y_k)$ treated in Sec.~\ref{app:sec:3var}, where order 3 is the only synergy that remains. In every such case, the corresponding synergy is obtained exactly as above: by marginalising down to the relevant subset of sources and targets, and applying \defs to the conditional-independence submanifolds of that marginal system.

For clarity and ease of notation, we specialise this general picture to the smallest system that is genuinely multivariate on both sides, two sources $X_1,X_2$ and two targets $Y_1,Y_2$, which is also the setting relevant for \phiid. The extension to larger systems follows the identical logic, at the cost of a combinatorially larger number of terms to track.

Let $M$ be the manifold containing every joint distribution $P_{X_1X_2Y_1Y_2}$, parametrised as
\begin{align} \label{app:eq:M4var}
M := \{ P_{X_1 X_2 Y_1 Y_2} \text{ s.t. } \Pff = \exp\!\left( \sum_{i \subseteq \{X_1,X_2,Y_1,Y_2\}} \theta_i F_i(i) - \psi(\theta) \right) \Bigg\} \,. 
\end{align}
Because this system has $2\times2$ source-target pairs, four distinct three-variable marginals, and one genuinely four-way interaction, the analysis proceeds in three stages of increasing resolution: first the three-variable marginal subsystems, which contribute the order-3 synergies, then the full double-conditioning construction, which isolates the order-4 synergy, and finally an additional kind of synergy where either sources or targets are treated as a unique multivariate distribution.
During this procedure, we also show precisely why these components do not provide a hierarchical decomposition. 

\subsubsection{Three-variable marginal subsystems} \label{app:sec:3varmarg}

Since $M$ in Eq.~\eqref{app:eq:M4var} contains more than three variables, it also contains four distinct three-variable marginals, obtained by leaving out either one source or one target:
\begin{align} \label{app:eq:Mij}
\allowdisplaybreaks
M_{X_1 X_2 Y_1} \begin{aligned}[t] := \{ P_{X_1 X_2 Y_1} \text{ s.t. } P_{X_1 X_2 Y_1} = \exp\!\left( \sum_{i \subseteq \{X_1,X_2,Y_1\}} \theta_i F_i(i) - \psi(\theta) \right) \Bigg\} \end{aligned} \\[1ex]
M_{X_1 X_2 Y_2} \begin{aligned}[t] := \{ P_{X_1 X_2 Y_2} \text{ s.t. } P_{X_1 X_2 Y_2} = \exp\!\left( \sum_{i \subseteq \{X_1,X_2,Y_2\}} \theta_i F_i(i) - \psi(\theta) \right) \Bigg\} \end{aligned} \\[1ex]
M_{X_1 Y_1 Y_2} \begin{aligned}[t] := \{ P_{X_1 Y_1 Y_2} \text{ s.t. } P_{X_1 Y_1 Y_2} = \exp\!\left( \sum_{i \subseteq \{X_1,Y_1,Y_2\}} \theta_i F_i(i) - \psi(\theta) \right) \Bigg\} \end{aligned} \\[1ex]
M_{X_2 Y_1 Y_2} \begin{aligned}[t] := \{ P_{X_2 Y_1 Y_2} \text{ s.t. } P_{X_2 Y_1 Y_2} = \exp\!\left( \sum_{i \subseteq \{X_2,Y_1,Y_2\}} \theta_i F_i(i) - \psi(\theta) \right) \Bigg\}
\end{aligned}
\end{align}
so that, for instance, $M_{X_2 Y_1 Y_2}$ contains the distributions relevant to $I(X_2;Y_1,Y_2)$. A subtlety that distinguishes this multivariate setting from the elementary case of the previous section is that, because marginalisation is not an invertible operation, these manifolds are not submanifolds of $M$: a distribution constrained to lie in $M_{X_2 Y_1 Y_2}$ does not correspond to any single, well-defined subset of distributions in $M$, so each of the submanifolds must be treated as an independent space in its own right, with its own exponential-family parametrisation and its own set of projections, rather than as a slice of $M$.

Each submanifold is analysed exactly as in the previous section, treating it as a self-contained three-variable system. We work through the case of $M_{X_1 Y_1 Y_2}$ in full, since it fixes the notation for the remaining three marginals. Here the order-three synergy is the irreducible structure between the pair $(Y_1,Y_2)$ and $X_1$.
The relevant submanifolds of $M_{X_1 Y_1 Y_2}$, obtained by zeroing the interactions between $X_1$ and one or both targets, are
\begin{align} \label{app:eq:S112}
    E_{0,\, X_1 Y_1 Y_2} & := \{ P\in M_{X_1 Y_1 Y_2} \text{ s.t. } \theta_{X_1Y_1} = \theta_{X_1Y_2} = \theta_{X_1Y_1Y_2} = 0\} \\
    E_{1,\, X_1 Y_1 Y_2} & := \{ P\in M_{X_1 Y_1 Y_2} \text{ s.t. } \theta_{X_1Y_2} = \theta_{X_1Y_1Y_2} = 0\} \\
    E_{2,\, X_1 Y_1 Y_2} & := \{ P\in M_{X_1 Y_1 Y_2} \text{ s.t. } \theta_{X_1Y_1} = \theta_{X_1Y_1Y_2} = 0\} \,.
\end{align}
These represent, respectively, full independence of $X_1$ from $(Y_1,Y_2)$, and the conditional independences $X_1\perp Y_2\mid Y_1$ and $X_1\perp Y_1\mid Y_2$. Projecting $P_{X_1Y_1Y_2}$ onto each of these submanifolds gives the explicit closed forms
\begin{align} \label{eq:P0_112proj}
    \bar{P}_{0,\, X_1 Y_1 Y_2} & = \underset{Q\in E_{0,\, {X_1 Y_1 Y_2}}}{\arg\min}\,\D(P_{X_1Y_1Y_2}\|Q) = P_{Y_1Y_2}P_{X_1} \\ \label{eq:P1_112proj}
    \bar{P}_{1,\, X_1 Y_1 Y_2} & = \underset{Q\in E_{1,\, {X_1 Y_1 Y_2}}}{\arg\min}\,\D(P_{X_1Y_1Y_2}\|Q) = \frac{P_{Y_1X_1}P_{Y_1Y_2}}{P_{Y_1}} \\ \label{eq:P2_112proj}
    \bar{P}_{2,\, X_1 Y_1 Y_2} & = \underset{Q\in E_{2,\, {X_1 Y_1 Y_2}}}{\arg\min}\,\D(P_{X_1Y_1Y_2}\|Q) = \frac{P_{X_1Y_2}P_{Y_1Y_2}}{P_{Y_2}} \,.
\end{align}
As anticipated, this is exactly the projection structure studied above, applied to input $X_1$ and outputs $Y_1,Y_2$. The Pythagorean relation therefore gives, for $i=1,2$,
\begin{align} \label{eq:phiid-PIDcase1}
    \D(P_{X_1Y_1Y_2}\|\bar{P}_{0,\,X_1Y_1Y_2}) = \D(P_{X_1Y_1Y_2}\|\bar{P}_{i,\,X_1Y_1Y_2}) + \D(\bar{P}_{i,\,X_1Y_1Y_2}\|\bar{P}_{0,\,X_1Y_1Y_2})\,,
\end{align}
mirroring once again the information-theoretic chain rule
\begin{equation}
    I(X_1;Y_1,Y_2) = I(X_1;Y_{-i}\mid Y_i) + I(X_1;Y_i) \,.
\end{equation}

Finally, interpolating between $\bar{P}_{1,\,X_1Y_1Y_2}$ and $\bar{P}_{2,\,X_1Y_1Y_2}$ along the $e$-geodesic, as in \defs, and applying the Pythagorean relation once more to $\D(P_{X_1Y_1Y_2}\|\bar{P}_{i,\,X_1Y_1Y_2})$, gives the order-3 synergistic contribution of this subsystem:
\begin{align} \label{app:eq:Pstar112}
    t^*_{X_1Y_1Y_2} & := \underset{t\in\mathbb{R}}{\arg\min}\, \D\!\left(P_{X_1Y_1Y_2}\middle\|\,\mathcal{Z}\left(\bar{P}_{1,\,X_1Y_1Y_2}^{t}\,\bar{P}_{2,\,X_1Y_1Y_2}^{1-t}\right)\right) \\
    P^*_{X_1Y_1Y_2} & := \mathcal{Z}\left(\bar{P}_{1,\,X_1Y_1Y_2}^{\,t^*_{X_1Y_1Y_2}}\,\bar{P}_{2,\,X_1Y_1Y_2}^{\,1-t^*_{X_1Y_1Y_2}}\right) \\
    S_3(X_1;Y_1,Y_2) & := \D(P_{X_1Y_1Y_2}\|P^*_{X_1Y_1Y_2}) \,,
\end{align}
where $\mathcal{Z}$ is a normalising constant.

With this definition at hand, we can then complete the decomposition of Eq.~\eqref{eq:phiid-PIDcase1}:
\begin{equation}\label{eq:subsys_3var_decomp_S}
\begin{split}
    \underbrace{\D(P_{X_1Y_1Y_2}||\bar{P}_{0, \,{X_1 Y_1 Y_2}})}_{\text{\normalfont all dependencies}} =
    & \underbrace{\D(P_{X_1Y_1Y_2}||P^*_{X_1 Y_1 Y_2})}_{S_3(X_1;Y_1,Y_2)} + \underbrace{\D(P^*_{X_1Y_1 Y_2}||\bar{P}_{i, \,{X_1 Y_1 Y_2}}) + \D(\bar{P}_{i, \,{X_1 Y_1 Y_2}}||\bar{P}_{0, \,{X_1 Y_1 Y_2}})}_{\text{\normalfont pairwise dependencies}} \,,\\[-1em]
    \underbrace{\rowstrut\phantom{\D(P_{X_1Y_1Y_2}||\bar{P}_{0, \,{X_1 Y_1 Y_2}})}}_{I(X_1;Y_1,Y_2)} \phantom{=}
    & \underbrace{\rowstrut\phantom{\D(P_{X_1Y_1Y_2}||P^*_{X_1 Y_1 Y_2})} \phantom{{}+{}} \phantom{\D(P^*_{X_1Y_1 Y_2}||\bar{P}_{i, \,{X_1 Y_1 Y_2}})}}_{I(X_1;Y_{-i}|Y_i)}
    \phantom{{}+{}} \underbrace{\rowstrut\phantom{\D(\bar{P}_{i, \,{X_1 Y_1 Y_2}}||\bar{P}_{0, \,{X_1 Y_1 Y_2}})}}_{I(X_1;Y_i)} \phantom{\,,}
\end{split}
\end{equation}
mirroring Eq.~\eqref{eq:3var_decomp_S_app}.

Repeating this same procedure for the remaining three marginals $M_{X_1X_2Y_1}, M_{X_1X_2Y_2}, M_{X_2Y_1Y_2}$ gives the analogous order-3 synergies, which read
\begin{align} \label{app:eq:S3_others}
    S_3(X_1,X_2;Y_1) := \D(P_{X_1X_2Y_1}\|P^*_{X_1X_2Y_1}) \\
    S_3(X_1,X_2;Y_2) := \D(P_{X_1X_2Y_2}\|P^*_{X_1X_2Y_2}) \\
    S_3(X_2;Y_1,Y_2) := \D(P_{X_2Y_1Y_2}\|P^*_{X_2Y_1Y_2}) \,,
\end{align}
where each $P^*_{X_1 X_2 Y_j}$ and $P^*_{X_i Y_1 Y_2}$ is defined analogously to Eq.~\eqref{app:eq:Pstar112}.
These can then be used to decompose the interdependencies of the three remaining 3-variable subsystems, similarly to Eq.~\eqref{eq:subsys_3var_decomp_S}.

\subsubsection{Double conditioning and the order-4 synergy} \label{app:sec:4varcore}

Finally, to fully disentangle the remaining contributions, we must condition on one source and one target simultaneously. Consider the four submanifolds of $M$ in which a single source-target pair is left free while every other interaction is removed:
\begin{align} \label{app:eq:Sij}
    E_0 & \begin{aligned}[t] := \{& \Pff \in M \text{ s.t. } \\ & \theta_{X_iY_j}=\theta_{X_1X_2Y_{j}}
     = \theta_{X_iY_1Y_2}  = \theta_{X_1X_2Y_1Y_2}=0\}
    \end{aligned}
    & \hspace{-1cm}\xrightarrow{\hspace{0.8cm}}\hspace{0.6cm} &\; (X_1,X_2)\perp(Y_1,Y_2) \\[0.3em]
    E_1^1 & \begin{aligned}[t] & := \{ \Pff\in M \text{ s.t. } \\ & \theta_{X_2Y_2}
    = \theta_{X_1X_2Y_2} = \theta_{X_2Y_1Y_2} = \theta_{X_1X_2Y_1Y_2}=0\}
    \end{aligned}
    & \hspace{-1cm}\xrightarrow{\hspace{0.8cm}}\hspace{0.6cm} &\; X_2\perp Y_2\mid X_1,Y_1 \\[0.3em]
    E_1^2 & \begin{aligned}[t] & := \{ \Pff\in M \text{ s.t. } \\ & \theta_{X_2Y_1}
    = \theta_{X_1X_2Y_1} = \theta_{X_2Y_1Y_2} = \theta_{X_1X_2Y_1Y_2}=0\}
    \end{aligned}
    & \hspace{-1cm}\xrightarrow{\hspace{0.8cm}}\hspace{0.6cm} &\; X_2\perp Y_1\mid X_1,Y_2 \\[0.3em]
    E_2^1 & \begin{aligned}[t] & := \{ \Pff\in M \text{ s.t. } \\ & \theta_{X_1Y_2}
    = \theta_{X_1X_2Y_2} = \theta_{X_1Y_1Y_2} = \theta_{X_1X_2Y_1Y_2}=0\}
    \end{aligned}
    & \hspace{-1cm}\xrightarrow{\hspace{0.8cm}}\hspace{0.6cm} &\; X_1\perp Y_2\mid X_2,Y_1 \\[0.3em]
    E_2^2 & \begin{aligned}[t] & := \{ \Pff\in M \text{ s.t. } \\ & \theta_{X_1Y_1}
    = \theta_{X_1X_2Y_1} = \theta_{X_1Y_1Y_2} = \theta_{X_1X_2Y_1Y_2}=0\}
    \end{aligned}
    & \hspace{-1cm}\xrightarrow{\hspace{0.8cm}}\hspace{0.6cm} &\; X_1\perp Y_1\mid X_2,Y_2 \,,
\end{align}
which represent the corresponding spaces of double conditional independence.

The projections onto these submanifolds are again available in closed form:
\begin{align} \label{app:eq:PijProj}
    \bar{P}_0 & = \underset{Q\in E_0}{\arg\min}\,\D(\Pff\|Q) = P_{X_1X_2}P_{Y_1Y_2} \\
    \bar{P}_1^1 & = \underset{Q\in E_1^1}{\arg\min}\,\D(\Pff\|Q) = \frac{P_{X_1X_2Y_1}P_{X_1Y_1Y_2}}{P_{X_1Y_1}} \\
    \bar{P}_1^2 & = \underset{Q\in E_1^2}{\arg\min}\,\D(\Pff\|Q) = \frac{P_{X_1X_2Y_2}P_{X_1Y_1Y_2}}{P_{X_1Y_2}} \\
    \bar{P}_2^1 & = \underset{Q\in E_2^1}{\arg\min}\,\D(\Pff\|Q) = \frac{P_{X_1X_2Y_1}P_{X_2Y_1Y_2}}{P_{X_2Y_1}} \\
    \bar{P}_2^2 & = \underset{Q\in E_2^2}{\arg\min}\,\D(\Pff\|Q) = \frac{P_{X_1X_2Y_2}P_{X_2Y_1Y_2}}{P_{X_2Y_2}} \,.
\end{align}
Unlike every projection considered so far, which removed a single explanatory pathway captured by one conditional-independence statement, $\bar{P}_i^j$ removes the interaction between $X_{-i}$ and $Y_{-j}$ while conditioning on both $X_i$ and $Y_j$. 
This projection allows us to use the Pythagorean relation and orthogonally decompose the KL divergences as:
\begin{align} \label{eq:S4_KL_decomp}
    \D(\Pff\|\bar P_0) = \D(\Pff\|\bar P_i^j) + \D(\bar P_i^j\|\bar P_0) \,.
\end{align}
Analogously, the mutual information decomposes as
\begin{equation} \label{eq:MI_four_chain}
    I(X_1,X_2;Y_1,Y_2) = I(X_{-i};Y_{-j}|X_i,Y_j) + I(X_{-i};Y_j|X_i) + I(X_i;Y_j,Y_{-j}) \,, \qquad i,j=1,2 \,.
\end{equation}
Because there are now four such projections, rather than the two encountered in every previous construction, isolating the genuinely irreducible, order-4 contribution requires interpolating over the full three-parameter family spanned by all four:
\begin{align} \label{app:eq:Pstarstar}
    (r^*,s^*,t^*) & := \underset{r,s,t\,\in\,\mathbb{R}^3}{\arg\min}\, \D\!\left(\Pff\,\middle\|\,\mathcal{Z}\left((\bar P_1^1)^{1-r-s-t}\,(\bar P_2^1)^{r}\,(\bar P_1^2)^{s}\,(\bar P_2^2)^{t}\right)\right) \\
    P^*_* & := \mathcal{Z}\left((\bar P_1^1)^{1-r^*-s^*-t^*}\,(\bar P_2^1)^{r^*}\,(\bar P_1^2)^{s^*}\,(\bar P_2^2)^{t^*}\right) \\
    S_4(X_1,X_2;Y_1,Y_2) & := \D(\Pff \| P^*_*) \,,
\end{align}
which is the same construction as Eqs.~\eqref{eq:s4_def}--\eqref{eq:s4_def_2} of the main text, written here directly in terms of the explicit projections of Eq.~\eqref{app:eq:PijProj}. A final application of the Pythagorean relation to Eq.~\eqref{eq:S4_KL_decomp} then splits each double-conditional mutual information into the order-4 synergy and a remaining, lower-order piece:
\begin{equation} \label{eq:doubl_cond_MI}
\begin{split}
    \underbrace{\DAllFour}_{\text{\normalfont all dependencies}} & = \underbrace{\DFourStar}_{S_4(X_1,X_2;Y_1,Y_2)} + \underbrace{\DPstarPij + \DPijPzero}_{\text{\normalfont remaining contributions}} \\[-1em]
    \underbrace{\rowstrut\phantom{\DAllFour}}_{I(X_1,X_2;Y_1,Y_2)}
    & \phantom{=} \underbrace{\rowstrut\phantom{\DFourStar} \phantom{{}+{}} \phantom{\DPstarPij}}_{I(X_{-i};Y_{-j}\mid X_i,Y_j)}
    \phantom{{}+{}} \underbrace{\rowstrut\phantom{\DPijPzero}}_{\mathclap{I(X_{-i};Y_j\mid X_i) + I(X_i;Y_j,Y_{-j})}}
\end{split}
\end{equation}

\subsubsection{The impossibility of a fully hierarchical decomposition} \label{app:sec:nonhier}

We can now make precise the claim that the total dependency of the system is not simply the sum of the order-4 synergy, the four order-3 synergies, and the remaining pairwise contributions. To avoid introducing yet another family of submanifolds, we directly manipulate $\D(\Pff||\bar{P}_0)$ purely in the language of information theory via chain rules, obtaining
\begin{equation} \label{eq:full_mi_decomp}
\begin{split}
\underbrace{\D(\Pff||\bar{P}_0)}_{I(X_1,X_2;Y_1,Y_2)} & = \underbrace{I(X_{-i};Y_{-j}|X_i,Y_j)}_{\Large \ \raisebox{0.15cm}{\rotatebox{-45}{\footnotesize $\ni$}} S_4(X_1,X_2;Y_1,Y_2)} + \underbrace{I(X_i;Y_1,Y_2)}_{\Large \ \raisebox{0.15cm}{\rotatebox{-45}{\footnotesize $\ni$}} S_3(X_i;Y_1,Y_2)} + \\ & + \underbrace{I(X_1,X_2;Y_j)}_{\Large \ \raisebox{0.15cm}{\rotatebox{-45}{\footnotesize $\ni$}} S_3(X_1,X_2;Y_j)} - \underbrace{I(X_i;Y_j)}_{\substack{\text{\normalfont pairwise} \\ \text{\normalfont contributions}}} \,,
\end{split}
\end{equation}
valid for any fixed choice of $i,j\in\{1,2\}$. The order-3 synergies $S_3(X_i;Y_1,Y_2)$ and $S_3(X_1,X_2;Y_j)$ that appear in the second and third term are precisely those identified in Eqs.~\eqref{app:eq:Pstar112}--\eqref{app:eq:S3_others}, a direct consequence of applying the three-term decomposition of Eq.~\eqref{eq:subsys_3var_decomp_S} to the corresponding marginal. The order-4 synergy, on the other hand, sits inside the first term, since Eq.~\eqref{eq:doubl_cond_MI} showed that
\begin{equation} \label{eq:double_cond_DKL}
    I(X_{-i};Y_{-j}|X_i,Y_j) = \underbrace{\D(\Pff||P^*_*)}_{\Large \!S_4(X_1,X_2;Y_1,Y_2)} + \underbrace{\D(P^*_*||\bar{P}_i^j)}_{\text{\normalfont remaining contributions}} \,.
\end{equation}
Crucially, Eq.~\eqref{eq:full_mi_decomp} involves only two of the four order-3 synergies -- those attached to the specific $i,j$ chosen -- while the remaining two, $S_3(X_{-i};Y_1,Y_2)$ and $S_3(X_1,X_2;Y_{-j})$, do not appear as a separate additive term at all. They must therefore be accounted for either inside the ``remaining contributions'' of Eq.~\eqref{eq:doubl_cond_MI}, by overlapping with the two order-3 synergies that do appear explicitly in Eq.~\eqref{eq:full_mi_decomp}, or by some combination of the two. This is the geometric root of the non-orthogonality mentioned in the main text: because the marginal manifolds $M_{X_iY_1Y_2}$ and $M_{X_1X_2Y_j}$ are not submanifolds of $M$ and are not nested within one another, the order-3 synergies they support cannot all be simultaneously read off as orthogonal projections of the same distribution $\Pff$. 

To further study the explicit relations among these synergies, we complement this geometric framework with an information-theoretic decomposition approach. However, before that, a further step is needed.

\subsubsection{3-variable joint-input/output decompositions} \label{app:sec:jointdecomp}

The four order-3 synergies studied above only resolve dependencies within a single three-variable marginal. Similarly, the order-4 synergy accounts for dependencies among single components when a double conditioning is applied. Thus, both scenarios say nothing yet about how the two sources or the two targets interact together at the level of the full four-variable system.
Consider the following submanifolds of $M$, where $i=1,2$ ranges over the two indices wherever it appears as a free subscript or superscript below:
\begin{align} \label{app:eq:E0S1S2}
    E_0 & \begin{aligned}[t] := \{& \Pff \in M \text{ s.t. } \\ & \theta_{X_iY_j}=\theta_{X_1X_2Y_{j}}
     = \theta_{X_iY_1Y_2}  = \theta_{X_1X_2Y_1Y_2}=0\}
    \end{aligned}
    & \hspace{-1cm}\xrightarrow{\hspace{0.8cm}}\hspace{0.6cm} &\; (X_1,X_2)\perp(Y_1,Y_2) \\[0.3em]
    E_1 & \begin{aligned}[t] := \{& \Pff\in M \text{ s.t. } \\ & \theta_{X_2Y_i}
     = \theta_{X_2Y_1Y_2} = \theta_{X_1X_2Y_i}=\theta_{X_1X_2Y_1Y_2}=0\}
    \end{aligned}
    & \hspace{-1cm}\xrightarrow{\hspace{0.8cm}}\hspace{0.6cm} &\; X_2\perp (Y_1,Y_2)\mid X_1 \\[0.3em]
    E_2 & \begin{aligned}[t] := \{& \Pff\in M \text{ s.t. } \\ & \theta_{X_1Y_i}
     = \theta_{X_1Y_1Y_2} = \theta_{X_1X_2Y_i}=\theta_{X_1X_2Y_1Y_2}=0\}
    \end{aligned}
    & \hspace{-1cm}\xrightarrow{\hspace{0.8cm}}\hspace{0.6cm} &\; X_1\perp(Y_1,Y_2)\mid X_2 \\[0.3em]
    E^1 & \begin{aligned}[t] := \{& \Pff\in M \text{ s.t. } \\ & \theta_{X_iY_2}
     = \theta_{X_1X_2Y_2} = \theta_{X_iY_1Y_2}=\theta_{X_1X_2Y_1Y_2}=0\}
    \end{aligned}
    & \hspace{-1cm}\xrightarrow{\hspace{0.8cm}}\hspace{0.6cm} &\; (X_1,X_2)\perp Y_2\mid Y_1 \\[0.3em]
    E^2 & \begin{aligned}[t] := \{& \Pff\in M \text{ s.t. } \\ & \theta_{X_iY_1}
     = \theta_{X_1X_2Y_1} = \theta_{X_iY_1Y_2} =\theta_{X_1X_2Y_1Y_2}=0\}
    \end{aligned}
    & \hspace{-1cm}\xrightarrow{\hspace{0.8cm}}\hspace{0.6cm} &\; (X_1,X_2)\perp Y_1\mid Y_2 \,.
\end{align}
where, for each submanifold, we indicated the family of distributions it contains. For convenience, we also reported $E_0$ which has already been introduced in the previous section (Eq.~\eqref{app:eq:Sij}). Hence, the corresponding projections onto $E_0,E_1,E_2,E^1,E^2$ take the explicit closed forms
\begin{align} \label{app:eq:P0P1P2}
    \bar{P}_0 & = \underset{Q\in E_0}{\arg\min}\,\D(\Pff\|Q) = P_{X_1X_2}P_{Y_1Y_2} \\
    \bar{P}_1 & = \underset{Q\in E_1}{\arg\min}\,\D(\Pff\|Q) = \frac{P_{X_1X_2}P_{X_1Y_1Y_2}}{P_{X_1}} \\
    \bar{P}_2 & = \underset{Q\in E_2}{\arg\min}\,\D(\Pff\|Q) = \frac{P_{X_1X_2}P_{X_2Y_1Y_2}}{P_{X_2}} \\
    \bar{P}^1 & = \underset{Q\in E^1}{\arg\min}\,\D(\Pff\|Q) = \frac{P_{X_1X_2Y_1}P_{Y_1Y_2}}{P_{Y_1}} \\
    \bar{P}^2 & = \underset{Q\in E^2}{\arg\min}\,\D(\Pff\|Q) = \frac{P_{X_1X_2Y_2}P_{Y_1Y_2}}{P_{Y_2}} \,.
\end{align}
Since $\bar{P}_0$ is the unique projection shared by both families, the Pythagorean relation now offers two equally valid ways of decomposing the full mutual information, depending on whether the sources or the targets are resolved first:
\begin{align}
    I(X_1,X_2; Y_1,Y_2) & = \D(\Pff\|\bar{P}_0) \notag \\
    & = \D(\Pff\|\bar{P}_i)+\D(\bar{P}_i\|\bar{P}_0) \label{eq:fullMI-PID1} \\
    & = \D(\Pff\|\bar{P}^i)+\D(\bar{P}^i\|\bar{P}_0) \label{eq:fullMI-PID2} \,,
\end{align}
where Eq.~\eqref{eq:fullMI-PID1} isolates the dependencies between the sources, with the two targets held jointly, and Eq.~\eqref{eq:fullMI-PID2} isolates the dependencies between the targets, with the two sources considered jointly. 
As usual, we can write these in the more classical information-theoretical language as:
\begin{equation} \label{eq:full_MI_cond_joint}
\begin{split}
    I(X_1,X_2;Y_1,Y_2) & = I(X_{-i};Y_1,Y_2\mid X_i) + I(X_i; Y_1,Y_2) \\
    & = I(X_1,X_2;Y_{-i}\mid Y_i) + I(X_1, X_2; Y_i) \,,
\end{split}
\end{equation}
which represent Eqs.~\eqref{eq:fullMI-PID1}-~\eqref{eq:fullMI-PID2}, respectively.

Applying \defs to each family, i.e.\ interpolating between $\bar{P}_1,\bar{P}_2$ on one hand and between $\bar{P}^1,\bar{P}^2$ on the other, yields two further, coarser synergistic quantities:
\begin{align} \label{app:eq:PstarSources}
    r^* & := \underset{r\in\mathbb{R}}{\arg\min}\, \D\!\left(\Pff\middle\|\,\mathcal{Z}\left(\bar{P}_1^{r}\,\bar{P}_2^{1-r}\right)\right) \\
    P_* & := \mathcal{Z}\left(\bar{P}_1^{\,r^*}\,\bar{P}_2^{\,1-r^*}\right) \\
    S_{\text{joint}}(X_1,X_2;Y_1,Y_2) & := \D(\Pff\|P_*) \\ \label{app:eq:PstarTargets}
    s^* & := \underset{s\in\mathbb{R}}{\arg\min}\, \D\!\left(\Pff\middle\|\,\mathcal{Z}\left((\bar{P}^1)^{s}\,(\bar{P}^2)^{1-s}\right)\right) \\
    P^* & := \mathcal{Z}\left((\bar{P}^1)^{\,s^*}\,(\bar{P}^2)^{\,1-s^*}\right) \\
    S^{\text{joint}}(X_1,X_2;Y_1,Y_2) & := \D(\Pff\|P^*) \,.
\end{align}
In practice, these two quantities are not particularly useful for the study of dependencies in a large system. In fact, they aggregate, respectively, the synergistic mode in which the sources act jointly on the targets and every mode in which the targets act jointly on the sources. Hence, they are larger than the order-3 synergies defined in Sec.~\ref{app:sec:3varmarg}, comprising additional dependencies. We will clarify how these different synergies relate to each other in the next section.

In the meantime, these last ``joint'' synergies provide the following decompositions of Eqs.~\eqref{eq:fullMI-PID1}-~\eqref{eq:fullMI-PID2} as:
\begin{equation} \label{eq:S_joint_down}
\begin{split}
    \underbrace{\DAllS}_{\text{\normalfont all dependencies}} & = \underbrace{\DSstar}_{S_{\text{joint}}(X_1,X_2;Y_1,Y_2)} + \underbrace{\DPstarPi + \DPiPzero}_{\text{\normalfont remaining dependencies}} \\[-1em]
    \underbrace{\rowstrut\phantom{\DAllS}}_{I(X_1,X_2;Y_1,Y_2)}
    & \phantom{=} \underbrace{\rowstrut\phantom{\DSstar} \phantom{{}+{}} \phantom{\DPstarPi}}_{I(X_{-i};Y_1,Y_2\mid X_i)}
    \phantom{{}+{}} \underbrace{\rowstrut\phantom{\DPiPzero}}_{I(X_i;Y_1,Y_2)}
\end{split}
\end{equation}
\begin{equation} \label{eq:S_joint_up}
\begin{split}
    \underbrace{\DAllT}_{\text{\normalfont all dependencies}} & = \underbrace{\DSstarT}_{S^{\text{joint}}(X_1,X_2;Y_1,Y_2)} + \underbrace{\DPstarPI + \DPIPzero}_{\text{\normalfont remaining dependencies}} \\[-1em]
    \underbrace{\rowstrut\phantom{\DAllT}}_{I(X_1,X_2;Y_1,Y_2)}
    & \phantom{=} \underbrace{\rowstrut\phantom{\DSstarT} \phantom{{}+{}} \phantom{\DPstarPI}}_{I(X_1,X_2;Y_{-i}\mid Y_i)}
    \phantom{{}+{}} \underbrace{\rowstrut\phantom{\DPIPzero}}_{I(X_1,X_2;Y_i)}
\end{split}
\end{equation}

With this formalism in hand, we can now perform the final information decomposition identification with \phiid.

\subsection{The information-geometric decomposition \phiidig} \label{app:sec:bridge}

Integrated information decomposition (\phiid) is an information-theoretic framework that analyses how the mutual information of the system can be broken down into distinct informational modes, describing how multiple sources jointly carry information about multiple targets~\cite{mediano2025toward}.

For a system with two sources (inputs) $(X_1,X_2)$ and two targets (outputs) $(Y_1,Y_2)$, the model tracks not only how much redundant, unique, and synergistic information the sources hold about the targets, but also how each of these modes, once it reaches the targets, is itself expressed redundantly, uniquely, or synergistically across them. This double resolution is encoded in sixteen informational \textit{atoms}, each denoted $\alpha\to\beta$ (with $\alpha,\beta\in\{R,U_1,U_2,S\}$), read as ``information that is $\alpha$ among the sources becomes $\beta$ among the targets''. For instance, \xtx denotes information uniquely carried by $X_1$ that remains uniquely associated with $Y_1$, while \rts denotes information redundant across the sources that becomes synergistic across the targets. 

\begin{figure}
    \centering
    \includegraphics[width=0.5\linewidth]{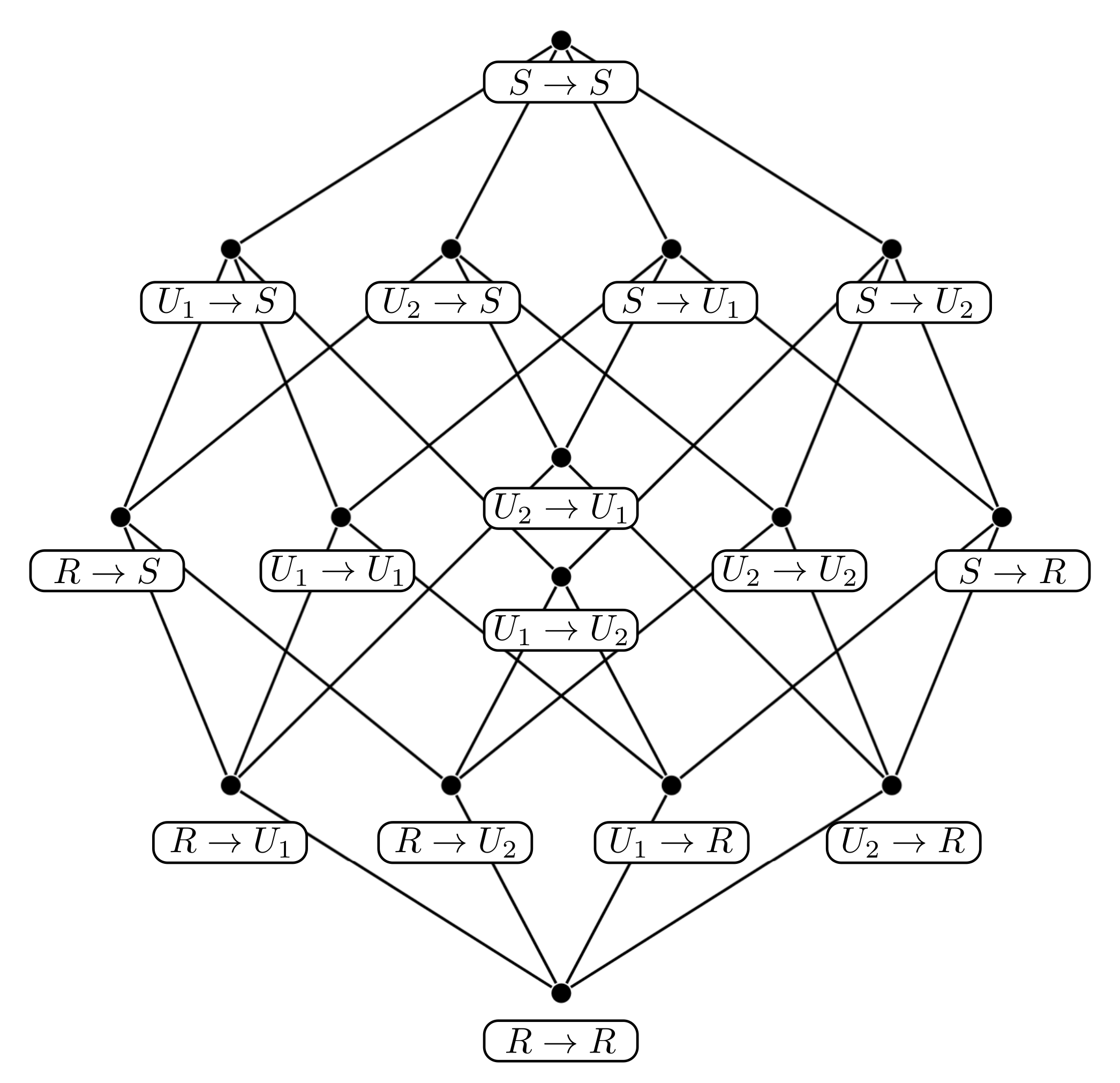}
    \caption{The \phiid double redundancy lattice.}
    \label{fig:phiid_lattice}
\end{figure}

The full structure of the \phiid atoms is encoded in the so-called ``double redundancy'' lattice~\cite{mediano2025toward} (Fig.~\ref{fig:phiid_lattice}). The sixteen atoms are ordered from the fully redundant atom \rtr at the bottom (lowest-order information) to the fully synergistic atom \sts at the top (highest-order information). Each mutual information term corresponds to a ``downward-closed'' region of this lattice: the sum of the atoms that lie on or below a given node is precisely the corresponding mutual information. For instance, $I(X_1;Y_1)$ sums the four atoms reachable from below \xtx (namely \rtr, \rtx, \xtr, \xtx, see Eq.~\eqref{eq:MI_X1Y1} below), while $I(X_1,X_2;Y_1,Y_2)$ comprises every atom in the lattice. The nine mutual informations of the next section are precisely nine such downward closures, one for each non-trivial subset of sources and targets.

Because there are sixteen atoms but only nine independent mutual informations to constrain them, the system of equations is underdetermined, with seven further constraints needed to pin down every atom uniquely. The usual route is to postulate a redundancy function, from which the remaining atoms follow~\cite{mediano2025toward, liardi2025scalable}. However, a large and growing number of such functions have been proposed~\cite{williams2010nonnegative, harder2013bivariate, griffith2014intersection, bertschinger2014quantifying, griffith2014quantifying, griffith2015quantifying, barrett2015exploration, goodwell2017temporal, ince2017measuring, james2018unique, kay2018exact, james2018uniquekey, finn2018pointwise, niu2019measure, kay2024partial, sigtermans2020partial, sigtermans2020towards, makkeh2021introducing, schick2021partial, ehrlich2024partial, kolchinsky2022novel, venkatesh2022partial, mages2023decomposing, lyu2024explicit}, each formalising a different intuition about what redundancy should mean, with no unique consensus on which should be universally employed~\cite{liardi2026mathematical}. 
Here we take a different route entirely: rather than adopting any particular redundancy function, we use only the structural properties of the \phiid lattice, showing that the geometric synergies constructed in Sec.~\ref{app:sec:4var} have a natural and clear interpretation within this formalism. Crucially, this identification provides exactly the missing constraints to invert the system of equations mentioned above, yielding a complete, synergy-first determination of the \phiid atoms.

\subsubsection{The nine mutual informations} \label{app:sec:bridge:9mi}

The first ingredient is the set of nine mutual informations between every non-empty subset of sources and every non-empty subset of targets, each expressed as a sum of \phiid atoms:
\begin{align}
\allowdisplaybreaks
I(X_1; Y_1) &= \rtr + \rtx + \xtr + \xtx \label{eq:MI_X1Y1} \\[1ex]
I(X_2; Y_2) &= \rtr + \rty + \ytr + \yty \\[1ex]
I(X_1; Y_2) &= \rtr + \rty + \xtr + \xty \label{eq:MI_X1Y2} \\[1ex]
I(X_2; Y_1) &= \rtr + \rtx + \ytr + \ytx \\[1ex]
I(X_1, X_2; Y_1) &=
  \begin{aligned}[t]
    &\rtr + \rtx + \xtr + \xtx + \\
    & \ytr + \ytx + \str + \stx
  \end{aligned} \\[1ex]
I(X_1, X_2; Y_2) &=
  \begin{aligned}[t]
    &\rtr + \rty + \xtr + \xty+ \\
    & \ytr + \yty + \str + \sty
  \end{aligned} \\[1ex]
I(X_1; Y_1, Y_2) &=
  \begin{aligned}[t]
    &\rtr + \xtr + \rtx + \xtx +\\
    & \rty + \xty + \rts + \xts
  \end{aligned} \label{eq:MI_X1Y1Y2} \\[1ex]
I(X_2; Y_1, Y_2) &=
  \begin{aligned}[t]
    &\rtr + \ytr + \rtx + \ytx +\\
    &\rty + \yty + \rts + \yts
  \end{aligned}  \\[1ex]
I(X_1, X_2; Y_1, Y_2) &=
\begin{aligned}[t]
    &\rtr + \xtr + \ytr + \str + \\
    & \rtx + \xtx + \ytx + \stx + \\
    & \rty + \xty + \yty + \sty + \\
    & \rts + \xts + \yts + \sts \,. \label{eq:phiid16}
\end{aligned}
\end{align}

However, as mentioned above, these do not allow a complete determination of the \phiid atoms. To achieve this, we now draw the connections between \phiid and the geometrical concepts introduced in the previous sections.

\subsubsection{Geometric constraints on \phiid} \label{app:sec:bridge:redundancy}

Let's first consider the 3-variable subsystem $(X_1,Y_1,Y_2)$. As we have seen in Sec.~\ref{app:sec:3varmarg}, the total dependencies of the system, as captured by the mutual information, can be either written as a sum of order-three synergy and pairwise contributions, or via the usual chain rule of mutual information (Eq.~\eqref{eq:subsys_3var_decomp_S}). Making use of the latter, we can investigate which \phiid atoms are contained in each term by using Eqs.~\eqref{eq:MI_X1Y1}-~\eqref{eq:MI_X1Y2}-~\eqref{eq:MI_X1Y1Y2}:
\begin{align}
    \begin{split}
    I(X_1;Y_{-i}|Y_i) & = I(X_1;Y_1,Y_2) - I(X_1;Y_i) = \\
        & = \rtmi + \xtmi + \rts + \xts \label{app:eq:cond_mi_3var} \,.
    \end{split}
\end{align}
At the same time, from Eq.~\eqref{eq:subsys_3var_decomp_S} we have
\begin{equation}
\begin{split}
    I(X_1;Y_{-i}|Y_i) = \D(P_{X_1Y_1Y_2}||P^*_{X_1 Y_1 Y_2}) + \D(P^*_{X_1Y_1 Y_2}||\bar{P}_{i, \,{X_1 Y_1 Y_2}})\,.
\end{split}
\end{equation}
Evidently, the common terms that do not depend on $i$, i.e.\ are present in both expressions for $i=1,2$, are $\rts + \xts$ and $\D(P_{X_1Y_1Y_2}||P^*_{X_1 Y_1 Y_2})$. This provides a natural identification between the two terms, leading to the connection between order-3 synergy and synergies in the \phiid framework:
\begin{equation}
    \rts + \xts = \D(P_{X_1Y_1Y_2}\|P^*_{X_1Y_1Y_2}) = S_3(X_1;Y_1,Y_2)  \,,
\end{equation}
where the last step follows from the definition of $S_3(X_1;Y_1,Y_2)$ (Eq.~\eqref{app:eq:Pstar112}).

The same argument applied to the remaining order-3 synergies of Eq.~\eqref{app:eq:S3_others} and the corresponding conditional mutual information provides analogous identities: 
\begin{align}
\str + \stx & = \D(P_{X_1X_2Y_1}||P^*_{X_1X_2Y_1}) = S_3(X_1,X_2;Y_1) \label{eq:2th_phiid_def} \\ 
\str + \sty & = \D(P_{X_1X_2Y_2}||P^*_{X_1X_2Y_2}) = S_3(X_1,X_2;Y_2) \label{eq:3th_phiid_def} \\
\rts + \yts & = \D(P_{X_2Y_1Y_2}||P^*_{X_2Y_1Y_2}) = S_3(X_2;Y_1,Y_2) \label{eq:4th_phiid_def} \,.
\end{align}

We now focus on the order-4 synergy term. Applying similar reasoning as before, we now consider the full mutual information $I(X_1,X_2;Y_1,Y_2)$ and use the chain rule decomposition of Eq.~\eqref{eq:MI_four_chain}. Performing some substitutions in terms of the \phiid atoms, we obtain
\begin{equation} \label{eq:double_cond_mi_atoms}
\begin{split}
    I(X_i;Y_j|X_{-i},Y_{-j}) & = I(X_1,X_2;Y_1,Y_2) - I(X_{-i};Y_j,Y_{-j}) - I(X_i;Y_{-j}|X_{-i}) \\
    & = \itj + \its + \stj + \sts \,.
\end{split}
\end{equation}
Importantly, we notice that the only term that does not depend on the choice of the specific inputs $i$ and outputs $j$ is $\sts$. Thus, comparing the double conditional mutual information with Eq.~\eqref{eq:double_cond_DKL}, this leads to the following identification:
\begin{equation}
    \sts = \D(\Pff||P^*_*) = S_4(X_1,X_2;Y_1,Y_2) \,.
\end{equation}
This is aligned with the conceptual meaning of order-4 synergy. Since, by definition, $S_4$ captures the higher-order dependencies present only at the level of the whole system and not in any strict subsystem, it must be the unique atom common to each of the four double-conditional mutual informations of Eq.~\eqref{eq:MI_four_chain}.

Finally, we consider the scenarios in which pairs of variables are treated as a single multivariate component. In such a case too, we consider a chain rule on the full mutual information, but this time written as Eq.~\eqref{eq:full_MI_cond_joint}, instead of the double conditioning. Applying it to the system $((X_1,X_2),Y_1,Y_2)$, this yields the following expression in terms of \phiid atoms:
\begin{equation}
\begin{split}
    I(X_i;Y_1,Y_2\mid X_{-i}) & = I(X_1,X_2;Y_1,Y_2) - I(X_{-i}; Y_1,Y_2) \\
    & = \begin{aligned}[t]
    & \itr + \itx + \ity + \its + \\
    & \str + \stx + \sty + \sts \,,
    \end{aligned}
\end{split}
\end{equation}
with a similar identity holding for $(X_1,X_2,(Y_1,Y_2))$:
\begin{equation}
\begin{split}
    I(X_1,X_2;Y_i\mid Y_{-i}) & = I(X_1,X_2;Y_1,Y_2) - I(X_1, X_2; Y_{-i}) \\
    & = \begin{aligned}[t]
    & \rti + \xti + \yti + \sti + \\
    & \rts + \xts + \yts + \sts \,.
    \end{aligned}
\end{split}
\end{equation}
Comparing these expressions with Eqs.~\eqref{eq:S_joint_down}-~\eqref{eq:S_joint_up}, we can equate the terms that do not depend on the index $i$, obtaining
\begin{align}
    \str + \stx + \sty + \sts & = \D(\Pff\|P_*) = S_{\text{joint}}(X_1,X_2;Y_1,Y_2) \\
    \rts + \xts + \yts + \sts & = \D(\Pff\|P^*) = S^{\text{joint}}(X_1,X_2;Y_1,Y_2) \,.
\end{align}
These two equations complete the link between the geometric synergies and the \phiid atoms. As expected, these joint synergies contain additional informational atoms that cannot be decoupled from univariate conditioning. 

Therefore, we have now completed the identification of geometric synergies in terms of \phiid atoms, which provided seven additional constraints: four from the order-3 synergies, one from the order-4 synergy, and the last two from the joint synergies.
Hence, with these identities and the nine mutual information expressions at hand, the \phiid system of sixteen equations is fully determined, and every atom can be solved for explicitly.

\subsection{Geometric synergy in general multivariate systems} \label{app:sec:general_nm}

The geometrical construction above specialises to the multivariate case with $n=2$ sources and $m=2$ targets.
However, our approach is far more general and can be easily applied to systems of any dimensions. Hence, we now briefly discuss how these geometric principles scale to an arbitrary system of $n$ sources $\Xs$ and $m$ targets $\Ys$ with underlying distribution $P_{\Xs \Ys}$.

As usual, we start by defining the full manifold of the system as
\begin{equation}
    M := \{ P_{\Xs \Ys} \text{ s.t. } P_{\Xs\Ys} = \exp\!\left( \sum_{i \subseteq \{\Xs,\Ys\}} \theta_i F_i(i) - \psi(\theta) \right) \Bigg\} \,. 
\end{equation}
For ease of notation, we denote $P_{\Xs \Ys}$ simply as $P$.
For every choice of one source $X_i$ ($i=1,\dots,n$) and one target $Y_j$ ($j=1,\dots,m$), define the submanifold
\begin{equation} \label{app:eq:Eij_general}
    E_{-i}^{-j} := \left\{P \in M \text{ s.t. } X_i \perp Y_j \;\middle|\; X_{-i} Y_{-j} \right\} \,,
\end{equation}
where $X_{-i}= \{X_k\}_{k\ne i}$ and $Y_{-j} = \{Y_l\}_{l\ne j}$.
In other words, this is the submanifold obtained by conditioning on every source and target other than the pair $(X_i,Y_j)$, and removing precisely the direct interaction between this one remaining pair. This gives $n\times m$ such submanifolds, of which the four submanifolds of Eq.~\eqref{app:eq:Sij} are the special case $n=m=2$. Projecting $P$ onto each $E_{-i}^{-j}$ gives the closed-form projection $\bar P_{-i}^{-j}$ as
\begin{equation}
    \bar P_{-i}^{-j} = \frac{P_{X_iX_{-i}Y_{-j}}P_{X_{-i}Y_jY_{-j}}}{P_{X_{-i}Y_{-j}}} \,,
\end{equation}
generalising Eq.~\eqref{app:eq:PijProj}. 

The highest-order synergy of the full system, $S_{n+m}$, is then defined by interpolating over all $n\times m$ projections:
\begin{align} \label{app:eq:Sfull_general}
    \boldsymbol\lambda^* := \underset{\boldsymbol\lambda\,:\,\sum_{i,j}\lambda_{ij}=1}{\arg\min} \; & \D\!\left(P \,\middle\|\, \mathcal{Z}\prod_{i=1}^n\prod_{j=1}^m \left(\bar P_{-i}^{-j}\right)^{\lambda_{ij}}\right) \,, \\
    S_{n+m}(\Xs;\Ys) := & \D\!\left(P\,\middle\|\,\mathcal{Z}\prod_{i=1}^n\prod_{j=1}^m\left(\bar P_{-i}^{-j}\right)^{\lambda^*_{ij}}\right) \,.
\end{align}
As in the $n=m=2$ case, this is the residual dependency that cannot be explained by any combination of pairwise, or higher but not fully joint, source-target pathways: every $\bar P_{-i}^{-j}$ retains all interactions of the system except the single one between $X_i$ and $Y_j$, so any point on the interpolating family remains explicable without a genuinely $(n+m)$-way contribution.

Lower-order synergies then follow recursively. Marginalising the system down to $n'\le n$ sources and $m'\le m$ targets, i.e.\ by dropping one source or one target and repeating as needed, yields a smaller system to which the same definition, Eq.~\eqref{app:eq:Sfull_general}, applies verbatim, giving its own top-order synergy $S_{n'+m'}$ for that subsystem. Repeating this by dropping one variable at a time from either side eventually reduces every subsystem to size three, at which point $n'+m'=3$ and exactly one of $n'=1$ or $m'=1$ holds. Then the elementary construction of Sec.~\ref{app:sec:3var} can be used, applied to the corresponding marginal as in Sec.~\ref{app:sec:3varmarg}. 
As shown in Sec.~\ref{app:sec:nonhier} for the simpler case $n=m=2$, synergies obtained along different branches of this recursion are generally not mutually orthogonal, so the total dependency of the system is not simply the sum of the synergies obtained at every order. Nonetheless, each such synergy remains individually well-defined and computable from Eq.~\eqref{app:eq:Sfull_general}, providing a scalable characterisation of higher-order structure in arbitrarily large systems.

\section{Gaussian implementation of \phiidig} \label{app:sec:gaussian}

The construction of Secs.~\ref{app:sec:3var}--\ref{app:sec:bridge} holds for arbitrary exponential-family distributions, but is not, in general, available in closed form: solving for a projection onto a conditional-independence submanifold, and then minimising the KL divergence along the $e$-geodesic connecting two or more such projections, requires numerical optimisation over the natural parameters $\theta$. When the system under study is jointly Gaussian, however, both steps reduce to simple linear algebra on the covariance matrix, and the minimisation collapses to a low-dimensional, analytically simplified optimisation~\cite{kay2018exact, kay2024partial}. This section makes that reduction explicit, following the same order-3-then-order-4 structure as Sec.~\ref{app:sec:4var}, and is the theoretical basis of the Gaussian solver used throughout the main text.

In the following treatment, all variables are taken to be zero-mean (in practice, each time series is standardised before the covariance matrix is estimated). For two zero-mean, $d$-dimensional Gaussian distributions $P=\mathcal{N}(\mathbf{0},\Sigma_P)$ and $Q=\mathcal{N}(\mathbf{0},\Sigma_Q)$, the KL divergence reads
\begin{equation} \label{app:eq:gauss_KL}
    \D(P\|Q) = \frac{1}{2}\left[\operatorname{tr}\!\left(\Sigma_Q^{-1}\Sigma_P\right) - d + \ln\frac{\det\Sigma_Q}{\det\Sigma_P}\right] \,.
\end{equation}
Every projection and every interpolation appearing below is a zero-mean Gaussian, so every KL divergence required by the construction of Secs.~\ref{app:sec:3varmarg}--\ref{app:sec:4varcore} is an instance of Eq.~\eqref{app:eq:gauss_KL}.

\subsection{Order-3 synergy} \label{app:sec:gauss_order3}

Consider one of the order-3 constructions of Sec.~\ref{app:sec:3varmarg}, with two source groups $\mathbf{A}_1,\mathbf{A}_2$ and a target group $\mathbf{B}$ jointly Gaussian with zero-mean covariance $\Sigma^{(3)}$ -- for instance $(\mathbf{A}_1,\mathbf{A}_2,\mathbf{B})=(X_1,X_2,Y_i)$ or $(\mathbf{A}_1,\mathbf{A}_2,\mathbf{B})=(Y_1,Y_2,X_i)$, as in Eqs.~\eqref{eq:P0_112proj}--\eqref{eq:P2_112proj}. The same construction applies unchanged when $\mathbf{B}$ (or $\mathbf{A}_1,\mathbf{A}_2$) is taken to be a joint pair rather than a single variable, which is the Gaussian realisation of the two coarser ``joint'' synergies of Sec.~\ref{app:sec:jointdecomp}.

The submanifold on which $\mathbf{A}_{-k}\perp\mathbf{B}\mid\mathbf{A}_k$ (i.e.\ the analogue of $E_k$ in Eq.~\eqref{app:eq:E3var}, for $k=1,2$) is realised, in the Gaussian case, by leaving every covariance block of $\Sigma^{(3)}$ unchanged except the cross-covariance between $\mathbf{A}_{-k}$ and $\mathbf{B}$, which is replaced by the value implied by conditioning on $\mathbf{A}_k$:
\begin{equation} \label{app:eq:gauss_proj3}
    \bar{\Sigma}^{(k)}_{\mathbf{A}_{-k}\mathbf{B}} = \Sigma_{\mathbf{A}_{-k}\mathbf{A}_k}\,\Sigma_{\mathbf{A}_k\mathbf{A}_k}^{-1}\,\Sigma_{\mathbf{A}_k\mathbf{B}} \,,
\end{equation}
with the transposed block $\bar\Sigma^{(k)}_{\mathbf{B}\mathbf{A}_{-k}}$ set to the transpose of Eq.~\eqref{app:eq:gauss_proj3} and every other block of $\bar\Sigma^{(k)}$ equal to the corresponding block of $\Sigma^{(3)}$. This is simply the Gaussian statement that, once conditioned on $\mathbf{A}_k$, the residual (partial) covariance between $\mathbf{A}_{-k}$ and $\mathbf{B}$ vanishes; Eq.~\eqref{app:eq:gauss_proj3} therefore gives $\bar{\Sigma}^{(k)}$ as the unique Gaussian distribution satisfying the conditional-independence constraint that defines the corresponding submanifold, playing the role of the explicit projections $\bar P_0,\bar P_i$ flagged throughout Secs.~\ref{app:sec:3var}--\ref{app:sec:3varmarg} for the general exponential-family case.

The natural parameter of a zero-mean Gaussian is the precision matrix, so the $e$-geodesic interpolation of \defs -- the general-case interpolation of, e.g.\ Eq.~\eqref{app:eq:Pstar112} -- is linear in precision space. Writing $K_k=(\bar\Sigma^{(k)})^{-1}$, the interpolating family is
\begin{equation} \label{app:eq:gauss_interp3}
    K(u) = (1-u)K_1 + uK_2 \,, \qquad P_u = \mathcal{N}\!\left(\mathbf{0},K(u)^{-1}\right) \,,
\end{equation}
restricted to the set $u\in\mathcal{U}$ of values for which $K(u)$ remains positive definite (an interval, generally a strict superset of $[0,1]$, found by root-finding on the boundary of positive-definiteness). The order-3 synergy is then the minimum KL divergence between $P=\mathcal{N}(\mathbf{0},\Sigma^{(3)})$ and this family,
\begin{equation} \label{app:eq:gauss_S3_full}
    S_3(\mathbf{A}_1,\mathbf{A}_2;\mathbf{B}) = \min_{u\in\mathcal{U}} \D\!\left(P\,\|\,P_u\right) \,,
\end{equation}
the Gaussian counterpart of the general interpolation-minimisation of \defs, e.g.\ Eq.~\eqref{app:eq:Pstar112}.

Substituting $P_u=\mathcal{N}(\mathbf{0},K(u)^{-1})$ into Eq.~\eqref{app:eq:gauss_KL} gives
\begin{equation}
    \D(P\|P_u) = \frac{1}{2}\left[\operatorname{tr}\!\left(K(u)\Sigma^{(3)}\right) - d + \ln\frac{\det K(u)^{-1}}{\det\Sigma^{(3)}}\right] \,,
\end{equation}
with $d$ the total dimension of $(\mathbf{A}_1,\mathbf{A}_2,\mathbf{B})$. Since each projection $\bar\Sigma^{(k)}$ differs from $\Sigma^{(3)}$ only in the one block set by Eq.~\eqref{app:eq:gauss_proj3}, and that block is precisely the one annihilated by $K_k$'s corresponding zero pattern, $\operatorname{tr}(K_k\Sigma^{(3)})=d$ for $k=1,2$; by linearity of the trace, $\operatorname{tr}(K(u)\Sigma^{(3)})=d$ for every $u$. The trace and dimensional terms therefore cancel identically along the whole interpolation, leaving the order-3 synergy as a one-dimensional minimisation of a single log-determinant ratio,
\begin{equation} \label{app:eq:gauss_S3}
    S_3(\mathbf{A}_1,\mathbf{A}_2;\mathbf{B}) = \min_{u\in\mathcal{U}}\; \frac{1}{2}\ln\frac{\det K(u)^{-1}}{\det\Sigma^{(3)}} \,.
\end{equation}
Evaluating Eq.~\eqref{app:eq:gauss_S3} at $(\mathbf{A}_1,\mathbf{A}_2,\mathbf{B})=(X_1,X_2,Y_i)$ and at $(Y_1,Y_2,X_i)$, for $i=1,2$, gives the four Gaussian order-3 synergies of Eqs.~\eqref{app:eq:Pstar112} and~\eqref{app:eq:S3_others} in closed form.

\subsection{Order-4 synergy} \label{app:sec:gauss_order4}

The order-4 construction of Sec.~\ref{app:sec:4varcore} follows the identical pattern, with the single conditioning of Eq.~\eqref{app:eq:gauss_proj3} replaced by the double conditioning that defines the submanifolds $E_i^j$ of Eq.~\eqref{app:eq:Sij}. Fix a source-target pair $(X_i,Y_j)$ and write $A=X_{-i}$, $B=Y_{-j}$, $\mathbf{C}=(X_i,Y_j)$ for the retained pair, so that $\Sigma$ denotes the full covariance of $(X_1,X_2,Y_1,Y_2)$ partitioned according to $(A,\mathbf{C},B)$.

The submanifold $E_i^j$, on which $A\perp B\mid\mathbf{C}$, is realised by the Gaussian projection $\bar\Sigma^{(ij)}$ that agrees with $\Sigma$ everywhere except the $A$--$B$ cross-covariance block, which is set to its value implied by conditioning on $\mathbf{C}$:
\begin{equation} \label{app:eq:gauss_proj4}
    \bar{\Sigma}^{(ij)}_{AB} = \Sigma_{A\mathbf{C}}\,\Sigma_{\mathbf{C}\mathbf{C}}^{-1}\,\Sigma_{\mathbf{C}B} \,, \qquad \bar{\Sigma}^{(ij)}_{BA} = \left(\bar{\Sigma}^{(ij)}_{AB}\right)^{\!\top} \,,
\end{equation}
which is the closed-form realisation of the projections $\bar P_i^j$ of Eq.~\eqref{app:eq:PijProj}, for each of the four choices $i,j\in\{1,2\}$, giving $\bar\Sigma^{(11)},\bar\Sigma^{(12)},\bar\Sigma^{(21)},\bar\Sigma^{(22)}$.

Writing $K_{ij}=(\bar\Sigma^{(ij)})^{-1}$, the precision-linear interpolation over all four projections -- the Gaussian realisation of the three-parameter interpolation of Eq.~\eqref{app:eq:Pstarstar} -- is
\begin{equation} \label{app:eq:gauss_interp4}
    K(\boldsymbol\lambda) = \lambda_{11}K_{11} + \lambda_{12}K_{12} + \lambda_{21}K_{21} + \lambda_{22}K_{22} \,, \qquad \textstyle\sum_{i,j=1}^{2}\lambda_{ij}=1 \,,
\end{equation}
restricted to the admissible set $\boldsymbol\lambda\in\Lambda$ for which $K(\boldsymbol\lambda)\succ0$, with $Q_{\boldsymbol\lambda}=\mathcal{N}(\mathbf{0},K(\boldsymbol\lambda)^{-1})$. The order-4 synergy is
\begin{equation}
    S_4(X_1,X_2;Y_1,Y_2) = \min_{\boldsymbol\lambda\in\Lambda} \D\!\left(P\,\|\,Q_{\boldsymbol\lambda}\right) \,,
\end{equation}
matching Eq.~\eqref{app:eq:Pstarstar} with $\Pff=\mathcal{N}(\mathbf{0},\Sigma)$.

By the same trace argument as in Sec.~\ref{app:sec:gauss_order3} -- each $\bar\Sigma^{(ij)}$ again differs from $\Sigma$ only in the block annihilated by $K_{ij}$, so $\operatorname{tr}(K_{ij}\Sigma)=4$ for every $i,j$, and hence $\operatorname{tr}(K(\boldsymbol\lambda)\Sigma)=4$ throughout $\Lambda$ by linearity -- the trace and dimensional terms of Eq.~\eqref{app:eq:gauss_KL} again cancel identically, and the order-4 synergy reduces to a three-dimensional minimisation of a single log-determinant ratio,
\begin{equation} \label{app:eq:gauss_S4}
    S_4(X_1,X_2;Y_1,Y_2) = \min_{\boldsymbol\lambda\in\Lambda}\; \frac{1}{2}\ln\frac{\det K(\boldsymbol\lambda)^{-1}}{\det\Sigma} \,.
\end{equation}

\subsection{Solving the Gaussian \phiidig} \label{app:sec:gauss_solve}

Once the covariance matrix $\Sigma$ of $(X_1,X_2,Y_1,Y_2)$ has been estimated, the nine mutual informations of Eqs.~\eqref{eq:MI_X1Y1}-~\eqref{eq:phiid16} follow directly from Eq.~\eqref{app:eq:gauss_KL} applied to the appropriate marginal and joint covariance blocks (equivalently, from the standard Gaussian mutual-information formula $I(\mathbf A;\mathbf B)=\tfrac12\ln\det\Sigma_{\mathbf A}\det\Sigma_{\mathbf B}/\det\Sigma_{\mathbf A,\mathbf B}$), and the remaining seven geometric constraints of Sec.~\ref{app:sec:bridge} -- the four order-3 synergies and the two coarser joint synergies of Sec.~\ref{app:sec:jointdecomp}, all instances of the single log-determinant minimisation of Eq.~\eqref{app:eq:gauss_S3}, together with the closing order-4 synergy $\sts$ of Eq.~\eqref{app:eq:gauss_S4} -- complete the system. Together, these sixteen quantities form a closed linear system for the sixteen \phiid atoms, exactly as in the general case of Sec.~\ref{app:sec:bridge}, but with every one of its constituent projections and interpolations available in closed form as elementary operations on $\Sigma$: matrix inversion and multiplication for the projections of Eqs.~\eqref{app:eq:gauss_proj3} and~\eqref{app:eq:gauss_proj4}, and a one- or three-dimensional log-determinant minimisation, respectively, for the order-3/joint and order-4 synergies of Eqs.~\eqref{app:eq:gauss_S3} and~\eqref{app:eq:gauss_S4}. This is the construction underlying the Gaussian implementation of \phiidig used throughout the main text.

\section{Additional results} \label{app:add_res}

\subsection{Discrete state transitions} \label{app:sec:discr_trans}

In the validation of the discrete procedure in the main text, we reported the behaviour of five specific transition matrices and the corresponding \phiidig computational mode maximised, each time showing that the findings match the theoretical expectations.

\begin{figure}[ht]
    \centering
    \includegraphics[width=\linewidth]{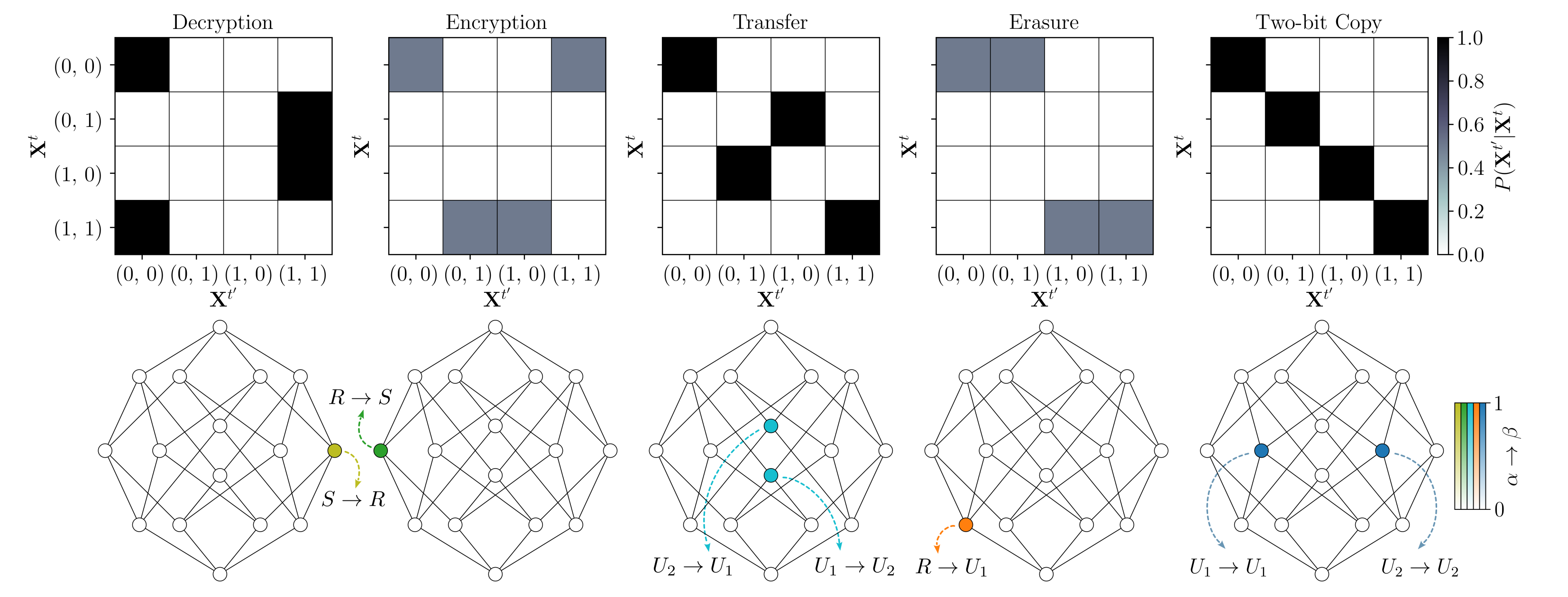}
    \caption{\textbf{Remaining \phiidig computational modes on discrete state transitions.} Transition probability matrices for five distinct logical gates, and corresponding values of the \phiidig atoms computed on them.}
    \label{fig:discrete_remaining}
\end{figure}

Here, we report five analogous logical gates that aim to test the sensitivity of the remaining computational modes. Specifically, we consider Decryption, which transforms synergistic information into redundant; Encryption, which does the opposite; Transfer, where information is exchanged between the two gates; Erasure, where redundant information in the sources is erased at the target level, becoming unique; and finally the Two-Bit Copy gate, where each component remains in its own state independently of what the other does. As expected, \phiidig applied to these systems provides specific computational signatures that only involve the atoms \str, \rts, \itj, \rti, and \iti, respectively, coherently with the theoretical instantiation of the models (Fig.~\ref{fig:discrete_remaining}).

\FloatBarrier

\subsection{Coupled map lattices} \label{app:add_res_cmls}
\begin{figure}[ht]
    \centering
    \includegraphics[width=\linewidth]{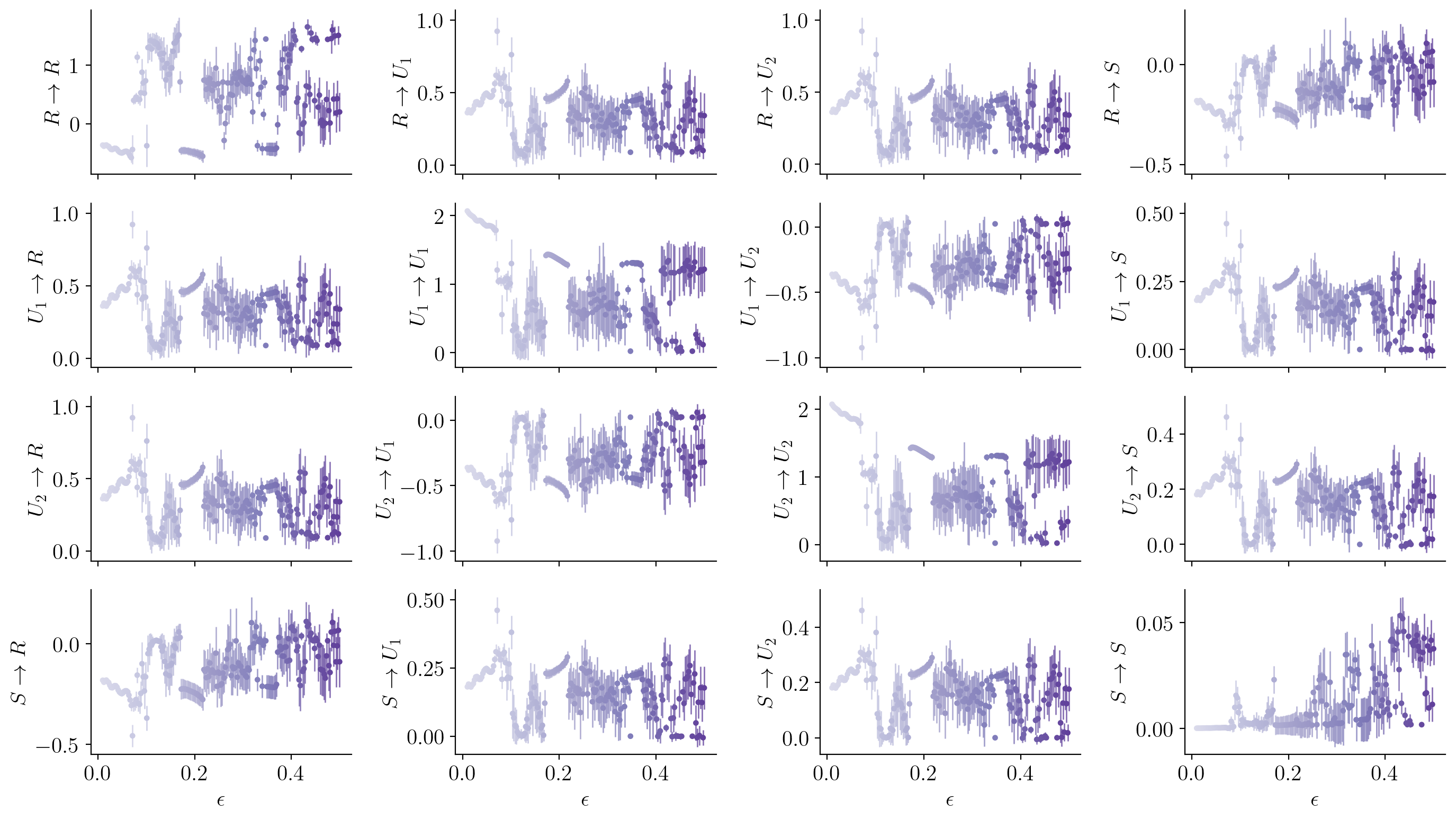}
    \caption{\textbf{All sixteen \phiidig atoms across the coupled-map-lattice sweep.} Each panel shows one atom as a function of the coupling strength $\epsilon\in(0,0.5)$. Markers are means over sampled node pairs and vertical bars their dispersion; colour encodes $\epsilon$, from light at weak coupling to dark at strong coupling. %
    }
    \label{fig:cmls_all_atoms}
\end{figure}

In this section, we provide complementary results regarding the coupled map lattices. First, we present the sweep over $\epsilon\in(0,0.5)$ for all atoms (Fig.~\ref{fig:cmls_all_atoms}). Interestingly, these provide complementary characterisations of the regimes, in the sense that different atoms respond to the coupling along different axes. The self-persistence of the unique information, \xtx and \yty, is largest in the weak-coupling limit and decays as $\epsilon$ grows, consistent with each site becoming progressively less predictable from its own past alone. The redundancy transfer \rtr behaves in the opposite way and, in addition, changes sign: it is negative for $\epsilon\lesssim0.1$, becoming strongly positive beyond it, so that the shared past acquires predictive power over the shared future only once the sites are appreciably coupled. The pure synergy \sts, by contrast, remains indistinguishable from zero over roughly the first half of the range and grows only for $\epsilon\gtrsim0.2$, reaching its largest values in the most strongly coupled regimes. The mixed terms interpolate between these behaviours, and all of them acquire a visibly larger dispersion across pairs for $\epsilon\gtrsim0.35$, where the lattice supports several coexisting spatiotemporal patterns. Taken together, no single atom orders the five regimes on its own, which is precisely the sense in which the decomposition is informative beyond the total mutual information.

\begin{figure}[ht]
    \centering
    \includegraphics[width=\linewidth]{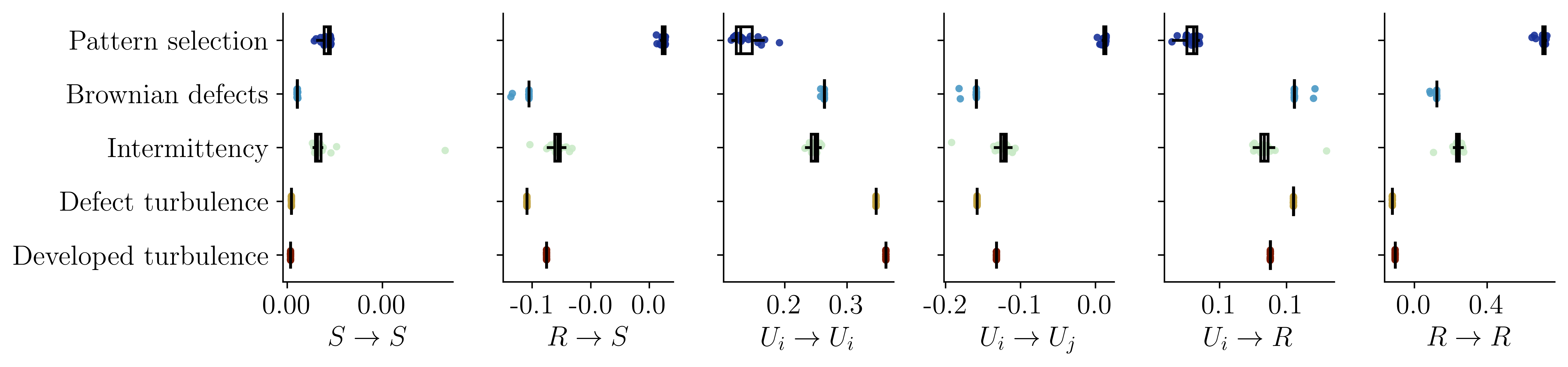}
    \caption{\textbf{Regime comparison for the coupled map lattices with normalised \phiidig atoms.} As Fig.~\figsubref{fig:chaos}{d} of the main text, but with every atom divided by the time-delayed mutual information of the corresponding past and future pair.}
    \label{fig:cmls_normalised}
\end{figure}

Moreover, we show that the comparison across regimes presented in Fig.~\figsubref{fig:chaos}{d} of the main manuscript is not purely driven by the amount of total information carried by each system. In fact, if we normalise each atom with respect to its mutual information~\cite{liardi2024null}, we obtain comparable results (Fig.~\ref{fig:cmls_normalised}). This suggests that the distinction across regimes is genuinely due to the different ways these systems compute information, as opposed to solely how much information is processed. The complete statistics of the regime comparisons are reported in Sec.~\ref{app:full_stats} (Tables~\ref{tab:cml_lme}--\ref{tab:cml_pairwise}).

\FloatBarrier

\subsection{Hopf model} \label{app:add_res_hopf}

\begin{figure}[ht]
    \centering
    \includegraphics[width=\linewidth]{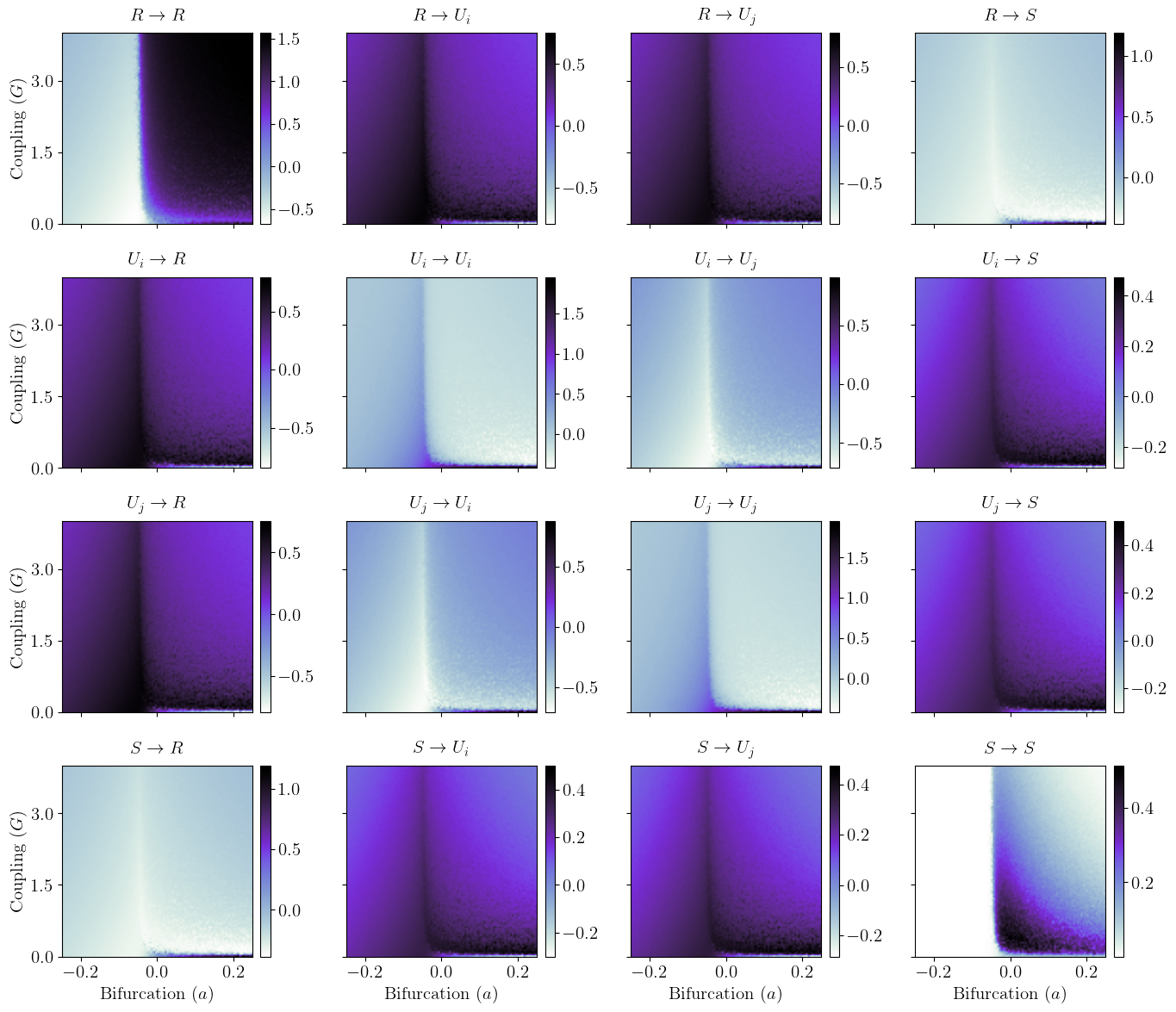}
    \caption{\textbf{All sixteen \phiidig atoms across the Hopf sweep.} Each panel shows one atom over the plane spanned by the bifurcation parameter $a\in[-0.25,0.25]$ and the global coupling $G\in[0,4]$. Values are means over the $300$ node pairs of the $25$-node small-world network, computed at $\tau=1$ with the Spearman correlation estimator.}
    \label{fig:sweep_all}
\end{figure}

In the Hopf model considered in the main manuscript, we showed how specific atoms of \phiidig can differentiate before and after the model's bifurcation point ($a<0$ and $a>0$). In particular, we noted that both \rtr and \sts exhibit strongly different behaviour. Here we complement these findings by showing the rest of the atoms (Fig.~\ref{fig:sweep_all}). We notice that \rtr and \sts yield the strongest differentiations, with the rest of the atoms being weakly dependent on $a$, except for the point $a=0$. In particular, we observe two interesting behaviours.
First, the boundary separating the two phases is essentially vertical in the $(a,G)$ plane for every atom: the coupling strength modulates the magnitude of the atoms but not the location of the transition, so within this model the decomposition is controlled by the local dynamical regime rather than by the strength of the interaction. 
Second, the atoms that most sharply separate the two phases are those defined on the shared past or the shared future, with \rtr increasing and the unique storage and transfer terms \iti, \itj decreasing after the bifurcation, while \sts is indistinguishable from zero for $a<0$ and becomes appreciable only for $a\geq0$. This is consistent with the onset of a coherent collective oscillation, in which the future of each node is predictable from the joint past of the pair rather than from its own past alone. The mixed transfer terms, instead, do not differ markedly between the two phases but exhibit a narrow band of extreme values localised at $a\approx0$. Such a localised enhancement at the critical point would be consistent with the critical slowing-down expected near a Hopf bifurcation, where the relaxation timescale grows and neither the fixed-point nor the limit-cycle description applies cleanly. However, further analyses that more accurately resolve critical exponents would be needed to corroborate these interpretations.

\begin{figure}[ht]
    \centering
    \includegraphics[width=\linewidth]{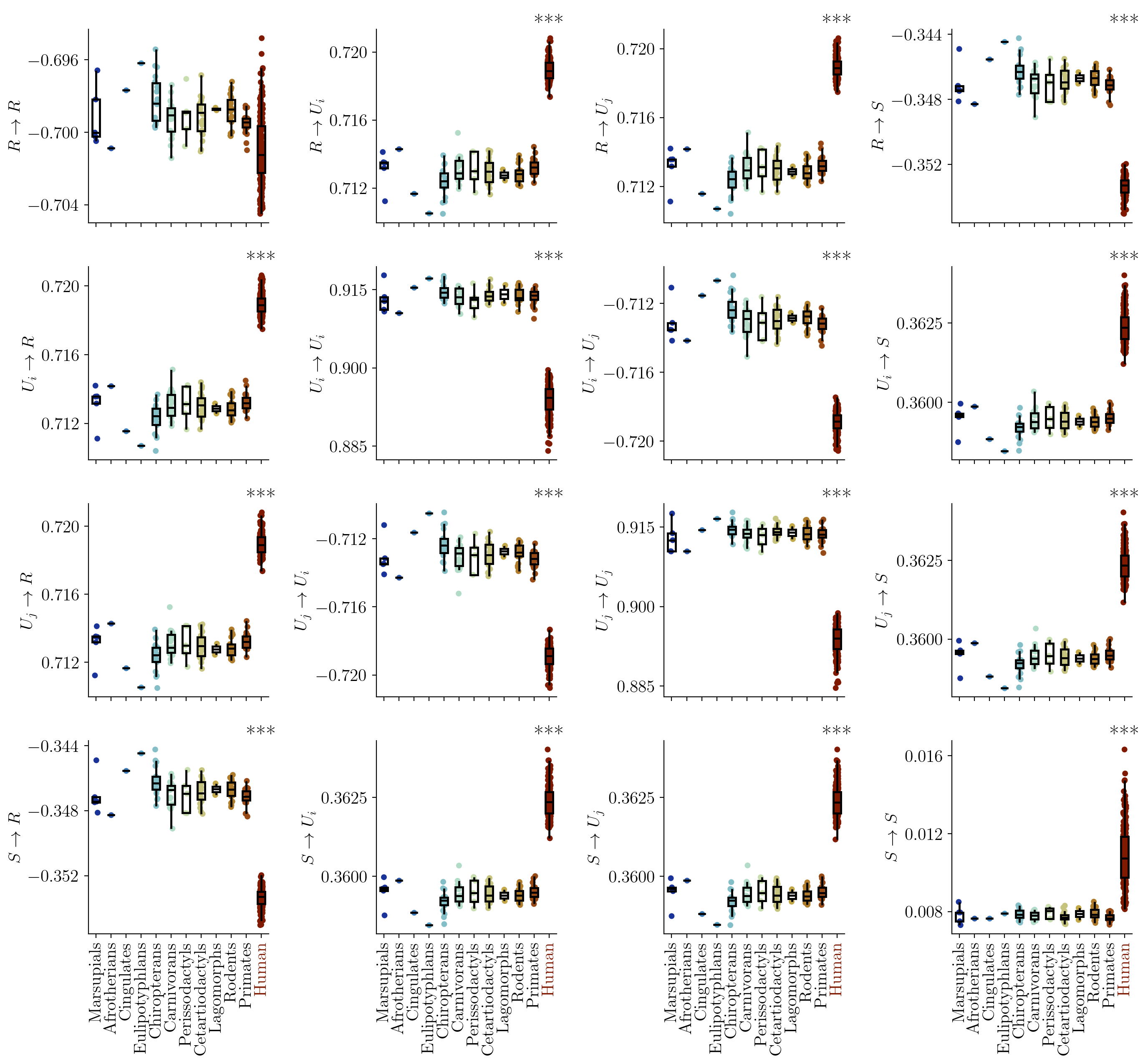}
    \caption{\textbf{The full \phiidig decomposition across the mammalian connectomes.} Each panel shows one \phiidig atom for the $114$ mammalian species grouped by taxonomic order, with the individual human connectomes shown separately as the rightmost group. Boxes give the median and interquartile range across species, and points represent the individual species. Asterisks mark the atoms for which the human group differs from the majority of the other orders at $p_{\mathrm{FDR}}<0.05$ under a Mann--Whitney $U$ test with Benjamini--Hochberg correction.}
    \label{fig:all_mammals}
\end{figure}

Finally, considering the \phiidig atoms' behaviour across the mammalian connectomes ($114$ species across $11$ orders, see Methods in the main text), and complementing the \sts findings shown in the main text, here we provide the outcomes of the full decomposition (Fig.~\ref{fig:all_mammals}). Interestingly, the major differences again involve the human connectome, which exhibits lower unique storage and transfer (\iti and \itj, $-12.0$ and $-7.3$ mammalian standard deviations), but higher synergistic modes, including \sts, \its, and \sti ($12.1$, $9.3$, and $9.3$ standard deviations), with each of these atoms differing from $8$ of the $11$ taxonomic orders (Table~\ref{tab:mammals}). \rtr is also lower in humans ($-1.7$ standard deviations), but differs from only $5$ of the $11$ orders. As discussed in the main paper, this suggests that human evolution has favoured the emergence of connections that would support collective computations, which require higher-order coordination across regions of the brain. 

\begin{figure}[ht]
    \centering
    \includegraphics[width=\linewidth]{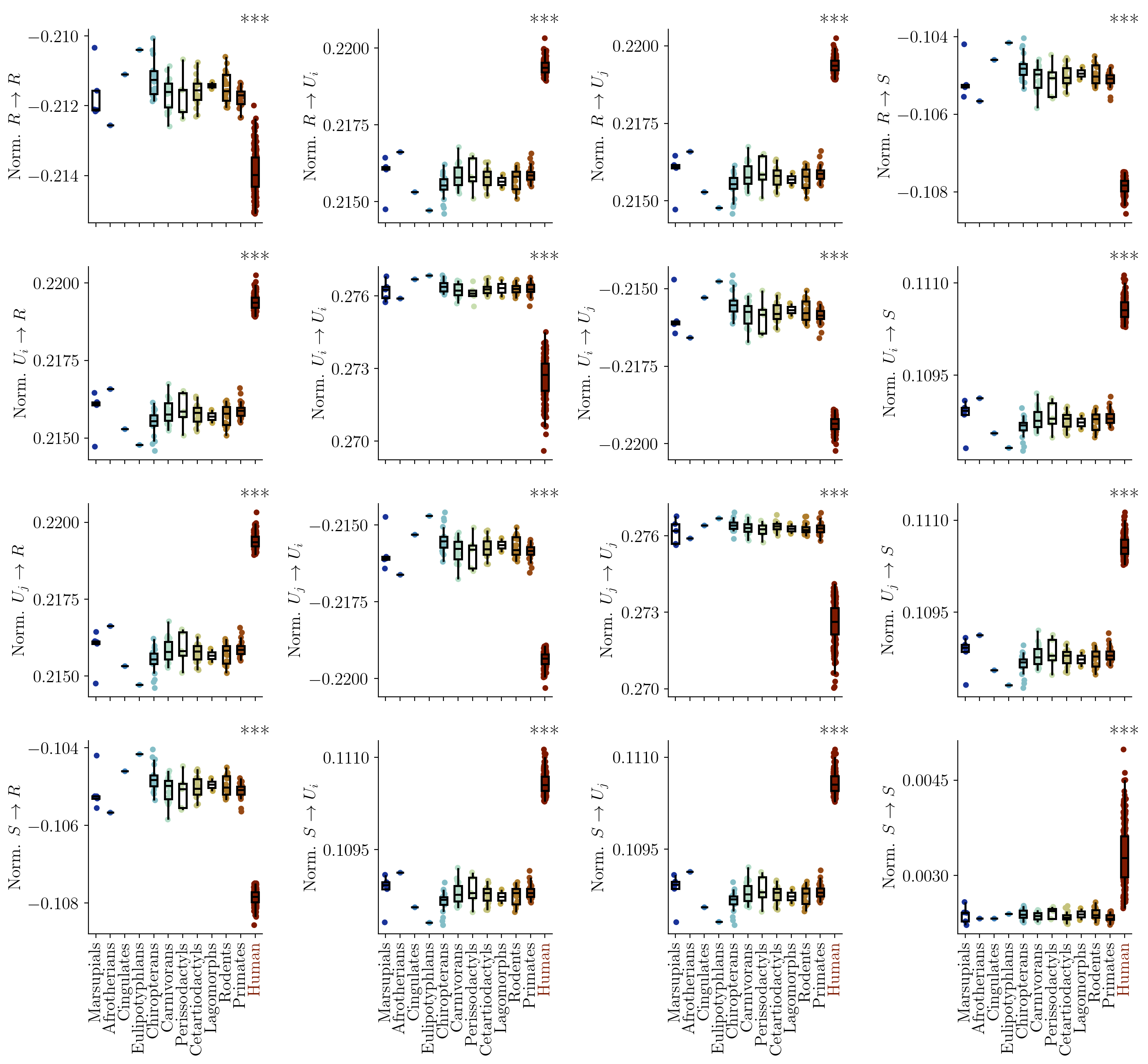}
    \caption{\textbf{The full \phiidig decomposition across the mammalian connectomes, normalised by the total mutual information.} As Fig.~\ref{fig:all_mammals}, but with every atom divided by the time-delayed mutual information of the corresponding past and future pair.}
    \label{fig:all_mammals_norm}
\end{figure}

Moreover, these results are not dictated by an overall difference in the amount of information processed, as normalising the \phiidig atoms w.r.t.\ the mutual information of the system yields the same results (Fig.~\ref{fig:all_mammals_norm}): all human deviations retain their sign (e.g.\ $12.4$ standard deviations for \sts), and after normalisation \rtr also differs from $8$ of the $11$ orders (Table~\ref{tab:mammals}).

We further verified that the differences between human and mammalian connectomes do not hinge on the specific value of the bifurcation parameter. We repeated the simulations of all $314$ connectomes ($200$ humans and $114$ non-human species) for seven further values of $a$ between $-0.1$ and $-0.0125$, keeping all other parameters as in the main analysis ($a=-0.05$). Since the mammalian standard deviation includes the variability across simulations, deviations expressed in standard deviations depend on the number of simulations, and to make all runs directly comparable we restrict the main analysis to its first five simulations here (which yields, e.g.\@, a deviation of $8.5$ instead of $12.1$ standard deviations for \sts).

\begin{figure}[ht]
    \centering
    \includegraphics[width=0.6\linewidth]{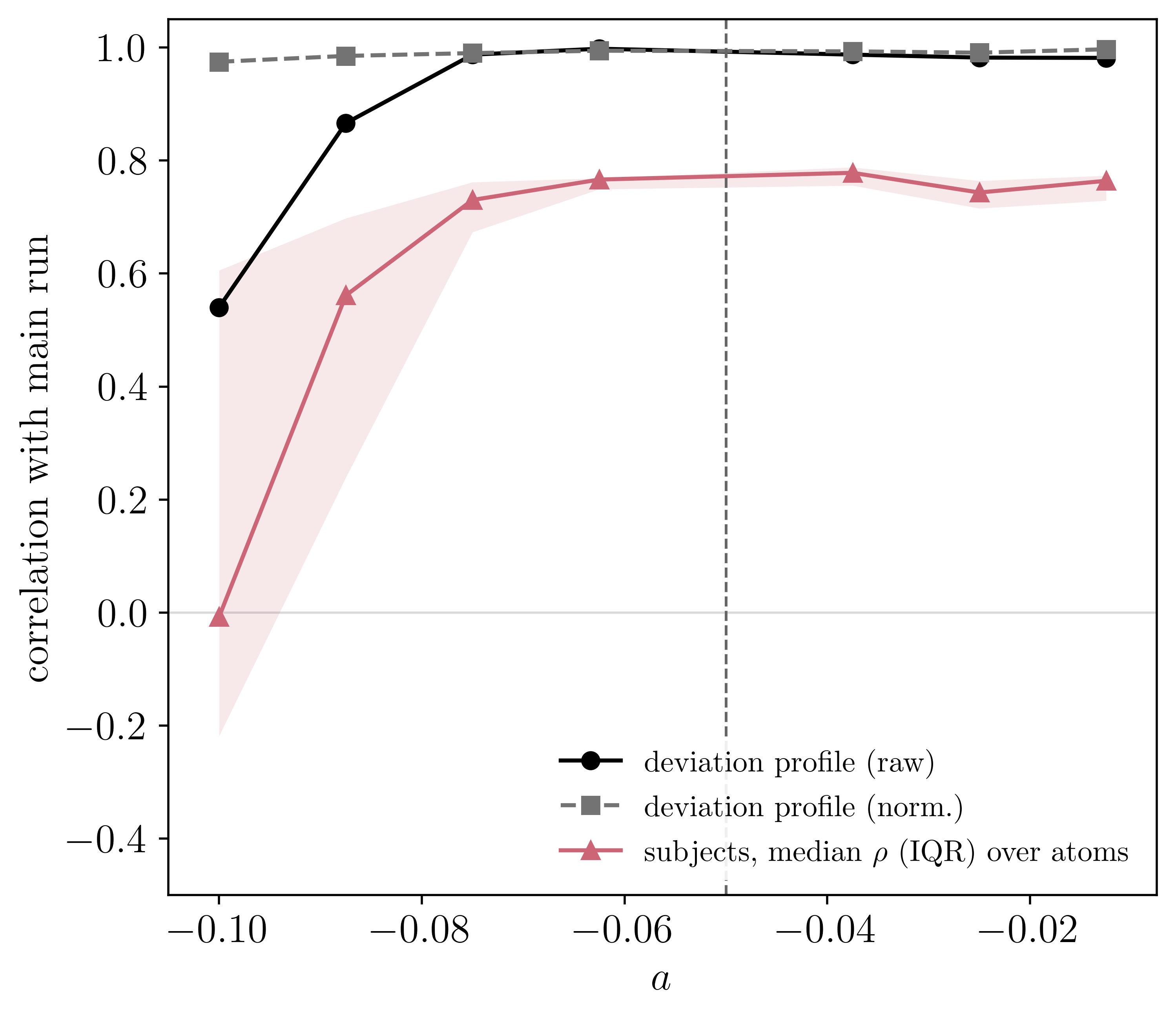}
    \caption{\textbf{Robustness of the human--mammal differences across bifurcation parameters.} Pearson correlation between the profile of the human deviations from the mammalian mean of the $10$ atoms at each value of the bifurcation parameter $a$ and that of the main analysis ($a=-0.05$, dashed line, first five simulations), for raw (circles) and normalised (squares) atoms. Triangles give the median across the $10$ atoms of the Spearman correlation between the values of the $314$ connectomes at each $a$ and those of the main analysis, with the shaded area showing the interquartile range across atoms. The full statistics are reported in Table~\ref{tab:hopf_a}.}
    \label{fig:hopf_a}
\end{figure}

For $-0.075\leq a\leq-0.0125$, the results of the main text are fully reproduced (Fig.~\ref{fig:hopf_a}): all ten atoms differ between humans and mammals in the same direction as for $a=-0.05$ ($p_{FDR}\leq2.0\cdot10^{-19}$, Mann--Whitney $U$ test), and the human deviation profile correlates with that of the main analysis at $r\geq0.98$ (Table~\ref{tab:hopf_a}). In particular, humans present higher \sts than mammals for every value of $a$ considered ($r_{rb}\geq0.24$, $p_{FDR}\leq4.5\cdot10^{-4}$). Further from the bifurcation, the raw atoms become less consistent: \its and \sti change sign at $a=-0.0875$, and \rti, \itr, and \itj also do so at $a=-0.1$, where the deviation profile correlates with the main one only at $r=0.54$. In contrast, the normalised atoms are robust over the whole range: all ten differ from mammals in the same direction as in the main analysis for every value of $a$ ($p_{FDR}\leq2.1\cdot10^{-9}$), with deviation profiles correlating with the main one at $r\geq0.97$. This difference stems from the total information of the system, which more than halves between $a=-0.05$ and $a=-0.1$, and is slightly but consistently lower in humans than in mammals, by $0.85\%$ at $a=-0.05$, increasing to $2.9\%$ at $a=-0.1$ and decreasing to $0.1\%$ at $a=-0.0125$. Beyond the group-level differences, the ranking of the individual connectomes is also moderately preserved: for $a\geq-0.075$, the median across atoms of the Spearman correlation between the values of the $314$ connectomes at each $a$ and those of the main analysis ranges between $0.73$ and $0.78$, dropping to $0.56$ at $a=-0.0875$ and to $-0.01$ at $a=-0.1$ (Fig.~\ref{fig:hopf_a}).

\FloatBarrier

\subsection{Market dynamics} \label{app:add_res_finance}

\begin{figure}[h]
    \centering
    \includegraphics[width=\linewidth]{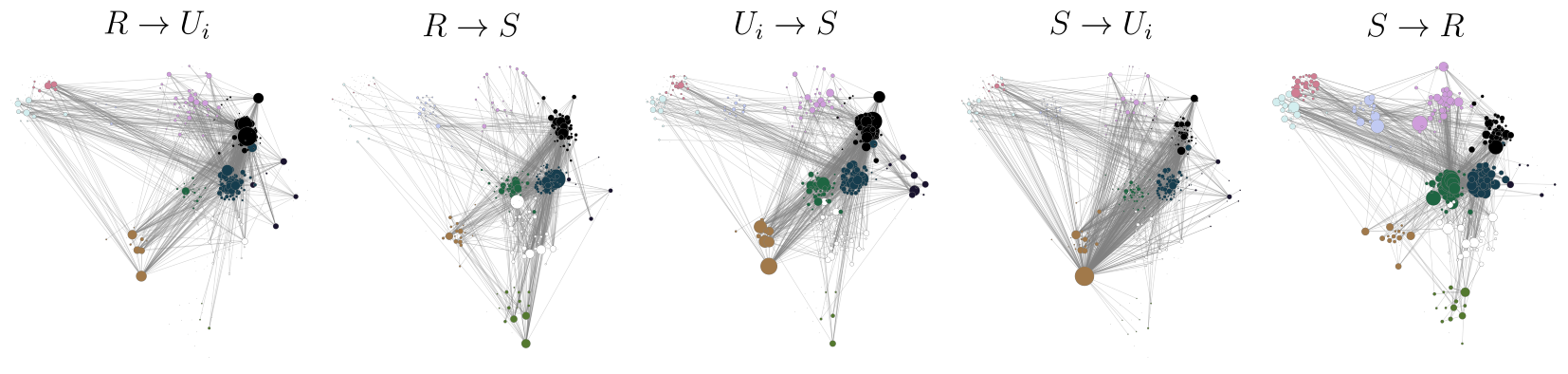}
    \caption{\textbf{Market networks for the remaining five computational modes.} From left to right, the networks induced by \rti, \rts, \its, \sti and \str, complementing the five modes reported in the main text. All panels share the layout and node positions, node colour encodes the economic sector, and node size is proportional to the node's total strength in the corresponding network.}
    \label{fig:remaining_nets}
\end{figure}

In the main text, we presented the market networks emerging from five computational modes. Here, we present the behaviour of the remaining five (Fig.~\ref{fig:remaining_nets}). All five are drawn with the same layout and the same node positions as those of the main text, so that they can be compared directly. With respect to the sector partition (Table~\ref{tab:finance_mod}), only \rts is more modular than the null networks ($Z=6.97$, $p_{\mathrm{perm}}=0.002$), whereas \its and \sti are less modular ($Z=-2.75$, $p_{\mathrm{perm}}=0.004$, and $Z=-5.73$, $p_{\mathrm{perm}}=0.002$, respectively), and neither \rti ($Z=-2.03$, $p_{\mathrm{perm}}=0.054$) nor \str ($Z=0.99$, $p_{\mathrm{perm}}=0.336$) differs significantly from them. The modularity of all networks is also shown in Fig.~\figsubref{fig:finance}{b}.

\FloatBarrier

\section{Full statistical results} \label{app:full_stats}

In this section, we report the complete statistics underlying the analyses of the main text. Unless stated otherwise, Mann--Whitney $U$ tests are two-sided, $Z$ is the tie- and continuity-corrected normal approximation of $U$, $r_{rb}$ is the rank-biserial correlation (positive values of both indicate that the first group is stochastically larger than the second), and $p_{FDR}$ denotes $p$-values corrected with the Benjamini--Hochberg procedure within the family of comparisons stated for each table.

\subsection{Coupled map lattices}

Table~\ref{tab:cml_lme} reports the linear mixed-effects models summarised in Fig.~\figsubref{fig:chaos}{c}. Each model is fitted on the atom values of all $4950$ node pairs, for both halves of each time series, and for the $20$ simulations of each of the five regimes, with regime and time-series half as fixed effects, a random intercept for each simulation, and a random effect for each node, shared by all the pairs that contain it. Atoms that map onto each other under the exchange of the two nodes (e.g.\ \xtx and \yty) are averaged before fitting, so that the results do not depend on the labelling of the nodes.

Table~\ref{tab:cml_regimes} reports the regime-level values shown in Fig.~\figsubref{fig:chaos}{d}, together with the weakest of the four contrasts between the regime in which each atom is maximal and the remaining ones, while Table~\ref{tab:cml_pairwise} lists all pairwise contrasts. All $100$ contrasts are significant ($p_{FDR}\leq1.2\cdot10^{-6}$). With $20$ realisations per regime, $|Z|=5.40$ corresponds to the complete separation of the two groups, i.e.\ the largest attainable value.

\begin{table}[ht]
\centering
\small
\renewcommand{\arraystretch}{1.3}
\setlength{\tabcolsep}{6pt}
\begin{tabular}{lccccccc}
\toprule
Atom & Wald $\chi^2(4)$ & $\log_{10}p_{FDR}$ & $R^2_{m}$ & $R^2_{c}$ & $\sigma^2_{\mathrm{sim}}$ & $\sigma^2_{\mathrm{node}}$ & $\sigma^2_{\mathrm{res}}$ \\
\midrule
\rtr & $25\,948$ & $-5629.7$ & $0.96$ & $0.98$ & $1.8\times10^{-3}$ & $4.7\times10^{-3}$ & $1.0\times10^{-2}$ \\
\iti & $90\,587$ & $-19665.0$ & $0.95$ & $0.98$ & $1.6\times10^{-4}$ & $8.0\times10^{-3}$ & $8.3\times10^{-3}$ \\
\sti & $14\,126$ & $-3063.3$ & $0.89$ & $0.95$ & $6.8\times10^{-5}$ & $3.8\times10^{-4}$ & $6.9\times10^{-4}$ \\
\itj & $14\,847$ & $-3219.8$ & $0.88$ & $0.94$ & $2.7\times10^{-4}$ & $1.6\times10^{-3}$ & $3.1\times10^{-3}$ \\
\its & $13\,924$ & $-3019.6$ & $0.88$ & $0.95$ & $6.9\times10^{-5}$ & $3.8\times10^{-4}$ & $7.3\times10^{-4}$ \\
\itr & $10\,661$ & $-2311.3$ & $0.86$ & $0.94$ & $2.9\times10^{-4}$ & $1.6\times10^{-3}$ & $2.5\times10^{-3}$ \\
\rti & $10\,671$ & $-2313.3$ & $0.86$ & $0.94$ & $2.9\times10^{-4}$ & $1.7\times10^{-3}$ & $2.5\times10^{-3}$ \\
\rts & $20\,069$ & $-4353.6$ & $0.85$ & $0.90$ & $4.8\times10^{-5}$ & $3.7\times10^{-4}$ & $1.5\times10^{-3}$ \\
\str & $20\,220$ & $-4386.3$ & $0.85$ & $0.90$ & $4.7\times10^{-5}$ & $3.7\times10^{-4}$ & $1.5\times10^{-3}$ \\
\sts & $129$ & $-26.1$ & $0.03$ & $0.11$ & $5.7\times10^{-7}$ & $5.1\times10^{-7}$ & $2.0\times10^{-5}$ \\
\bottomrule
\end{tabular}
\caption{\textbf{Linear mixed-effects models of the regime dependence of each \phiidig atom in the coupled map lattices.} Wald $\chi^2$ statistic (4 degrees of freedom) of the regime fixed effect and its $p$-value, FDR-corrected across the $10$ atoms. $R^2_m$ and $R^2_c$ are the marginal (fixed effects) and conditional (fixed and random effects) coefficients of determination, and $\sigma^2_{\mathrm{sim}}$, $\sigma^2_{\mathrm{node}}$, and $\sigma^2_{\mathrm{res}}$ are the variances of the simulation and node random effects and of the residuals. Atoms are sorted by $R^2_m$.}
\label{tab:cml_lme}
\end{table}

\begin{table}[ht]
\centering
\footnotesize
\renewcommand{\arraystretch}{1.3}
\setlength{\tabcolsep}{2.5pt}
\begin{tabular}{lcccccccccc}
\toprule
Atom & PS & BMWD & STI & DT & FDT & Max & vs. & $Z$ & $r_{rb}$ & $p_{FDR}$ \\
\midrule
\rtr & $\mathbf{1.280\,(0.064)}$ & $0.495\,(0.044)$ & $0.706\,(0.093)$ & $-0.579\,(<0.001)$ & $-0.495\,(<0.001)$ & PS & BMWD & $5.40$ & $1.00$ & $7.6\times10^{-8}$ \\
\rti & $0.063\,(0.013)$ & $0.533\,(0.025)$ & $0.309\,(0.049)$ & $\mathbf{0.606\,(<0.001)}$ & $0.496\,(<0.001)$ & DT & BMWD & $4.86$ & $0.90$ & $1.2\times10^{-6}$ \\
\rts & $\mathbf{0.013\,(0.003)}$ & $-0.258\,(0.013)$ & $-0.143\,(0.018)$ & $-0.300\,(<0.001)$ & $-0.247\,(<0.001)$ & PS & BMWD & $5.40$ & $1.00$ & $6.8\times10^{-8}$ \\
\itr & $0.063\,(0.013)$ & $0.533\,(0.025)$ & $0.310\,(0.049)$ & $\mathbf{0.606\,(<0.001)}$ & $0.496\,(<0.001)$ & DT & BMWD & $4.86$ & $0.90$ & $1.2\times10^{-6}$ \\
\iti & $0.099\,(0.055)$ & $1.012\,(0.007)$ & $0.668\,(0.046)$ & $1.777\,(<0.001)$ & $\mathbf{1.851\,(<0.001)}$ & FDT & PS & $5.40$ & $1.00$ & $6.8\times10^{-8}$ \\
\itj & $\mathbf{0.017\,(0.005)}$ & $-0.533\,(0.025)$ & $-0.303\,(0.050)$ & $-0.605\,(<0.001)$ & $-0.496\,(<0.001)$ & PS & BMWD & $5.40$ & $1.00$ & $7.6\times10^{-8}$ \\
\its & $0.001\,(0.001)$ & $0.267\,(0.013)$ & $0.153\,(0.025)$ & $\mathbf{0.303\,(<0.001)}$ & $0.248\,(<0.001)$ & DT & BMWD & $4.86$ & $0.90$ & $1.2\times10^{-6}$ \\
\str & $\mathbf{0.013\,(0.003)}$ & $-0.258\,(0.013)$ & $-0.142\,(0.017)$ & $-0.300\,(<0.001)$ & $-0.247\,(<0.001)$ & PS & BMWD & $5.40$ & $1.00$ & $6.8\times10^{-8}$ \\
\sti & $0.001\,(0.001)$ & $0.266\,(0.013)$ & $0.154\,(0.025)$ & $\mathbf{0.303\,(<0.001)}$ & $0.248\,(<0.001)$ & DT & BMWD & $4.86$ & $0.90$ & $1.2\times10^{-6}$ \\
\sts ($\times10^{3}$) & $1.869\,(0.283)$ & $1.103\,(0.033)$ & $\mathbf{3.054\,(2.393)}$ & $0.538\,(0.001)$ & $0.423\,(<0.001)$ & STI & PS & $4.99$ & $0.93$ & $6.0\times10^{-7}$ \\
\bottomrule
\end{tabular}
\caption{\textbf{\phiidig atoms across the regimes of the coupled map lattices.} Mean (standard deviation) across the $20$ realisations of each regime of the atom averaged over all node pairs, with the largest value in bold. The last columns report the weakest of the four contrasts between the regime of maximal value and the other regimes (two-sided Mann--Whitney $U$ test, FDR-corrected across the $10$ regime pairs of each atom). The values of \sts are multiplied by $10^{3}$.}
\label{tab:cml_regimes}
\end{table}

\begin{table}[ht]
\centering
\footnotesize
\renewcommand{\arraystretch}{1.3}
\setlength{\tabcolsep}{2.5pt}
\begin{tabular}{lcccccccccc}
\toprule
Atom & PS--BMWD & PS--STI & PS--DT & PS--FDT & BMWD--STI & BMWD--DT & BMWD--FDT & STI--DT & STI--FDT & DT--FDT \\
\midrule
\rtr & $5.40$ & $5.40$ & $5.40$ & $5.40$ & $-4.86$ & $5.40$ & $5.40$ & $5.40$ & $5.40$ & $-5.40$ \\
\rti & $-5.40$ & $-5.40$ & $-5.40$ & $-5.40$ & $5.40$ & $-4.86$ & $5.40$ & $-5.40$ & $-5.40$ & $5.40$ \\
\rts & $5.40$ & $5.40$ & $5.40$ & $5.40$ & $-5.40$ & $5.40$ & $-5.40$ & $5.40$ & $5.40$ & $-5.40$ \\
\itr & $-5.40$ & $-5.40$ & $-5.40$ & $-5.40$ & $5.40$ & $-4.86$ & $5.40$ & $-5.40$ & $-5.40$ & $5.40$ \\
\iti & $-5.40$ & $-5.40$ & $-5.40$ & $-5.40$ & $5.40$ & $-5.40$ & $-5.40$ & $-5.40$ & $-5.40$ & $-5.40$ \\
\itj & $5.40$ & $5.40$ & $5.40$ & $5.40$ & $-5.40$ & $4.86$ & $-5.40$ & $5.40$ & $5.40$ & $-5.40$ \\
\its & $-5.40$ & $-5.40$ & $-5.40$ & $-5.40$ & $5.40$ & $-4.86$ & $5.40$ & $-5.40$ & $-5.40$ & $5.40$ \\
\str & $5.40$ & $5.40$ & $5.40$ & $5.40$ & $-5.40$ & $5.40$ & $-5.40$ & $5.40$ & $5.40$ & $-5.40$ \\
\sti & $-5.40$ & $-5.40$ & $-5.40$ & $-5.40$ & $5.40$ & $-4.86$ & $5.40$ & $-5.40$ & $-5.40$ & $5.40$ \\
\sts & $5.32$ & $-4.99$ & $5.40$ & $5.40$ & $-5.40$ & $5.40$ & $5.40$ & $5.40$ & $5.40$ & $5.40$ \\
\bottomrule
\end{tabular}
\caption{\textbf{All pairwise regime contrasts of the \phiidig atoms in the coupled map lattices.} $Z$ statistic of the two-sided Mann--Whitney $U$ test between each pair of regimes ($20$ realisations each), positive when the first regime has larger values. All contrasts survive FDR correction across the $10$ regime pairs of each atom ($p_{FDR}\leq1.2\cdot10^{-6}$).}
\label{tab:cml_pairwise}
\end{table}

\FloatBarrier

\subsection{Neural computation across mammals}

Table~\ref{tab:mammals} reports the comparison between the human and the mammalian connectomes shown in Fig.~\figsubref{fig:hopf}{d}, contrasting the $200$ individual human connectomes with the $114$ non-human species assigned to one of the $11$ taxonomic orders (see Methods in the main text). The deviation is the difference between the human and the mammalian means, in units of the mammalian standard deviation. For each atom, we also report the number of the $11$ taxonomic orders from which the human group differs at $p_{FDR}<0.05$ (all $66$ pairwise comparisons between the $12$ groups, FDR-corrected, as in Fig.~\ref{fig:all_mammals}). Afrotherians, cingulates, and eulipotyphlans are represented by a single species each, and cannot reach significance after correction. Table~\ref{tab:pc1} reports the loadings of each atom on the first principal component of the simulated human topography (Fig.~\figsubref{fig:hopf}{e}), and Table~\ref{tab:hopf_a} summarises the robustness of the human--mammal comparison to the bifurcation parameter (Fig.~\ref{fig:hopf_a}).

\begin{table}[ht]
\centering
\small
\renewcommand{\arraystretch}{1.3}
\setlength{\tabcolsep}{6pt}
\begin{tabular}{lccccccc}
\toprule
Atom & Human & Mammals & Deviation & $Z$ & $r_{rb}$ & $p_{FDR}$ & Orders \\
\midrule
\sts & $0.0109\,(0.0015)$ & $0.0078\,(0.0003)$ & $12.1$ & $14.69$ & $0.997$ & $8.6\times10^{-49}$ & 8/11 (8/11) \\
\iti & $0.8939\,(0.0029)$ & $0.9138\,(0.0017)$ & $-12.0$ & $-14.74$ & $-1.000$ & $4.8\times10^{-49}$ & 8/11 (8/11) \\
\its & $0.3624\,(0.0005)$ & $0.3594\,(0.0003)$ & $9.3$ & $14.74$ & $1.000$ & $4.8\times10^{-49}$ & 8/11 (8/11) \\
\sti & $0.3624\,(0.0005)$ & $0.3594\,(0.0003)$ & $9.3$ & $14.74$ & $1.000$ & $4.8\times10^{-49}$ & 8/11 (8/11) \\
\rts & $-0.3533\,(0.0005)$ & $-0.3468\,(0.0008)$ & $-7.8$ & $-14.74$ & $-1.000$ & $4.8\times10^{-49}$ & 8/11 (8/11) \\
\str & $-0.3533\,(0.0005)$ & $-0.3468\,(0.0008)$ & $-7.8$ & $-14.74$ & $-1.000$ & $4.8\times10^{-49}$ & 8/11 (8/11) \\
\rti & $0.7189\,(0.0006)$ & $0.7129\,(0.0008)$ & $7.5$ & $14.74$ & $1.000$ & $4.8\times10^{-49}$ & 8/11 (8/11) \\
\itr & $0.7189\,(0.0006)$ & $0.7129\,(0.0008)$ & $7.3$ & $14.74$ & $1.000$ & $4.8\times10^{-49}$ & 8/11 (8/11) \\
\itj & $-0.7189\,(0.0006)$ & $-0.7129\,(0.0008)$ & $-7.3$ & $-14.74$ & $-1.000$ & $4.8\times10^{-49}$ & 8/11 (8/11) \\
\rtr & $-0.7009\,(0.0019)$ & $-0.6990\,(0.0011)$ & $-1.7$ & $-9.13$ & $-0.619$ & $7.0\times10^{-20}$ & 5/11 (8/11) \\
\bottomrule
\end{tabular}
\caption{\textbf{\phiidig atoms of humans versus other mammals.} Mean (standard deviation) of each atom for the human and the mammalian connectomes, deviation of the human mean in units of mammalian standard deviations, and two-sided Mann--Whitney $U$ test between the two groups, FDR-corrected across the $10$ atoms. The last column gives the number of taxonomic orders from which humans differ significantly, for the raw atoms and, in parentheses, for the atoms normalised by the time-delayed mutual information. Atoms are sorted by the absolute deviation.}
\label{tab:mammals}
\end{table}

\begin{table}[ht]
\centering
\small
\renewcommand{\arraystretch}{1.3}
\setlength{\tabcolsep}{6pt}
\begin{tabular}{lc}
\toprule
Atom & $r$ with $PC_1$ \\
\midrule
\str & $0.997$ \\
\rts & $0.997$ \\
\itj & $0.990$ / $0.992$ \\
\iti & $0.956$ / $0.946$ \\
\rtr & $-0.784$ \\
\sts & $-0.985$ \\
\itr & $-0.990$ / $-0.992$ \\
\rti & $-0.992$ / $-0.990$ \\
\sti & $-0.999$ / $-0.998$ \\
\its & $-0.999$ / $-0.998$ \\
\bottomrule
\end{tabular}
\caption{\textbf{Loadings of the \phiidig atoms on the first principal component of the simulated human topography.} Pearson correlation between the z-scored regional map of each atom and the $PC_1$ map. For atoms with a counterpart under the exchange of the two regions, both values are reported (e.g.\ \xtx\,/\,\yty).}
\label{tab:pc1}
\end{table}

\begin{table}[ht]
\centering
\small
\renewcommand{\arraystretch}{1.3}
\setlength{\tabcolsep}{5pt}
\begin{tabular}{cccccc}
\toprule
 & \multicolumn{2}{c}{Profile $r$} & Atoms & \multicolumn{2}{c}{$r_{rb}$ of \sts} \\
\cmidrule(lr){2-3} \cmidrule(lr){5-6}
$a$ & Raw & Norm. & Raw (Norm.) & Raw & Norm. \\
\midrule
$-0.1$    & $0.54$ & $0.97$ & 5/10 (10/10)  & $0.242$ & $0.407$ \\
$-0.0875$ & $0.87$ & $0.99$ & 8/10 (10/10)  & $0.489$ & $0.578$ \\
$-0.075$  & $0.99$ & $0.99$ & 10/10 (10/10) & $0.641$ & $0.704$ \\
$-0.0625$ & $1.00$ & $0.99$ & 10/10 (10/10) & $0.815$ & $0.839$ \\
$-0.05$ & -- & -- & 10/10 (10/10) & $0.918$ & $0.925$ \\
$-0.0375$ & $0.99$ & $0.99$ & 10/10 (10/10) & $0.855$ & $0.865$ \\
$-0.025$  & $0.98$ & $0.99$ & 10/10 (10/10) & $0.905$ & $0.909$ \\
$-0.0125$ & $0.98$ & $1.00$ & 10/10 (10/10) & $0.862$ & $0.864$ \\
\bottomrule
\end{tabular}
\caption{\textbf{Robustness of the human--mammal comparison across bifurcation parameters.} For each value of $a$, Pearson correlation between the profile of the human deviations of the $10$ atoms and that of the main analysis, number of atoms that differ between humans and mammals at $p_{FDR}<0.05$ (two-sided Mann--Whitney $U$ test, FDR-corrected across the $10$ atoms) in the same direction as in the main analysis, and rank-biserial correlation of \sts, for raw and normalised atoms. The rank-biserial correlation ranges from $-1$ to $1$, with $\pm1$ indicating that all human connectomes lie above (below) all mammalian ones.}
\label{tab:hopf_a}
\end{table}

\FloatBarrier

\subsection{Financial markets}

Table~\ref{tab:finance_mod} reports the modularity of each network of information dynamics with respect to the sector partition (Fig.~\figsubref{fig:finance}{b}), compared against $1000$ strength- and degree-preserving null networks. With this number of null networks, $p_{\mathrm{perm}}=0.002$ is the smallest attainable two-sided $p$-value. Table~\ref{tab:finance_sectors} reports the sector comparisons for the unique-to-redundant (\itr) and redundant-to-unique (\rti) information of the financial sector, computed on the company-level values, i.e.\ the atoms of each company averaged over all its pairings with the other companies.

\begin{table}[ht]
\centering
\small
\renewcommand{\arraystretch}{1.3}
\setlength{\tabcolsep}{6pt}
\begin{tabular}{lcccc}
\toprule
Atom & $Q$ ($\times10^{3}$) & Null $Q$ ($\times10^{3}$) & $Z$ & $p_{\mathrm{perm}}$ \\
\midrule
\rtr & $1.49$ & $-2.12\,(0.10)$ & $36.29$ & $0.002$ \\
\rtx & $-5.01$ & $-2.96\,(1.01)$ & $-2.03$ & $0.054$ \\
\rty & $-3.69$ & $-2.82\,(1.00)$ & $-0.88$ & $0.398$ \\
\rts & $-1.49$ & $-2.13\,(0.09)$ & $6.97$ & $0.002$ \\
\xtr & $-8.73$ & $-2.71\,(0.97)$ & $-6.21$ & $0.002$ \\
\xtx & $-7.16$ & $-2.18\,(0.27)$ & $-18.78$ & $0.002$ \\
\xty & $-1.68$ & $-2.13\,(0.10)$ & $4.55$ & $0.002$ \\
\xts & $-4.98$ & $-2.34\,(0.96)$ & $-2.75$ & $0.004$ \\
\ytr & $-8.96$ & $-2.70\,(1.04)$ & $-6.04$ & $0.002$ \\
\ytx & $-1.62$ & $-2.12\,(0.09)$ & $5.62$ & $0.002$ \\
\yty & $-6.57$ & $-2.17\,(0.28)$ & $-15.95$ & $0.002$ \\
\yts & $-5.24$ & $-2.18\,(0.92)$ & $-3.34$ & $0.002$ \\
\str & $-2.05$ & $-2.13\,(0.08)$ & $0.99$ & $0.336$ \\
\stx & $-7.22$ & $-2.19\,(0.88)$ & $-5.73$ & $0.002$ \\
\sty & $-8.42$ & $-2.29\,(0.97)$ & $-6.29$ & $0.002$ \\
\sts & $26.49$ & $-2.14\,(3.10)$ & $9.23$ & $0.002$ \\
\midrule
$I(\mathbf{X}^t;\mathbf{X}^{t^\prime})$ & $-10.67$ & $-3.21\,(0.71)$ & $-10.58$ & $0.002$ \\
\bottomrule
\end{tabular}
\caption{\textbf{Modularity of the market networks of information dynamics.} Modularity $Q$ of each network with respect to the sector partition, mean (standard deviation) of $Q$ across $1000$ strength- and degree-preserving null networks, $Z$-score of the empirical value with respect to the null distribution, and two-sided permutation $p$-value.}
\label{tab:finance_mod}
\end{table}

\begin{table}[ht]
\centering
\small
\renewcommand{\arraystretch}{1.3}
\setlength{\tabcolsep}{6pt}
\begin{tabular}{lccccccc}
\toprule
 & & \multicolumn{3}{c}{\itr} & \multicolumn{3}{c}{\rti} \\
\cmidrule(lr){3-5}\cmidrule(lr){6-8}
Financials vs. & $n$ & $Z$ & $r_{rb}$ & $p_{FDR}$ & $Z$ & $r_{rb}$ & $p_{FDR}$ \\
\midrule
Information Technology & 53 & $1.98$ & $0.21$ & $0.114$ & $2.17$ & $0.23$ & $0.060$ \\
Communication Services & 16 & $4.19$ & $0.68$ & $3.0\times10^{-4}$ & $3.81$ & $0.62$ & $0.001$ \\
Industrials & 66 & $5.97$ & $0.61$ & $1.3\times10^{-7}$ & $6.75$ & $0.69$ & $7.9\times10^{-10}$ \\
Energy & 16 & $4.62$ & $0.75$ & $7.8\times10^{-5}$ & $4.84$ & $0.79$ & $1.4\times10^{-5}$ \\
Materials & 22 & $3.13$ & $0.45$ & $0.010$ & $3.61$ & $0.52$ & $0.002$ \\
Utilities & 29 & $1.86$ & $0.24$ & $0.124$ & $3.31$ & $0.43$ & $0.005$ \\
Health Care & 53 & $3.15$ & $0.34$ & $0.010$ & $4.89$ & $0.53$ & $1.4\times10^{-5}$ \\
Real Estate & 29 & $4.60$ & $0.60$ & $7.8\times10^{-5}$ & $5.56$ & $0.72$ & $7.4\times10^{-7}$ \\
Consumer Staples & 33 & $2.27$ & $0.28$ & $0.068$ & $2.78$ & $0.35$ & $0.018$ \\
Consumer Discretionary & 40 & $4.41$ & $0.52$ & $1.4\times10^{-4}$ & $4.85$ & $0.57$ & $1.4\times10^{-5}$ \\
\bottomrule
\end{tabular}
\caption{\textbf{Financial sector versus the other market sectors.} Two-sided Mann--Whitney $U$ tests between the company-level values of the financial sector ($n=64$ companies) and those of each other sector ($n$ companies), FDR-corrected across the $55$ sector pairs of each atom.}
\label{tab:finance_sectors}
\end{table}

\FloatBarrier

\section{Validation of \phiidig pipelines on empirical data}

\subsection{Estimating \phiidig on continuous data with Gaussian-copula} \label{app:est:copula}

The Gaussian implementation of Sec.~\ref{app:sec:gaussian} makes \phiidig available in closed form in the case of jointly Gaussian variables. Concretely, for a pair of variables $X_i,X_j$ and a prediction horizon $\tau=\tp-t$, it assumes that the lag-stacked vector
\begin{equation} \label{eq:app:est:fourvector}
    \mathbf{X} \;=\; \big( X_i^t,\, X_j^t,\, X_i^{t^\prime},\, X_j^{\tp} \big)^\top
\end{equation}
is drawn from a distribution of the form $\sim\mathcal{N}(0,\Sigma)$, so that all dependencies in the systems are captured by the $4\times4$ covariance matrix $\Sigma$. 
Neither of the two continuous datasets analysed in the main text -- the whole-brain model and the financial markets -- satisfies this assumption exactly. In this section, we hence describe how this limitation can be directly resolved using a Gaussian-copula approach, reporting the diagnostics that justify doing so in each case.

The key observation is that what the Gaussian solver actually requires is not that the data is Gaussian, but that their dependence structure is. We therefore replace each raw variable by its normal scores: for a series $x$ of length $T$,
\begin{equation} \label{eq:app:est:gaussianize}
    u_t = \frac{\mathrm{rank}(x_t) - 1/2}{T} \,, \qquad z_t = \Phi^{-1}(u_t) \,,
\end{equation}
with $\Phi$ the standard normal CDF and ties resolved by average rank. This amounts to adopting a \emph{Gaussian-copula} model of the system. Since Eq.~\eqref{eq:app:est:gaussianize} is a monotone, per-variable transformation, it leaves the copula, i.e.\ the entire dependence structure, unchanged, and discards only the shape of the marginals, which is precisely the part that invalidates the Gaussian parametrisation in the first place.

Importantly, this leaves every quantity entering \phiidig unaffected: writing $T$ for the componentwise map of Eq.~\eqref{eq:app:est:gaussianize} and $\mathbf{Z}=T(\mathbf{X})$, this follows from three key properties. First, the KL divergence is invariant under any diffeomorphism applied to both of its arguments, as the Jacobians cancel between the log-ratio and the measure
\begin{equation}
    \D(P_{\mathbf{Z}}\|Q_{\mathbf{Z}}) = \int p_{\mathbf{X}}(\mathbf{x})\,
    \log\frac{p_{\mathbf{X}}(\mathbf{x})/|J|}{q_{\mathbf{X}}(\mathbf{x})/|J|}\;d\mathbf{x}
    = \D(P_{\mathbf{X}}\|Q_{\mathbf{X}}) \,.
\end{equation}
Second, because $T$ acts on each variable separately, it preserves every conditional-independence statement, and therefore maps each submanifold of Sec.~\ref{app:sec:4var} onto its counterpart in $\mathbf{Z}$-coordinates. Thus, since the projections of Eq.~\eqref{app:eq:PijProj} minimise an invariant objective over a set that maps bijectively, they are equivariant. 
Third, since densities transform as $p_{\mathbf{Z}} = p_{\mathbf{X}}/|J|$, the $e$-geodesics are equivariant too:
\begin{equation} \label{eq:app:est:jacobian}
    p^t_{1,\,\mathbf{Z}}\,p^{1-t}_{2,\,\mathbf{Z}} = \frac{p^t_{1,\,\mathbf{X}}\,p^{1-t}_{2,\,\mathbf{X}}}{|J|^{\,t+(1-t)}} \,,
\end{equation}
and the single residual power of $|J|$ is exactly what the change of measure absorbs, leaving the normalising constant $\mathcal{Z}$ and the whole interpolating family unchanged. We remark that this last step relies on the interpolation weights summing to one, as imposed throughout Sec.~\ref{app:sec:general_nm}. Every synergy of \defs, and hence every \phiid atom, is therefore a functional of the copula alone.
As a consequence of these invariant properties, using a Gaussian copula for the closed-form solver of Sec.~\ref{app:sec:gaussian} is not an approximation of \phiidig for the original non-Gaussian data, but an exact evaluation of it. The diagnostics reported below, which probe the joint rather than the marginals, test precisely this assumption. 
Importantly, we remark that Eq.~\eqref{eq:app:est:gaussianize} is applied once per variable, and hence assumes stationary marginals, so that $X_i^t$ and $X_i^{\tp}$ are standardised by the same transform. %

Therefore, given the transformed vector of Eq.~\eqref{eq:app:est:gaussianize}, the $4\times4$ correlation matrix $\Sigma$ is estimated from the transformed variables $z$ and passed to the Gaussian \phiidig solver of Sec.~\ref{app:sec:gauss_solve}. We consider two such estimators: the \emph{normal-scores Pearson} matrix, i.e.\ the plain Pearson correlation of $z$, and the \emph{Spearman rank correlation}.

Throughout this validation, we check the goodness of these parametric estimates against non-parametric Kraskov--St\"ogbauer--Grassberger (KSG) nearest-neighbour mutual-information estimator~\cite{kraskov2004estimating}, which we consider as ground truth.  

\begin{figure}[t]
    \centering
    \includegraphics[width=\linewidth]{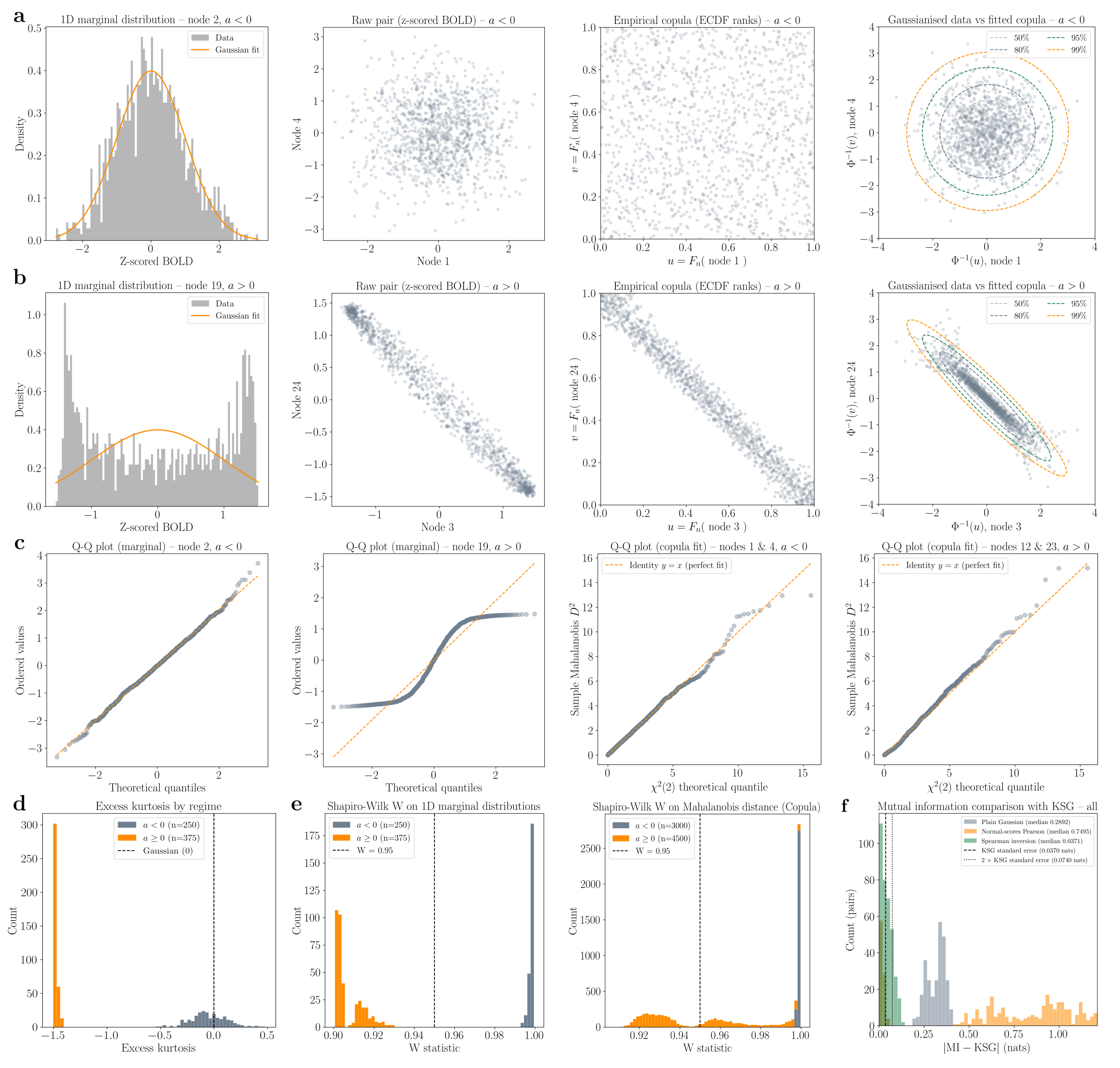}
    \caption{\textbf{Validation of Gaussian-copula construction on Hopf model.} \textbf{a} From left to right, for the regime $a<0$: 1D marginal distribution for a randomly selected node, 2D joint distribution between two randomly selected nodes, along with their copula CDF and the transformed original variables. \textbf{b} Same as \textbf{a}, but for the $a>0$ regime. 
    \textbf{c} Q-Q plots for 1D and 2D copula distributions for a sample node and pair of nodes, respectively, for both $a<0$ and $a\geq0$. \textbf{d} Distribution of the excess kurtosis of the marginals, with the bars colour-coded by regime ($a<0$ versus $a\geq0$) to expose the bimodality. The dashed line marks the Gaussian value of $0$.
    \textbf{e} Shapiro-Wilk distributions on the 1D marginals and the 2D joints after the copula transform, split by regime in the same way. \textbf{f} Comparison between KSG, Gaussian copula with both normal-scores Pearson correlation and Spearman correlation, and plain-Gaussian pairwise mutual information, as the absolute deviation from the KSG reference. The dashed and dotted lines mark one and two KSG block standard errors.}
    \label{fig:hopf_validation}
\end{figure}

\subsubsection{Whole-brain model} \label{app:est:hopf}

The simulations analysed here use the coupled Stuart--Landau system reported in the Refs.~\cite{hopf1942abzweigung, andronov2013theory} on a $25$-node small-world network, sweeping the global coupling $G\in[0,4]$ and the bifurcation parameter $a\in[-0.25,0.25]$ on a $100\times100$ grid. Additional parameters include a step size of $dt=1$, $N=10^3$ equilibration steps, and an additional $1200$ steps per trial. \phiidig is then computed at $\tau=1$ for all $300$ node pairs, and the quantities reported are means over pairs.

Interestingly, the need for the copula transform is confined to one side of the bifurcation, following directly from the nature of the local dynamics. For $a<0$, the system is a damped noise-driven fixed point, akin to an Ornstein--Uhlenbeck process, which is itself a linear Gaussian process (Fig.~\figsubref{fig:hopf_validation}{a}). Accordingly, the raw signals are already essentially Gaussian, with a median 1D marginal Shapiro--Wilk (SW) statistic of $W=0.9984\pm0.0010$
(Fig.~\figsubref{fig:hopf_validation}{e}). This reflects how close to Gaussianity ($W=1$) single variables are. On the other hand, for $a\geq0$, where the dynamics become a pure limit-cycle oscillation, the SW values decrease (median $W=0.9034\pm0.0068$), showing a median excess kurtosis of $-1.49\pm0.02$, against $-0.06\pm0.17$ for $a<0$ (Fig.~\figsubref{fig:hopf_validation}{d}). This thin-tailed distribution is exactly the signature one expects of a limit cycle, in which the trajectories in phase space often dwell near the turning points of the oscillation. %

In this post-bifurcation regime, the oscillators become strongly synchronised, with pairwise copula correlations approaching unity (Fig.~\figsubref{fig:hopf_validation}{b}). In this scenario, even small perturbations of the correlations largely affect the quantities estimated. For instance, even in the simplest case of a mutual information between two variables with correlation $\rho$, we have that ${I = -\tfrac12\log(1-\rho^2)}$, which hence becomes extremely steep as $|\rho|\to1$.
Thus, in this regime, the choice of correlation estimator becomes critical. To settle it, we compared three estimators against the KSG reference over $358$ pairs spanning $a\in\{-0.12,0,0.12,0.25\}$ and $G\in\{2,3,4\}$, calculating the mutual information between pairs of variables. Such estimators consist of plain Gaussian correlation, normal-score Pearson correlation, and Spearman correlation.

Results show that the rank-based estimators are markedly superior. Restricting to the post-bifurcation pairs ($a\geq0$, $n=268$), where the near-unit correlations make the comparison meaningful, the normal-scores Pearson estimator returns a median absolute deviation from KSG of $0.927$ nats and the plain Gaussian fit $0.331$ nats, whereas the Spearman inversion yields deviations of only $0.049$ nats. Pooled over the full set of sampled pairs, i.e.\ also including the $a<0$ regime, the corresponding medians decrease to $0.750$, $0.289$ and $0.037$ nats (Fig.~\figsubref{fig:hopf_validation}{f}).
Given the better accuracy of the rank-based estimator, we therefore adopt the Spearman inversion throughout the Hopf sweep. 
We finally note that for $a<0$, where the correlations are far from unity, these distinctions vanish, with all three estimators agreeing with the KSG reference to within $0.014$ nats. 

Finally, the Q-Q plots of Fig.~\figsubref{fig:hopf_validation}{c} make the same point graphically. For $a<0$, the quantiles of the squared Mahalanobis distances follow the $\chi^2_2$ reference line over their entire range, in both the raw and the transformed variables, consistent with the near-unit SW statistics reported above. For $a\geq0$ the agreement is good through the bulk of the distribution and departs from the identity only in the upper tail, where the finite number of extreme quantiles makes the comparison least reliable. The Gaussian-copula parametrisation therefore captures the dependence structure over the range that carries the bulk of the probability mass, with the residual misfit confined to the tail.

\FloatBarrier

\begin{figure}[t]
    \centering
    \includegraphics[width=\linewidth]{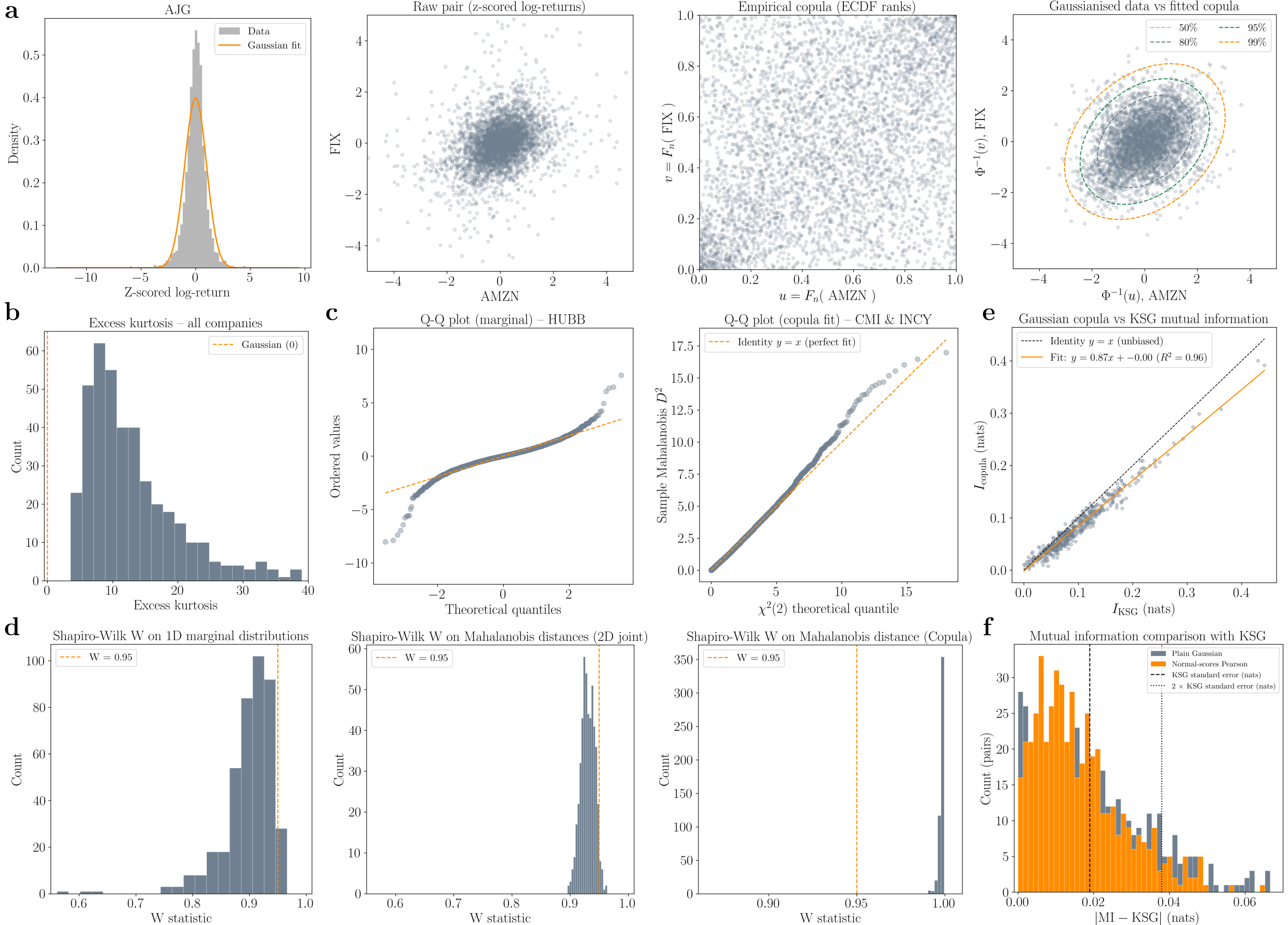}
    \caption{\textbf{Validation of Gaussian-copula construction on market data.} \textbf{a} From left to right: 1D marginal distribution for a randomly selected company, 2D joint distribution between two randomly selected companies, along with their copula CDF and the transformed original variables. \textbf{b} Kurtosis distribution across all companies. \textbf{c} Q-Q plots for 1D and 2D copula distributions for a sample company and pair of companies, respectively. \textbf{d} Shapiro-Wilk distributions on 1D marginal, 2D joint, and 2D copula distributions. \textbf{e} Relation between KSG and Gaussian-copula pairwise mutual information. \textbf{f} Comparison between KSG, Gaussian-copula, and plain-Gaussian pairwise mutual information.}
    \label{fig:finance_validation}
\end{figure}

\subsubsection{Financial markets} \label{app:est:finance}

The financial data were obtained from the historical records of Yahoo Finance~\cite{fyahoo}. The dataset consists of daily adjusted closing prices for the $421$ S\&P 500 constituents with complete coverage over 2010--2026, converted to log returns ${r_t = \log p_t - \log p_{t-1}}$ ($4028$ samples) and standardised per company. \phiidig is computed for all $88\,410$ company pairs at prediction horizons $\tau=1,\dots,19$.

Unlike the Hopf case, the pairwise correlations here are far from unity, and the Spearman correlation and normal-scores Pearson estimator provide comparable results. Thus, we use the latter for the rest of the analyses.
In this scenario, the departure from Gaussianity runs in the opposite direction to the Hopf case: the marginals are strongly heavy-tailed, with a median excess kurtosis of $11.6$ (Fig.~\figsubref{fig:finance_validation}{b}) and a median marginal and joint Shapiro--Wilk statistic of $W=0.907\pm0.046$ and $W=0.930\pm0.012$, respectively (Fig.~\figsubref{fig:finance_validation}{d}). However, after the copula transformation, the dependence structure is very well described by a Gaussian copula, with the Mahalanobis probability-integral-transform statistic rising to $W=0.9983\pm0.0013$ (Fig.~\figsubref{fig:finance_validation}{d}), and a clear linear relation appearing in the Q-Q plot (Fig.~\figsubref{fig:finance_validation}{c}). 

We additionally validate the copula construction by calculating the mutual information between $500$ randomly selected pairs, comparing it with the non-parametric KSG estimator introduced above. The median absolute deviation from KSG is $0.0135$ nats for the copula estimate, with a correlation between copula and KSG mutual information across pairs of $R^2=0.96$ (Fig.~\figsubref{fig:finance_validation}{e}). This finding indicates that, even without considering the KSG estimation errors, the only discrepancy between the copula approach and the KSG technique is supplied by a systematic error, which therefore cancels out when we compare our quantities across scenarios.
We now also consider the reference's own sampling error, which we estimate as the sampling error of each KSG value by splitting every pair into ten contiguous blocks, recomputing the estimate on each and scaling the resulting spread to the full series length. This yields a median standard error of $0.0188$ nats. The copula--KSG discrepancy of $0.0135$ nats is therefore of the same order as the reference's own bias against known truth, and smaller than its sampling error: $68.0\%$ of pairs fall within one standard error of the KSG estimate and $96.8\%$ within two. Most of the residual disagreement is thus attributable to the reference rather than to the copula approach. 

Finally, a comparison between the Gaussian-copula approach and the plain Gaussian shows that the former provides a more accurate description of the data (Fig.~\figsubref{fig:finance_validation}{f}). The plain-Gaussian median absolute deviation from KSG is $0.0156\pm0.0151$ nats, against $0.0135\pm0.0114$ nats for the copula estimate, which is the closer of the two for $60.6\%$ of pairs. The improvement is therefore systematic but modest in absolute terms, as one would expect given that the transform corrects the marginals, which are severely non-Gaussian here, while leaving the dependence structure, which is already well described by a Gaussian copula, essentially untouched.

\FloatBarrier

\begin{figure}[t]
    \centering
    \includegraphics[width=\linewidth]{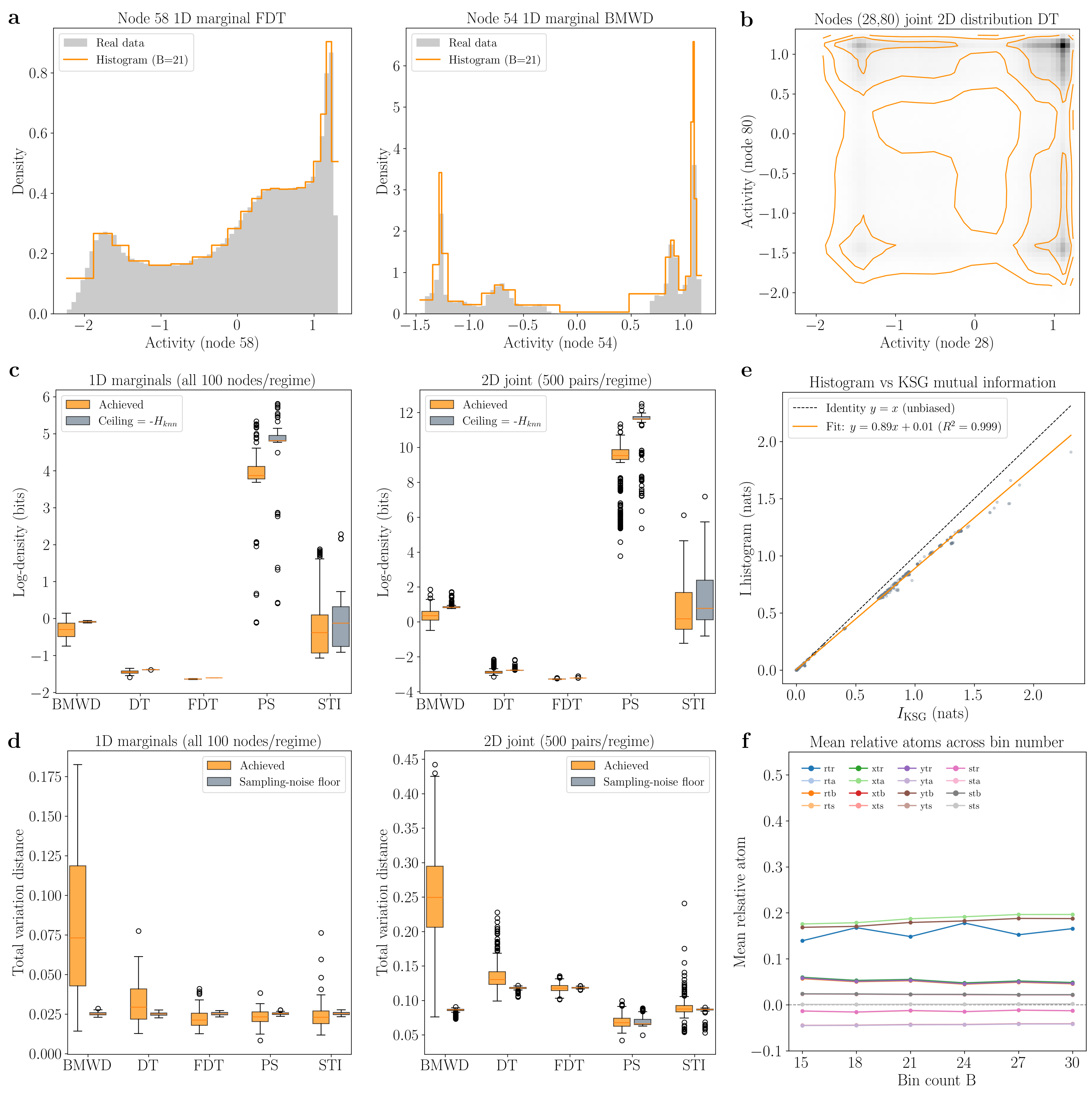}
    \caption{\textbf{Validation of the nonparametric discrete \phiidig construction on coupled map lattices.} \textbf{a} Two representative 1D marginals (node $58$ in FDT, node $54$ in BMWD), showing the empirical density (grey) against the quantile-binned histogram model at $B=21$ (orange). \textbf{b} A representative 2D joint distribution (nodes $28$ and $80$ in DT), with the empirical density in greyscale and the iso-density contours of the histogram model overlaid at the $50$th, $80$th and $95$th percentiles of the non-zero reference density. \textbf{c} Held-out log-density of the fitted histogram (orange) against the information-theoretic ceiling $-H(X)$ estimated by a $k$-nearest-neighbour entropy (grey), for the 1D marginals of all $100$ nodes per regime (left) and for $500$ node pairs per regime (right). \textbf{d} Held-out total variation distance of the same fits (orange) against the sampling-noise floor that a perfectly specified model would still incur on a finite test sample (grey). \textbf{e} Mutual information from the discretised distribution at $B=21$ against the KSG reference, over all $2500$ sampled pairs, with the identity (dashed) and a linear fit (orange). \textbf{f} Mean normalised atom for each of the sixteen \phiid atoms as a function of the bin count $B$, over $40$ sample pairs per regime.}
    \label{fig:cmls_validation}
\end{figure}

\subsection{Nonparametric estimation of \phiidig on coupled map lattices} \label{app:est:cml}

Coupled map lattices exhibit strongly chaotic and erratic behaviour, with their stationary marginal distributions being frequently sharply peaked, bounded, or bimodal near the turning points of the local map. For these reasons, a Gaussian approach, even with a copula, is fundamentally inappropriate for their description. As a consequence, we investigate these systems by applying a discrete nonparametric estimate of \phiidig, directly feeding a per-pair histogram discretisation into the discrete \phiid solver.  
In this section, we validate such an approach, employing again KSG estimators as the reference.

For a pair of nodes $(X_i,X_j)$, we discretise the lag-stacked vector of Eq.~\eqref{eq:app:est:fourvector} onto a $B^4$ grid using quantile bins, i.e.\ bins whose edges are placed at the empirical quantiles of each axis, so that every bin along a given axis does not span the same width but receives approximately the same number of samples. This adapts the resolution to the probability density, allocating more bins where the sharply peaked marginals of the local map concentrate their mass. The resulting empirical distribution is then passed to the discrete \phiid solver. 
Thus, here the only free parameter is the number of bins $B$. This is bounded from above by the $B^4$ computational cost of the discrete distribution and by its sample requirement: at a simulated series length $T$, each cell of the table, i.e.\ each of the $B^4$ hypercubes into which the four-dimensional space is partitioned, holds on average $T/B^4$ samples, and this occupancy must remain large enough for the table not to be dominated by sampling noise. With $T=2\cdot10^{6}$, we set $B=21$, corresponding to $\approx10.3$ samples per cell, providing a balance between cell occupancy and fitness accuracy. We validate this choice below.

We first ask how well the resulting histogram describes the data (Figs.~\figsubref{fig:cmls_validation}{a-b}). Held out on a tail slice never used for fitting, the histogram's log-density and total variation distance both approach the limits that a perfectly correct model would attain, i.e.\, the information-theoretic ceiling $-H(X)$ (with $H$ from KSG) (Fig.~\figsubref{fig:cmls_validation}{c}) and the sampling-noise floor set by the finite test size (Fig.~\figsubref{fig:cmls_validation}{d}), with Brownian motion with defects (BMWD) behaving the worst across regimes. %

Turning to the information-theoretic quantities themselves, we consider the mutual information derived from the discretised distribution at $B=21$. Across all $2500$ sampled pairs, it tracks the KSG reference closely, with a Spearman correlation of $\rho=0.957$ against KSG and a linear fit of slope $0.885$ and $R^2=0.999$ (Fig.~\figsubref{fig:cmls_validation}{e}). As in the finance case, what the discretisation introduces is therefore a small and systematic rescaling rather than noise, with the ranking of pairs and of regimes preserved almost exactly.

To check whether this stability extends to the full \phiidig decomposition, and not only to pairwise mutual information, we repeated the discrete solve at every $B\in\{15,18,21,24,27,30\}$ (corresponding to $\approx40$ down to $\approx2.5$ samples per cell) for specific samples of $40$ pairs per regime: $10$ pairs with the largest mutual information, $10$ with the smallest, and a further $20$ drawn at random. We stratify it in this way because bias and instability behave differently for near-independent pairs than for appreciably coupled ones, so a purely random sample would conflate the two. We observe that the atoms remain robust across the choice of $B$ (Fig.~\figsubref{fig:cmls_validation}{f}). Expressed as relative contributions to the decomposition, i.e.\ normalising each atom w.r.t.\ the total time-delayed mutual information, the mean value of every atom drifts by at most $0.039$ over the whole range $B\in\{15,\dots,30\}$, with a median drift across the sixteen atoms of only $0.004$.
In absolute terms, the same comparison gives a median change of $0.03$--$0.07$ bits per atom with respect to $B=21$, with no systematic trend in $B$. 
Hence, although raising $B$ would in principle reduce the discretisation bias, the occupancy $T/B^4$ would fall as the fourth power, so the table filled by a fixed number of samples becomes progressively noisier. At $T=2\times10^{6}$, moving from $B=21$ to $B=30$ takes the mean occupancy from $\approx10$ to $\approx2.5$ samples per cell, and the sampling noise this adds outweighs the bias it removes.
Thus, the binding constraint at fixed $T$ is statistical as opposed to computational. Although the $B^4$ cost of the solve is affordable well beyond $B=30$, the occupancy of the table is not. In fact, recovering at $B=30$ the same $\approx10$ samples per cell that $B=21$ already achieves would require a series roughly four times longer, and a further doubling of $B$ would demand a sixteen-fold increase, which is impractical for the long-run simulations used here. Since the atoms are in any case stable across the range of $B$ that the present $T$ supports, we adopt $B=21$ throughout, with the evidence above indicating that a larger $B$ would not practically change the results.

\subsubsection{Effect of the observation noise} \label{app:est:cml_noise}

As mentioned in the main manuscript, the coupled-map trajectories are analysed with a small amount of additive Gaussian noise with standard deviation $\sigma=10^{-3}$. This prevents the near-degenerate distributions of the more ordered regimes from making the discretisation ill-defined. In the following, noise amplitudes are quoted relative to $\sigma_x$, the standard deviation of the noise-free lattice activity pooled over all $100$ nodes and all time steps. For each regime, this takes the values $\sigma_x=0.497$ (BMWD), $0.489$ (DT), $0.495$ (FDT), $0.476$ (PS) and $0.452$ (STI). Therefore, the noise used here is $0.201\%$, $0.205\%$, $0.202\%$, $0.210\%$ and $0.221\%$ of $\sigma_x$, respectively, i.e.\ between two and three orders of magnitude below the scale of the activity it is added to. Nonetheless, it plays an important role for the PS regime. Here, with the production bin count $B=21$ and $T=2\times10^{6}$ samples per node, the noise-free PS activity is close to degenerate, and the $22$ quantile bin edges of the histogram estimator collapse to a median of $4$ distinct values, with repeated edges at all $100$ nodes and a minimum edge separation of exactly zero. With the noise, all $22$ edges are distinct at every node, enabling a more robust and numerically stable characterisation of its statistical activity. In the other four regimes, the edges are already well separated with or without it.

\begin{figure}[t]
    \centering
    \includegraphics[width=\linewidth]{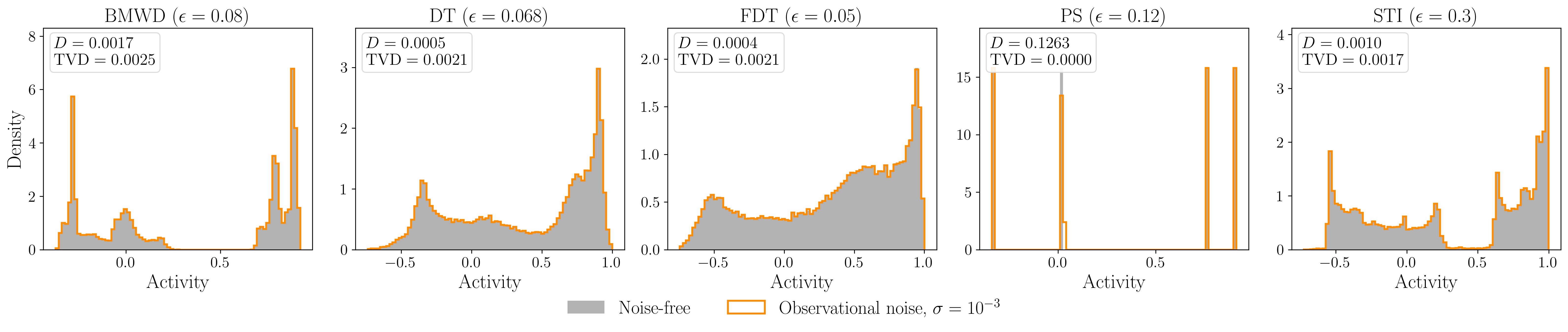}
    \caption{\textbf{The observation noise does not change the recorded activity distribution.} Activity distribution of a representative node in each of the five coupled-map regimes, from noise-free simulation (grey) and with the additive observation noise of $\sigma=10^{-3}$ used throughout (orange), both from the same initial condition and the same $5\times10^{4}$ time steps. Insets report the median over nodes of the Kolmogorov--Smirnov statistic $D$ and of the total variation distance between the two.}
    \label{fig:cmls_noise}
\end{figure}

Because the maps are in principle deterministic, we then verify explicitly that this noise does not alter what the system does. We remark once again that it is observational, i.e.\ it is added to the stored trajectory and never re-enters the map, so the dynamics themselves are untouched by construction, and the only question is whether the recorded activity is measurably different. Running each regime three times from the same initial condition, both noise-free and with observational noise, we compare the resulting activity distributions node by node. 
With the noise employed, the change is smaller than the baseline variability in every regime (Fig.~\ref{fig:cmls_noise}). Median over nodes, the noise-free and noisy activity distributions differ by a Kolmogorov--Smirnov statistic of $D=0.0004$--$0.0017$ and a total variation distance of $0.0017$--$0.0025$, whereas the split-half baseline of the noise-free run alone gives $D=0.003$--$0.019$. In other words, this indicates that the first half of a trajectory differs from its second half in the same way as when the observational noise is added, as supported by a two-sample test ($p\approx1$). After the addition of noise, the first four moments are likewise unchanged to five decimal places.

Taken together, these checks support treating the discrete-histogram \phiid atoms as meaningful, even though the histogram is only an approximation of the underlying distributions. Importantly, we remark that although the discretisation does not provide exactly the KSG estimates, it introduces a systematic linear rescaling ($R^2=0.999$). This means that comparisons between pairs and regimes, i.e.\ how we performed the analyses in the main text, remain reliable even where the absolute value has a slight bias.

\end{document}